\documentclass[pra,superscriptaddress,twocolumn]{revtex4}
\UseRawInputEncoding
\usepackage{amsmath}
\usepackage{amsfonts}
\usepackage{graphicx}
\usepackage{appendix}
\usepackage{dsfont}
\usepackage{amssymb}
\usepackage{bm}
\usepackage{color}
\usepackage{ulem}
\usepackage{soul}
\usepackage{natbib}
\usepackage{nccmath}
\usepackage{array}
\newcommand{\beq}{\begin{equation}}
\newcommand{\eneq}{\end{equation}}
\newcommand{\be}{\begin{equation}}
\newcommand{\eeq}{\end{equation}}
\newcommand{\bea}{\begin{eqnarray}}
\newcommand{\eeqa}{\end{eqnarray}}

\usepackage{multirow}
\usepackage{wasysym}
\usepackage{hyperref}

\hypersetup{
colorlinks=true,
citecolor=blue,
linkcolor=blue,
urlcolor=blue,
}


\begin{document}
\title{Interplay  between singlet and triplet pairings in multi-band two-dimensional oxide superconductors}

\author{L. Lepori}
\email[correspondence at: ]{llepori81@gmail.com}
\affiliation{Dipartimento di Fisica, Universit\'a della Calabria, Arcavacata di Rende I-87036, Cosenza, Italy.}
\affiliation{I.N.F.N., Gruppo collegato di Cosenza, Arcavacata di Rende I-87036, Cosenza, Italy.}

\author{D. Giuliano}
\affiliation{Dipartimento di Fisica, Universit\'a della Calabria, Arcavacata di Rende I-87036, Cosenza, Italy.}
\affiliation{I.N.F.N., Gruppo collegato di Cosenza, Arcavacata di Rende I-87036, Cosenza, Italy.}

\author{A. Nava}
\affiliation{Dipartimento di Fisica, Universit\'a della Calabria, Arcavacata di Rende I-87036, Cosenza, Italy.}
\affiliation{I.N.F.N., Gruppo collegato di Cosenza, Arcavacata di Rende I-87036, Cosenza, Italy.}

\author{C.  A. Perroni}
\affiliation{Dipartmento di Fisica "Ettore  Pancini'', Universit\`a  degli  Studi  di  Napoli  Federico  II, Complesso 
Universitario  Monte  S.  Angelo,  Via  Cintia,  I-80126  Napoli,  Italy.}
\affiliation{CNR-SPIN  c/o Universit\`a degli  Studi  di  Napoli  Federico  II, Complesso  Universitario  
Monte  S.  Angelo,  Via  Cintia,  I-80126  Napoli,  Italy.}

\begin{abstract}

We theoretically study the
superconducting properties of multi-band two-dimensional transition metal oxide superconductors by  
analyzing not only the role played by conventional singlet pairings, but also by the triplet order parameters, 
favored by the spin-orbit couplings present in these materials.  In particular, we focus 
on the two-dimensional electron gas at the (001) 
interface between $LaAlO_3$ and $SrTiO_3$ band insulators where the low electron densities and the sizeable spin-orbit couplings 
affect the superconducting features. Our theoretical study is based on  an extended superconducting mean-field analysis 
of the typical multi-band tight-binding Hamiltonian, as well as on a parallel analysis of the effective electronic bands 
in the low-momentum limit, including static on-site and inter-site intra-band attractive potentials 
under applied magnetic fields. The presence of triplet pairings is able to strongly reduce the singlet order parameters
which, as a result, are no longer a monotonic  function of the charge density. The interplay between the singlet and the triplet
pairings affects the dispersion of quasi-particle excitations in the Brillouin zone and also induces anisotropy in the 
superconducting behavior under the action of an in-plane and of an 
out-of-plane magnetic fields. Finally, non-trivial topological superconducting 
states become stable as a function of the charge density,  as well as of the magnitude and of the 
orientation of the magnetic field. 
In addition to the chiral, time-reversal breaking, topological superconducting phase, favored by the linear Rashba couplings 
and by the on-site attractive potentials in the presence of an out-of-plane magnetic field, we find that a time-reversal invariant
topological helical superconducting phase is promoted by  not-linear spin-orbit couplings and by the inter-site attractive 
interactions in the absence of magnetic field.

\end{abstract}

\maketitle

\section{Introduction}

The transition metal oxides represent a large class of materials with functional properties not only 
in the bulk but also in hetero-structures and nano-structures. In particular, the two-dimensional electron gas (2DEG)   
at the (001) interface between $LaAlO_3$ (LAO) and $SrTiO_3$ (STO) band insulators has gained a continuously growing 
interest in recent years, as an ideal playground to investigate the interplay between magnetism,  superconductivity, 
and spin-orbit coupling. Indeed, the 2DEG  hosts a complex phase diagram, depending on the electron density
\cite{Caviglia2,biscaras2}, on the temperature and {\color{black} on the applied  magnetic field}.   

It is well known that LAO/STO 2DEGs host an unconventional superconducting regime, achievable by tuning the applied gate voltage.  
The origin of the superconductivity is still not understood \cite{Gorkov2016,Edge2015,Ruhman2016,gariglio2016,tafuri2017}, and various 
open questions remain unsolved about  the role of quantum {\color{black} electronic correlations} \cite{maniv2015strong,Monteiro2019}, 
{\color{black} of} multiband effects \cite{Trevisan2018,Wojcik2020,zeg2020,jouan2021}, and of the spin-orbit coupling 
\cite{Khalsa2013,Diez2015,Zhong2013,Shalom2010,Rout2017}. Moreover, the possibility of topological
superconductivity is also under debate \cite{Scheurer,Mohanta,Loder,Fukaya,settino2020,santamaria2021,tafuri2017,perroni2019}.

Even more interestingly, the superconducting critical temperature, $T_c$, in the LAO/STO 2DEG exhibits a dome-shape behavior, 
as a function of the applied gate voltage \cite{Reyren,Rout2017,joshua2012universal,maniv2015strong,biscaras2,biscaras2010two,Caviglia2}. 
When the carrier density increases,  $T_c$ {\color{black} first} increases up to a maximum value, $T_c^\mathrm{max}$ $\simeq$ 300 mK, at an optimal effettive doping, 
then it starts to decrease. The resulting  phase diagram is qualitatively very similar to that of high $T_c$ cuprates,  of organic superconductors, 
of Fe-based superconductors, as well as of heavy fermions \cite{taillefer,keimer}. 
Recently, the shape of the superconducting dome has been qualitatively explained by assuming 
{\color{black} a  particular real-space potential, effectively attractive in suitable windows of momentum space, and resulting into an extended s-wave pairing \cite{zeg2020}. 
Moreover, a forthcoming insightful work 
\cite{paramekanti2020} showed that the formation of a similar dome (or even many of them, when multiband fermionic models are  considered) 
is related generically to an attractive potential with finite range.} The same work shed light on previous works, where the dome has been suggested 
instead as an effect of the spin-orbit coupling \cite{Shalom2010, Rout2017, Yin2019,singh2018gap}. Furthermore, an asymmetric
response in shear-resistivity to an applied magnetic field (in-plane or out-of-plane) has been observed \cite{caviglia2009}: a similar 
asymmetry can suggest a possible spatial asymmetry in the pairing. 

To {\color{black} address the open questions above, in this paper we discuss the superconductivity in 
LAO/STO 2DEG, by making a singlet-triplet mixed ansatz for the pairing and by studying its physical consequences}.  
This possibility looks  pretty natural, due to the inversion symmetry breaking term of the 
heterostructure  which gives rise to an effective Rashba-like coupling, already known to favour mixed pairings \cite{rashba2001,tafuri2017,alidoust2021}. Related notable effects are qualitative deviations of the standard BCS/BEC crossover \cite{pieri2019}. The same 
possibility has been corroborated quite recently, using a Monte-Carlo approach on a square lattice \cite{rosenberg2017}: there, even a 
local (Hubbard) interaction proved sufficient for singlet-triplet mixing, provided that a Rashba coupling is added. Interestingly, 
a singlet-triplet mixed pairing allows (but does not imply) edge excitations, protected by a nontrivial topology,  whose presence 
has not been ruled out so far by current transport experiments. Moreover, it determines an asymmetric response to an applied magnetic field,
qualitatively similar to that observed in  \cite{caviglia2009}. We finally notice that, {\color{black} while a pure triplet p-wave pairing  
is ruled out by previous experiments which did not detect the expected nodes in the superconducting gap \cite{hwang2018,sumita2020}, 
instead  a singlet-triplet mixed pairing would not contradict the experimental results.}

In {\color{black} this paper, we employ}  a tight-binding model including the low-energy electronic structure of {\color{black} the LAO/STO 2DEG} 
with the ${d_{xy},d_{xz},d_{yz}}$  orbitals of the Ti atoms \cite{popovic,delugas,scopigno,salluzzo,Khalsa2013}. Various papers have pointed out the close relation between 
{\color{black} the onset of the} superconductivity
and the filling of the degenerate $d_{xz/yz}$ sub-bands, at an high density of states \cite{valentinis,singh}.
Then, we adopt an attractive static potential, with both local and nearest-neighbour terms, able to host all the pairing  configurations mentioned above. 
In addition, we include  the atomic  spin-orbit and the inversion asymmetric potential associated with the orbital Rashba interaction. Finally,
we consider a magnetic field as a source of time reversal symmetry breaking.  To achieve our results, we perform a detailed analysis 
of the most favorable topological superconducting phases. This analysis is based on self-consistent computations of the order parameters,
minimizing the mean-field free energy in the Hamiltonian parameters space, set by the electron filling, by the attraction strengths, 
and by the amplitude and orientation of the magnetic field. 

We point out  how the interplay of singlet and triplet pairings is able to affect the superconducting properties of LAO/STO 2DEGs. 
First,  we show that  the singlet order parameters are strongly reduced with increasing the role of triplet pairings,  thus acquiring a non-monotonic dependence 
on the charge density.
Interestingly, some notable effects are found related with not-linear corrections to the effective Rashba spin-orbit coupling.
For instance, in the absence of magnetic fields, the not-linear spin-orbit terms, combined with the triplet pairings, 
favor a quite stable (time-reversal invariant) topological helical superconducting phase.
 The triplet pairings are also responsible
for an anisotropic behavior of the superconducting order parameters, when the magnetic field is applied in-plane and out-of-plane. 
Finally, in the presence of out-of-plane magnetic fields, we recover (time-reversal breaking) chiral topological superconducting phases,
also when the triplet pairings are vanishing.

The paper is organized as follows:

In Section \ref{normalstate}, we report the main electronic properties of the normal state.
 
In Section \ref{super}, we discuss the general set-up of the mean-analysis of superconductivity. 
 
In Section \ref{MFsol}, we analyze the superconducting solutions at zero magnetic field.
 
In Section \ref{supermagnetic},  {\color{black} we analyze} the effects of a magnetic field. 
 
Finally, we devote Section \ref{conclusion} to our conclusions and outlook.

%-----------------------------------------------------------------------------------------------------------------------------------------------------------------------

\section{Normal state}
\label{normalstate}

\subsection{Model Hamiltonian}

\label{modham}
Following the derivation of 
Ref. [\onlinecite{perroni2019}], we write down  the general tight-binding Hamiltonian for the two-dimensional LAO/STO-001 system by 
considering  the 2DEG  effectively confined on a square with lattice step $a = 3.9 \, \AA \equiv 1$.
The system has also a broken out-of-plane inversion symmetry, having only the $t_{2g}$-orbitals close to the Fermi level.
In LAO/STO systems, the transition metal (TM)-oxygen bond angle is almost ideal, and the three $t_{2g}$-bands are mainly decoupled in the momentum space ${\bf k}$. 
%%%%%%%%%%%%%%%%%%%%%%%%%%%%%%%%%%%%%%%%%%%
\begin{figure*} [t!]
\includegraphics[scale=0.28]{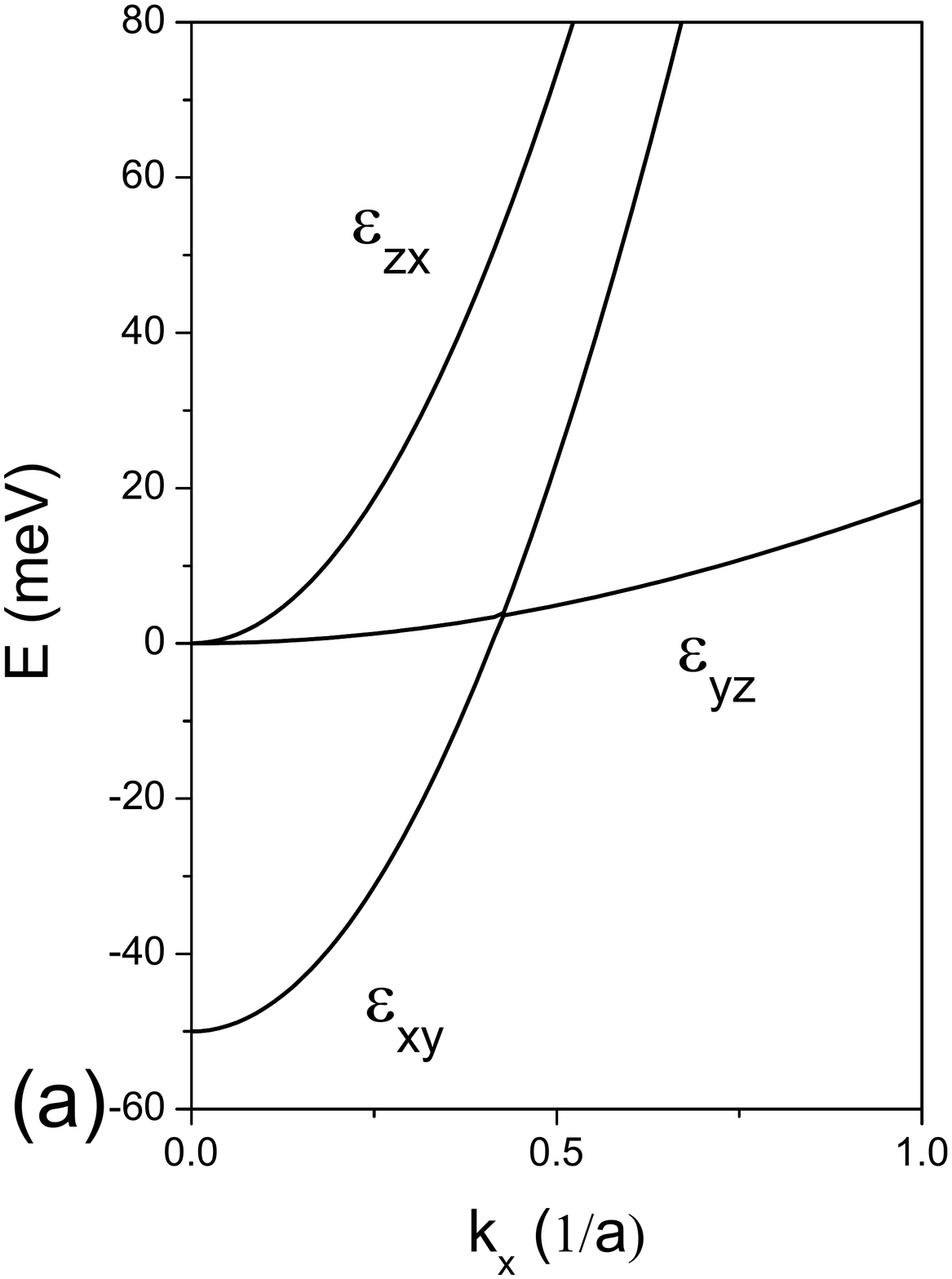}
\includegraphics[scale=0.28]{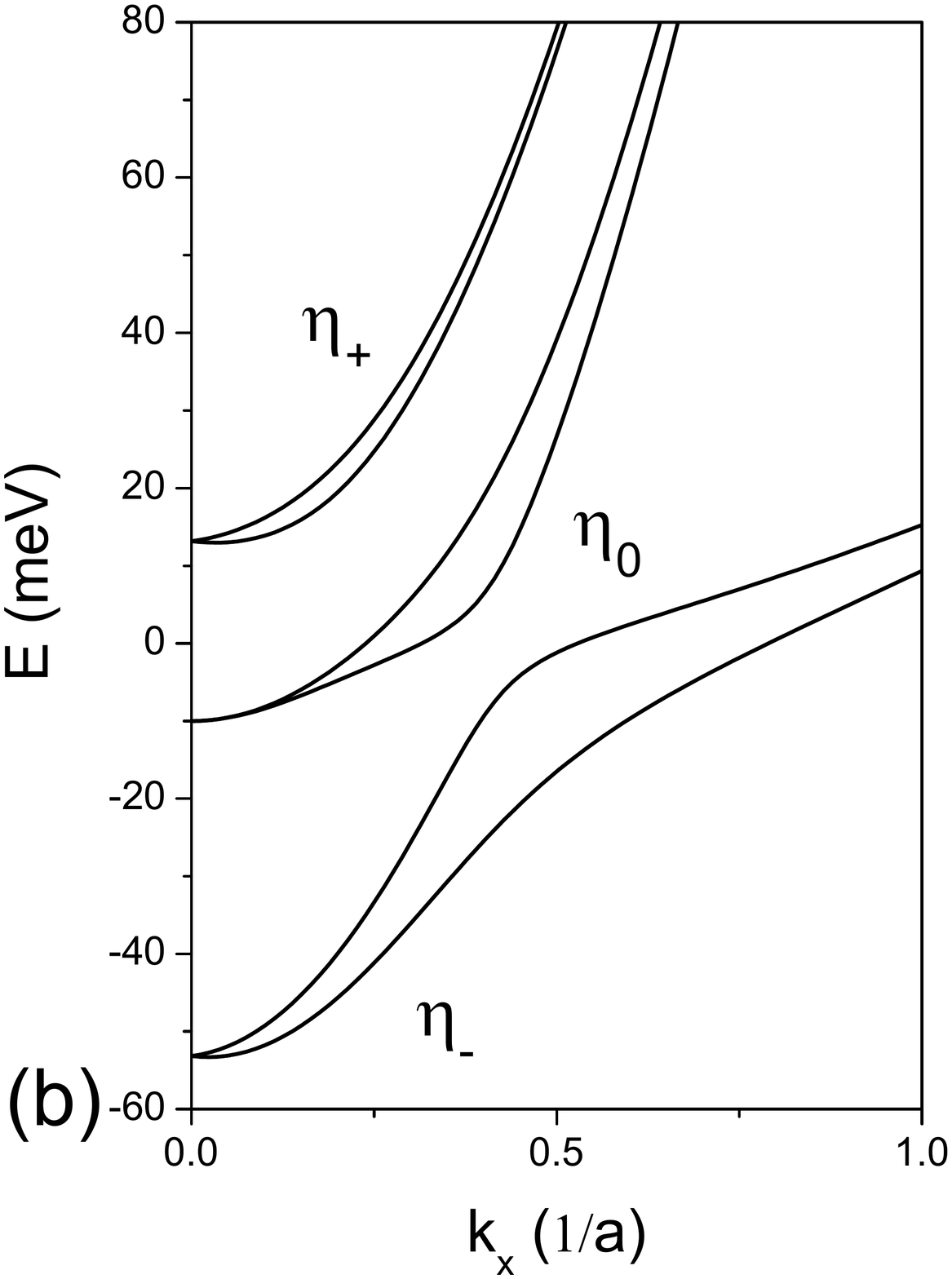}
\includegraphics[scale=0.28]{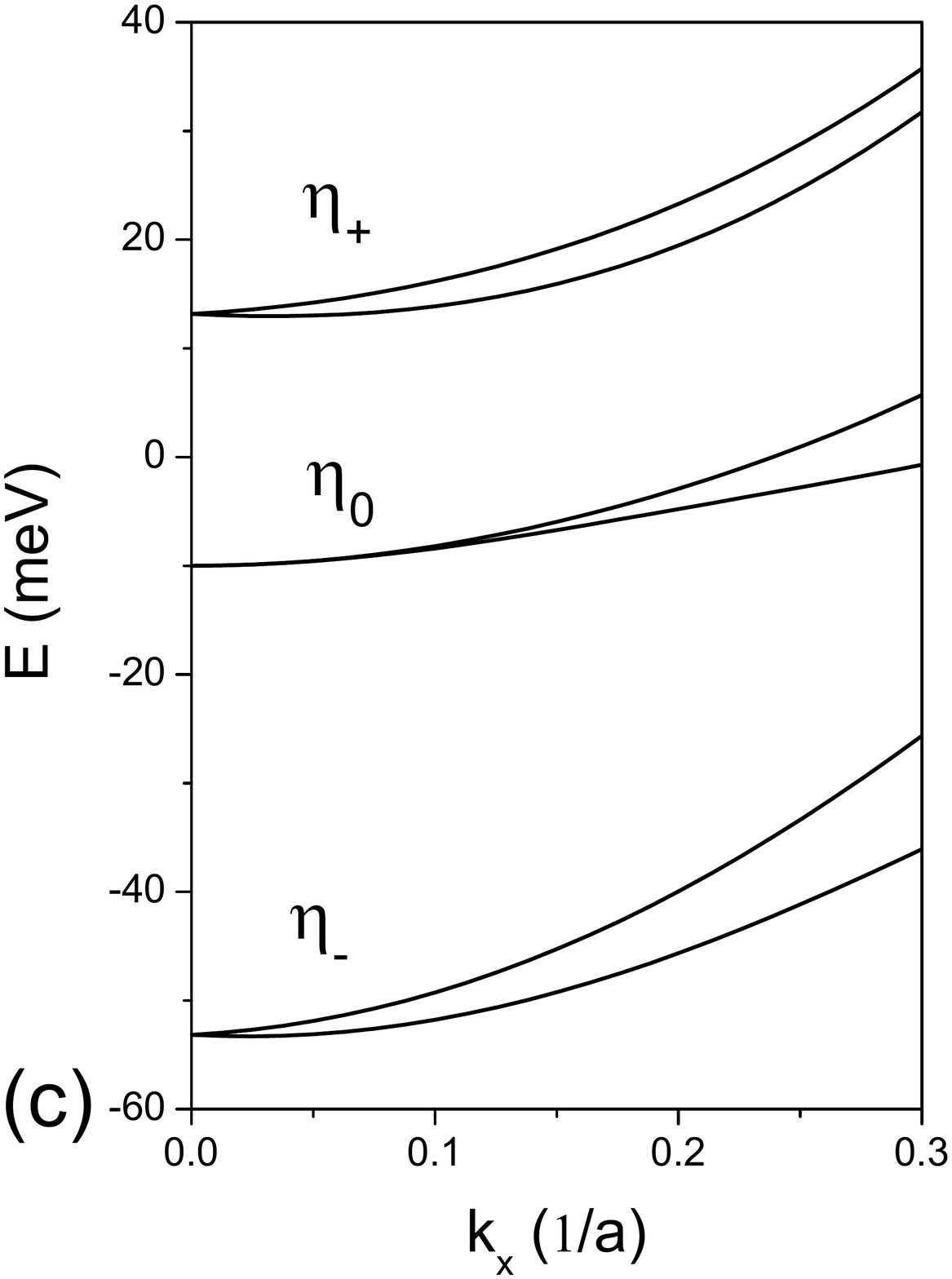}
\caption{Plot of $\epsilon_{yz}{\color{black} ( {\bf k} )}$, $\epsilon_{zx}{\color{black} ( {\bf k} )}$, and $\epsilon_{xy}{\color{black} ( {\bf k} )}$ (left panel, in meV), and of the mixed bands
$\epsilon_{-,0,+}{\color{black} ( {\bf k} )}$ (middle and right panels), in the absence of magnetic field, for $k_y = 0$, and as functions of 
$k_x$ {\color{black} (still around $k_x = 0$, then for $\nu \to 0$). In the middle panel, around $k_x \approx 0.35$, we
have avoided crossings, due to the role of the atomic spin-orbit coupling and the breaking inversion symmetry term.}}
\label{bande}
\end{figure*}
%%%%%%%%%%%%%%%%%%%%%%%%%%%%%%%%%%%%%%%%%%%%%%%%%%%%%%%

Moreover, the $d_{xy}$ band has a truly two-dimensional character, while the $d_{yz}$ and $d_{zx}$ bands are quasi one-dimensional.
In the following, we denote  with $H_0 ( {\bf k} )$ the corresponding normal-state contribution to 
the total system Hamiltonian in momentum space. Moreover, herewith we use the index $\tau = \{yz , zx, xy\}$ 
 to refer to the three different $t_{2g}$ orbitals $d_{yz}$, $d_{zx}$, and $d_{xy}$, respectively, while we label the spin with  $\sigma  =
\{ \uparrow , \downarrow \}$.  In addition, we add a term $H_{SO}$ to the total Hamiltonian, accounting for  the atomic spin-orbit coupling of the TM ions. 
Finally,  we include the 
microscopic couplings arising from the out-of-plane oxygen displacements around the TM, with the inversion asymmetry giving rise to an 
effective hybridization of $d_{xy}$ and $d_{yz}$ or $d_{zx}$-orbitals along the $y$ or $x$ directions, respectively. We denote this contribution  
as $H_Z ( {\bf k} )$.

In momentum space, we set 
\beq
{\cal H} = \hat{D}^\dagger ( {\bf k} ) H ( {\bf k} ) \hat{D} ( {\bf k} )
\:\: ,
\label{e.1}
\eneq
\noindent
with $\hat{D} ( {\bf k} )$ labelling the vector
\beq
[ c_{yz , \uparrow}  ( {\bf k} ), c_{yz , \downarrow}  ( {\bf k} ), c_{zx , \uparrow}  ( {\bf k} ), 
c_{zx,  \downarrow}  ( {\bf k} ), c_{xy , \uparrow}  ( {\bf k} ), c_{xy , \downarrow} ( {\bf k} ) ]_t \: , 
\label{e.2}
\eneq
(note the different grouping of the operators corresponding to the various orbitals with respect to \cite{perroni2019}), and 
\beq
H ( {\bf k} ) = \sum_{\tau} H_{\tau} ( {\bf k} ) = H_0 ( {\bf k} ) + H_{\rm SO} + H_Z ( {\bf k} ) + H_M 
\:\:\:\:.
\label{e.3}
\eneq

\noindent
The various terms at the right-hand side of Eq. (\ref{e.3}) are grouped as follows.
The first term $H_0 ( {\bf k} )$ is the lattice band term
\beq
H_0  ( {\bf k} )  = \left[ \begin{array}{ccc}
\epsilon_{yz}  {\color{black} ( {\bf k} )} {\bf I}_{2 \mathrm{x} 2} & {\bf 0} & {\bf 0 } \\
{\bf 0} & \epsilon_{zx}  {\color{black} ( {\bf k} )} {\bf I}_{2 \mathrm{x} 2} & {\bf 0} \\
{\bf 0} & {\bf 0} & \epsilon_{xy}  {\color{black} ( {\bf k} )} {\bf I}_{2 \mathrm{x} 2} 
                         \end{array} \right] \; ,                          
\label{e.4}
\eneq
\noindent
with 
\begin{eqnarray}
 \epsilon_{yz}{\color{black} ( {\bf k} )} &=& 2 t_{1y} [ 1 - \cos ( k_y ) ] + 2 t_{2x} [ 1 - \cos  ( k_x ) ] \nonumber \\
 \epsilon_{zx}{\color{black} ( {\bf k} )} &=&  2 t_{1x} [ 1 - \cos ( k_x ) ] + 2 t_{2y} [ 1 - \cos  ( k_y) ] \nonumber \\
 \epsilon_{xy}{\color{black} ( {\bf k} )} &=& 4 t_1 - 2 t_{1x} \cos ( k_x ) - 2 t_{1y} \cos ( k_y ) +  E_t \, ,
 \label{e.5}
\end{eqnarray}
\noindent
and the parameters set so that 
\begin{eqnarray}
 && t_{1x} = t_{1y} \equiv t_1 = 300 \: {\rm meV} \nonumber \\
 && t_{2x} = t_{2y} \equiv t_2 = 20 \: {\rm meV} \nonumber \\
 && E_t = -50 \: {\rm meV} \, .
 \label{e.6}
\end{eqnarray}
\noindent

The second term in Eq.(\ref{e.3}) is the atomic $l-s$ spin-orbit coupling term,
given by 
\begin{equation}
H_{\rm SO}=w_{{\rm SO}} \, \hat{{\bf l}}  \otimes {\bf \sigma},    
\end{equation}
with $\hat{{\bf l}}$ the vector operator describing the orbital angular momentum and ${\bf \sigma}$ (the Pauli matrices)
the spin angular momentum. In order to make the spin-orbit term explicit, we introduce the matrices $\Hat{l}_{x}$, $\Hat{l}_{y}$ and $\Hat{l}_{z}$, 
which are the projections of the $l=2$ angular momentum operator onto the $t_{2g}$ subspace, 
\begin{align}
\Hat{l}_{x}&=
\begin{pmatrix}
0 & 0 & 0 \\
0 & 0 & i \\
0 & -i & 0
\end{pmatrix}, \\
\Hat{l}_{y}&=
\begin{pmatrix}
0 & 0 & -i \\
0 & 0 & 0 \\
i & 0 & 0
\end{pmatrix}, \\
\Hat{l}_{z}&=
\begin{pmatrix}
0 & i & 0 \\
-i & 0 & 0 \\
0 & 0 & 0
\end{pmatrix},
\end{align}
assuming $\{yz, zx, xy\}$ as orbital basis. Using these operators,  we write the atomic spin-orbit coupling as
\beq
{\color{black}
H_{\rm SO} = i {\color{black} w_{{\rm SO}}} \left[ \begin{array}{ccc}
{\bf 0} & {\bf \sigma}_z & - {\bf \sigma}_y \\
- {\bf \sigma}_z & {\bf 0 } & {\bf \sigma}_x \\
{\bf \sigma}_y & - {\bf \sigma}_x & {\bf 0} 
                    \end{array} \right]
}
\:\:\:\: , 
\label{e.7}
\eneq
\noindent
with ${\color{black} w_{\rm SO}} = 10 \: {\rm meV}$. 

The third term in Eq. (\ref{e.3}) is the inversion symmetry breaking term $H_Z ( {\bf k} )$, given by
{\color{black}
\begin{equation}
H_{Z} ( {\bf k} )
=\gamma \left[ \Hat{l}_y \otimes {\bf I}_{2 \mathrm{x} 2} \sin{k_x}-\Hat{l}_x \otimes {\bf I}_{2 \mathrm{x} 2} \sin{k_y} \right],  
\end{equation}
}
or equivalently
\beq
H_Z ( {\bf k} ) = i \gamma \left[ \begin{array}{ccc}
{\bf 0} & {\bf 0} & - \sin ( k_x ) {\bf I}_{2 \mathrm{x} 2} \\
{\bf 0} & {\bf 0} & - \sin ( k_y ) {\bf I}_{2 \mathrm{x} 2} \\
\sin ( k_x ) {\bf I}_{2 \mathrm{x} 2} & \sin (k_y ) {\bf I}_{2 \mathrm{x} 2} & {\bf 0} 
                                  \end{array} \right]                              
\;\;\;\; , 
\label{e.8}
\eneq
\noindent
with $\gamma = 20 \: {\rm meV}$. {\color{black} This important term stems from the breakdown of a reflection
symmetry along a particular axis of the LAO/STO compounds, due to a corresponding lattice distortion. The net effect is the mixing
of the different orbitals $yz$, $zx$, and $xy$, having different parity under the mentioned reflection symmetry. More details are in \cite{perroni2019}. 
Furthermore, the effects of the coupling $\gamma$ on the stability of singlet and triplet superconducting order parameters will be discussed in Appendix 5}.

The last term in Eq. \eqref{e.3} describes the coupling of the electron spin and orbital moments {\color{black} with} 
an external magnetic field {\bf B}, whose direction is given by the vector 
${\bf M} =- \mu_B {\bf B}/ \hbar$, with $\mu_B$ Bohr magneton: 
{\color{black}
\beq
H_M 
=  g_s \, {\bf I}_{3 \mathrm{x} 3} \otimes  {\bf M} \cdot \frac{{\bf \sigma}}{2} +   {\bf M} \cdot {\bf \l}  \otimes  {\bf I}_{2 \mathrm{x} 2} \, ,
\label{e.9}
\eeq
 which   includes  the gyromagnetic factor $g_s$ assumed equal to 2.  
}
In the absence of  this term, the total Hamiltonian is time-reversal invariant:
\beq
 H ( {\bf k} ) = U_T^{-1}  H^* (- {\bf k} ) U_T \, , \quad  \quad \quad U_T = {\bf I}_{3 \mathrm{x} 3} \otimes \sigma_y 
 \label{defUT}
\eeq
 ($\sigma_y$ acting on the spin index $\sigma$, ${\bf I}_{3 \mathrm{x} 3}$ on the $\tau$ index, and $U_T \,  U_T^* =- {\bf I}_{3 \mathrm{x} 3}$), 
 as it can be straightforwardly  checked by expressing $H ( {\bf k} )$ in terms e. g. of Gell-Mann matrices  (see Appendix 1).
Therefore, in the absence of superconductive pairing, $ H ( {\bf k} )$ belongs to the class AII of the classification for topological 
insulators and superconductors, see e.g. \cite{ludwig2009, ryu2016, Bernevigbook}. 

In the following part of the Section, 
we will analyze the bands in the normal state in the absence of the external magnetic field. The external field weakly
affects the electronic structure of the normal state, but its role will be relevant in the analysis of the superconducting
phases since the low energy induced by the field competes with those due to the superconducting pairings.

\subsection{Band structure}

In  the left panel of Fig. \ref{bande} we show the "bare"   $ \tau$ bands, which we  derived by setting to zero
both  the spin-orbit coupling and the inversion symmetry breaking term. 
We notice that the  lowest band is {\color{black} the $xy$ one,} separated by the energy $|E_t|$ from the 
$yz$ and $zx$ bands, which are degenerate at $k_x=0$. In the same figure, 
we plot the bands, as a function of $k_x$, at $k_y=0$. We identify the almost flat band with the   $yz$ one 
and we also note that the dispersion of the $zx$ band is negligible as a function of $k_y$ and at fixed $k_x$.  

Introducing the spin-orbit coupling and the inversion symmetry breaking term, the  $\tau$-bands  are mixed together by the rotation matrix 
$M(\bf{k})$ that diagonalizes $H(\bf{k})$. Therefore, the orbital character is mixed, giving rise to a more complex spectrum \cite{perroni2019}   and to new,
``mixed'' bands, which in the following we label with the indices $\eta_{\{ -, 0, + \}}$.
 In fact, we see that the spin-orbit coupling and the inversion symmetry breaking 
mostly affect the electronic bands at low densities. To evidence this fact,
in the central and right panels of Fig. \ref{bande}, we report these bands, which at finite values of the momentum exhibit avoided crossing.
In particular, in the right panel of Fig. \ref{bande}, we focus on small values of the momenta. We see that 
the $\eta_{\pm}$-bands
display minima at finite momenta,  which is traced back  to an emerging effective linear Rashba coupling (see Ref. [\onlinecite{perroni2019}] and the next subsection). 
On the other  hand, $\eta_0$ shows a single minimum at ${\bf k} = 0$, indicating an enhanced importance of not-linear corrections.

 In the following, we will carefully investigate the behavior of the $\eta_0$-band at small values of ${\bf k}$. 
 In fact, the minima of the $\eta_0$-band mark 
 the density values  where superconductivity sets in.      
Indeed, theoretical and experimental studies suggest the presence of superconducting phases in the range for the lattice average filling $\nu$ approximately
between the onset of the $\eta_0$ band and that of the $\eta_+$ band (see e.g. \cite{gariglio2016}). 
%%%%%%%%%%%%%%%%%%%%%%%%%%%%%%%%%%%%%%%%%%%%%%%%%%%%%%%%%%
\begin{figure} [t]
\includegraphics[scale=0.37]{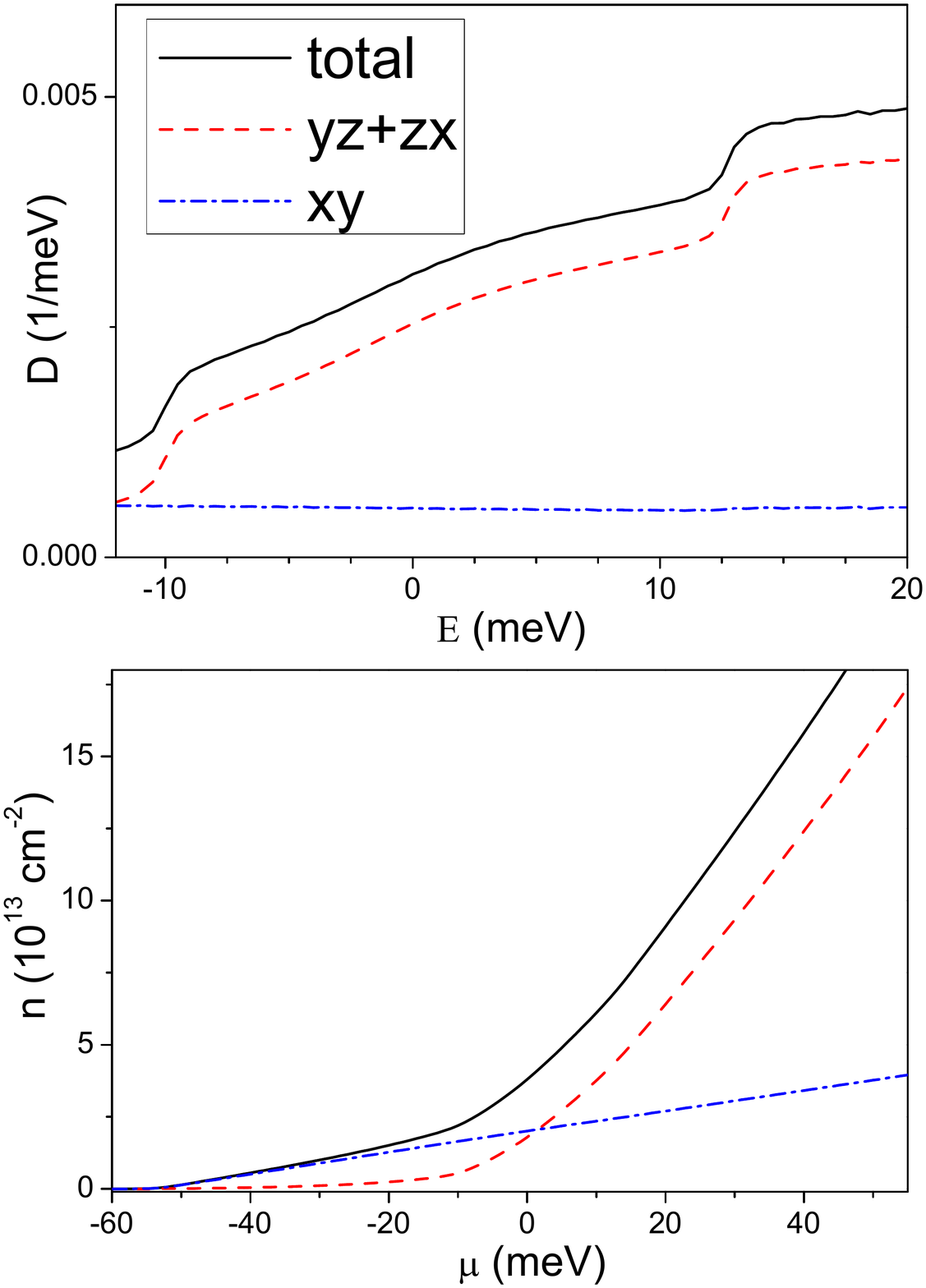}
\caption{{\color{black} Upper panel: density of states of the $\tau$-bands, as a function of the single-particle energy. Lower panel: 
carrier surface density, as a function of the chemical potential.}}
\label{DOS}
\end{figure} 
%%%%%%%%%%%%%%%%%%%%%%%%%%%%%%%%%%%%%%%%%%%%%%%%%%%%%%%%%%% 

Therefore, we focus on the electronic properties within this density range. In the upper panel of Fig. \ref{DOS}, we plot the density of states 
(DOS) of the $\eta$-bands, as a function of the single-particle energy  (note that the density has the dimensions of the inverse of an energy, 
since we  set the lattice step $a = 1$). The minima of the bands are characterized by steps in the density of states.  In the same panel, 
the projection of the DOS on the orbital basis $xy$, $yz$, and $zx$  is also reported, such to highlight their contributions. We remark that the 
quasi-one-dimensional bands $yz$ and $zx$  provide the most relevant contribution to the density of states. Actually, within 
this energy range, the density of states due to the $xy$ band is almost constant, after a step at much lower energy (of the order of $E_t$).  
This asymmetry in the density of states within this energy range will be fundamental for the interpretation of {\color{black} 
some superconducting properties, which are very sensitive to the magnitude of the available electronic states. Finally, 
within a slightly larger energy range, in the lower panel of Fig. \ref{DOS}, we plot the carrier space density,  as a function
of the chemical potential.} For the parameters of the tight-binding model used in this paper, the density at the minimum of 
{\color{black} $yz$ and $zx$ bands} is about $2 \cdot 10^{13}$ $cm^{-2}$. This value is in agreement with previous 
experimental results \cite{Stornaiuolo}. Moreover, we notice that, in analogy with the behavior of the 
density of states, as soon as the quasi-one-dimensional $yz$ and $zx$ bands start to be filled, {\color{black} they rapidly 
become predominant. This occurs around zero energy, where a crossover in the densities is visible. There, the carrier
density, of the order of $5 \cdot 10^{13}$ $cm^{-2}$, results half from the $xy$ band and half from {\color{black} the} $yz$ and $zx$ ones.}

\subsection{Effective low-energy theories}

In order to better  spell out the following results, we now present an effective theory for the   $\eta_-$- and $\eta_0$-bands, around their 
common minima, at $k_x = k_y = 0$. To derive the corresponding Hamiltonians we exploit the second order degenerate perturbation theory. We 
present the details of our derivation in Appendix 3. 
The results are respectively:
\beq
H_-^{(\mathrm{eff})}  ({\bf k}) = \epsilon_- ({\bf k}) \, {\bf I}  -  (a_1 \, k_x + a_2 \,  k_x^3)  \, \sigma_y +  (a_1 \, k_y + a_2 \, k_y^3)  \, \sigma_x \, ,
\label{eff-} 
\eeq 
with $a_1 = 8$ meV (in units of the lattice step $a \equiv 1$), $a_2 = 43.46$ meV, $ \epsilon_- ({\bf k}) = \big(- 54. + 280.8 \, (k_x^2 + k_y^2) \big)$
meV (so that  $t_-^{(\mathrm{eff})} = 280.8$ meV), and 
\beq
H_0^{(\mathrm{eff})} ({\bf k}) =  
\left[ 
\begin{array}{c}
\epsilon_0 ({\bf k})  \quad  \quad i \, a_3 \, k_- - i\, a_4 \, k_+^3 - a_5 \, k_x k_y \, k_+  \\
 {} \\
-i \, a_3 \, k_+ + i\, a_4 \, k_-^3 - a_5 \, k_x k_y \, k_-  \quad \quad \epsilon_0({\bf k}) \\
 \end{array} 
 \right] \, ,
 \label{eff0}
\eeq
with $k_{\pm} = k_x \pm i k_y$, $a_3= 0.8$ meV, $a_4 = 8.627$ meV, $a_5 = 22.8$ meV and  $\epsilon_0 ({\bf k}) = 
\big(-10.8 +157.2 \, (k_x^2 + k_y^2) \big)$ meV (so that $t_0^{(\mathrm{eff})} = 157.2$ meV).

We notice that, around the minima of the $\eta$ bands, $k_x= k_y = 0$, an effective linear Rashba-like spin-orbit 
coupling appears in both the effective Hamiltonians, together with a not-linear spin-orbit term.  This second term
is especially relevant for the first dome  on the $\eta_0$ band,  already at vanishing filling, as the central and right panels of  Fig. \ref{bande} suggest. 
It is known \cite{rashba2001} that  the linear coupling,  breaking space inversion symmetry (as $H_Z ( {\bf k} )$ 
it derives from), favours triplet components for superconducting 
pairings, inducing single-triplet mixings, see  e. g. \cite{annett, brydon2015}.
Instead, not-linear spin-orbit terms have been postulated to induce observable modifications on the spin
polarization in inversion symmetric (001) $SrTiO_3$ 
compounds \cite{nakamura2012} and for Josephson junctions \cite{alidoust2021}. Other notable effects will be described in the following. 

In Fig. \ref{bande2} of Appendix 3, we perform a  direct comparison of $H_-^{(\mathrm{eff}) ({\bf k})}$ and $H_0^{(\mathrm{eff})}({\bf k})$
with the exact spectrum of $H({\bf k})$ in Eq. \eqref{e.1}. We see that  
 the agreement is excellent up to $k_{x , y} \approx 0.2$. This value must be compared with the typical momenta where the $\eta_+$ 
 band starts to be populated, around $k_{x , y} = 0.4$. 
 The corresponding filling is estimated to be approximately the end of the superconductive dome \cite{gariglio2016}. 
 Therefore, we expect $H_0^{(\mathrm{eff})}({\bf k})$ to 
 describe approximately a low-density half of the first dome.
Importantly, $H_-^{(\mathrm{eff})} ({\bf k})$ and $H_0^{(\mathrm{eff})}({\bf k})$, that are a central result in 
the present work, still act on the $\sigma = \{ \uparrow , \downarrow \}$ 
indices, that are not mixed each others along the perturbation theory procedure.

\section{Mean-field analysis of superconductivity: general set-up}
\label{super}

In this Section we analyze the possible presence of superconducting phases, within the same energy and density ranges, 
 and focusing in particular on the interplay of singlet and triplet pairings.

\subsection{{\color{black} Setting up the effective interaction Hamiltonian}}
\label{supmulti}

In order to induce superconductivity in the 2DEG, in the following we {\color{black} consider a pertinent,  nearly realistic, attractive electronic 
interaction, together with its effects on  the bands of $H({\bf k})$. In particular, in the absence of more specific coupling mechanism related to, 
e.g., antiferromagnetic correlations, we  rely on the ``natural''   mechanism, 
based on phonon exchange plus screened Coulomb repulsion}. While the phonon attraction is expected to be
independent of the orbital index $\tau$, the Coulomb repulsion is expected  to be more important within the same orbital, 
due to the larger overlaps between the electronic wavefunctions.  Therefore, in the low density regime analyzed in this paper, 
{\color{black} we choose the}  realistic attractive potential given by:
\beq
W =  -  \sum_{i,   j ,  \tau, \sigma , \sigma^{\prime}}   U_{i,j} \, n_{i \tau  \sigma}  \, n_{j \tau \sigma^{\prime}} 
-  \sum_{i,j, \tau , \tau^{\prime} \neq \tau ,  \sigma , \sigma^{\prime}}  Z_{i,j}\, n_{i \tau \sigma}  \, n_{j  \tau^{\prime} \sigma^{\prime} }  \, ,
 \label{potgen}
\eeq
with {\color{black} the decay of $U_{i,j}$ and $Z_{i,j}$,  with the separation $i-j$, ruled by two} different lengths $\xi_U$ and $\xi_V$. 
The same lengths depend on the electronic density $\rho_e$, that is, {\color{black} 
on  the lattice average filling $\nu$, and are expected to be of the order of a}
 few lattice steps. Moreover, {\color{black} we assume a static potential, consistently with the low charge density in the regime of Fermi energies  that we are considering.}
Actually, the dynamic part of the potential from the phonon exchange can be extrapolated to the  low-energy limit, therefore the effective attractive potential at the Fermi energy can be assumes as static \cite{mazin2011,cataudella2021}.

The bands  $\tau$, $\tau^{\prime}$ are mixed together by the rotation matrix $M(\bf{k})$ that diagonalizes $H(\bf{k})$; {\color{black} $M(\bf{k})$ 
accordingly affects  the potential $W$ in Eq. \eqref{potgen}, as well.}
It is difficult to {\color{black} deal with the transformed potential}, since  $M(\bf{k})$ depends on $\bf{k}$ 
and the Fourier transform of $W$ contains 4 momenta $\bf{k}_i$, $i = 1, \dots, 4$.

Therefore, in this paper  we analyze the effects, especially on the topology, 
of the full potential $W$, focusing on its first term, diagonal on the orbital index $\tau$.  This choice 
is motivated by the energy gap between the $\eta_-$ and $\eta_0$ bands around ${\bf k} = 0$, 
where the  superconductive dome is located \cite{gariglio2016},
being  approximately $\delta E_{\eta} = 43$ meV. This gap is an obstruction for the zero-momentum pairings between 
(unbalanced species in) the bands, and forbids them from attractions such that the superconductive gap, calculated at vanishing unbalance, 
is under a threshold around $\delta E_{\eta}$ (see \cite{pethick,marchetti2007,rad2007} and references therein). Furthermore, 
nonzero-momentum balanced pairings are known to require at least a subtle fine-tuning between interaction and density \cite{pethick,marchetti2007,rad2007,pieri2019}.

Within the tight-binding model, we focus on the on-site and the nearest neighbor contribution of the attractive potential {\color{black} between   
electrons in states  with} the same orbital symmetry. The local interaction  necessarily couples {\color{black} electrons with} opposite spins favoring spin-singlet symmetry. 
On the other hand, we assume that the term due to the nearest neighbors takes into account the equal spin contribution controlling 
the spin-triplet instability.
Hence, {\color{black} we simplify  $W$ according to}
\begin{equation}
W = -U \sum_{i,\tau}   n_{i \tau \uparrow} \, n_{ i \tau \downarrow} 
- \frac{V}{2}  \sum_{i \delta \tau \sigma}  n_{i \tau \sigma} \, n_{ i+\delta \tau  \sigma} 
\label{hamilpair}
\end{equation} 
where  $i$ and $j$ label two dimensional vectors associated to the lattice sites,  $U$ and $V$ are
the local and nearest neighbor pairing energy, respectively, and $n_{i \tau \sigma} =c_{i \tau \sigma}^{\dagger} c_{i \tau \sigma}$
is the local density operator for the $\sigma$ spin polarization and the $\tau$ orbital, at a given position $i$, whose nearest neighbor sites are indicated by $\delta$. {\color{black} In Appendix 4, we discuss additional attractive terms which have not been introduced in Eq. (\ref{hamilpair}).}

{\color{black} As specified in the subsection title, we have
considered in Eq. \eqref{hamilpair} an effective interaction, in particular an 
attractive local Hubbard term. We note that in \cite{breit2010} a local repulsive Hubbard term, around 2 eV, has been predicted from  tunneling spectroscopy. In fact, this value of U is not large, but it is comparable with the electron bandwidth. Furthermore, in LAO/STO systems analyzed in this paper, the typical densities per spin polarization are low compared to, e.g. ,  the half-filling regime. Accordingly,  the net contribution to the total energy  from Hubbard interaction is limited. Moreover, it is known that, in these density regimes, the effects from polaron dynamics are relevant on the electronic states \cite{cancellieri2016,cataudella2021} giving rise to a net lowering of the  Hubbard interaction for the 2D quantum gas. Finally, our effective model for the local interaction has been widely used in the literature, for example \cite{Mohanta,Loder,Fukaya,perroni2019,settino2020} quoted in the introduction. }

In the following, we will  check that {\color{black} $W$ in Eq. \eqref{hamilpair}  qualitatively reproduces}  the most 
interesting features of superconductivity in LAO/STO compounds \cite{gariglio2016}, {\color{black} with the values of $U$ and $V$, yielding 
  the superconducting behavior discussed in this paper always, being} of the order of hundreds of meV.

\subsection{Mean field analysis}

We now analyze,  within mean-field approximation, the Hamiltonian in Eq. \eqref{e.1}, with the interaction in 
Eq. \eqref{hamilpair}, by mostly focusing onto the zero temperature case. 

 Due to the introduction of the nearest neighbor attractive term $V$, it becomes important to 
properly infer the profile of the superconducting order parameter in real space. 
In the following, we do so by encompassing within our mean-field approach both the spin  and the orbital degrees of freedom.  
In particular, we recover the appropriate pairing {\it ansatz} by analyzing the set of the irreducible representations of the point-group symmetry of the square lattice, by assuming 
 over-all  translational invariance (we provide the details in Appendix 2).
 Moreover, in the absence of an externally-applied magnetic coupling,  we retain time-reversal invariance. 
As a result, we obtain:
\begin{eqnarray}
W & \approx&  - U \sum_{i,\tau} D_{i \tau} \left[ c_{ i\tau \uparrow}^{\dagger} c_{i \tau,\downarrow}^{\dagger} + {\color{black} \mathrm{H. c.}} \right] -\nonumber \\
&& - \frac{V}{2} \sum_{i, \tau,\delta, \sigma} \left[  F_{i \tau \sigma}(\delta) c_{ i \tau \sigma}^{\dagger} c_{i+\delta \tau \sigma}^{\dagger} + 
{\color{black} \mathrm{H. c.}} \right] + \nonumber \\
&&+ \, U \sum_{i \tau} D^2_{i \tau} + \frac{V}{2} \sum_{i, \tau, \delta, \sigma } |F_{i \tau,\sigma}(\delta)|^2.
\label{hamilbogo}
\end{eqnarray}
In Eq. (\ref{hamilbogo}), $D_{i \tau}=\langle c_{i \tau \downarrow}  
c_{i \tau \uparrow}\rangle$ {\color{black} is} the singlet pairing amplitudes, depending on the orbital index $\tau$, with $\langle\rangle$ {\color{black} denoting}
the ground state average, and the {\color{black} over-all gauge choice is made so that the} singlet order parameters {\color{black} $ \Delta_{i s}^{\tau}=U 
D_{i \tau}$} are  real. The local s-wave pairing {\color{black} corresponds to} the most favored superconducting
instability \cite{Michaeli,Nakamura,Loder1,Mohanta1}. 
{\color{black} Finally, $F_{i \delta \sigma} (\delta) = \langle c_{i+\delta \tau \sigma}  
c_{i \tau \sigma}\rangle$} are the equal spin triplet pairing complex amplitudes, depending on both $\tau$ and $\sigma$. 
Due to the space inversion symmetry of the  square lattice, $F_{i \tau \sigma}(\delta)=F_{i \tau \sigma}(-\delta)$, and the triplet amplitudes along the $y$-axis have 
only a phase different from those along the $x$-axis: $F_{i \tau \sigma}(\delta_y)=\theta^{\sigma} F_{i \tau \sigma}(\delta_x)$, 
with $F_{i \tau \sigma}(\delta_x)=F_{i \tau \sigma}$ fixed real. 
{\color{black} In Appendix 4, we discuss additional pairings which could be considered in the system.}

{\color{black}
We point out that our ansatz for the superconducting mean field is diagonal in bare bands $\epsilon_i$, therefore inter-orbital pairing between the rotated bands $\eta_i$ is present
since we get {\color{black}  the corresponding gap parameters}  via a self-consistent procedure,
minimizing the (free) energy of the system, where 
the mixing between the $\epsilon_i$ bands is included.
The statement is reinforced by the fact that, in the absence of inversion symmetry breaking term $\gamma$,  the structure of the
gaps is changed. Indeed, as discussed in Appendix 5, the coupling $\gamma$ is relevant to control the interplay between singlet and triplet order parameters. }

 At the mean-field level, the triplet and the singlet-triplet mixed pairing  are  separately 
determined  by the nearest neighbour part of the potential in Eq. \eqref{hamilpair}, 
and by the spin-orbit (Rashba-like and not-linear) terms in Eqs. \eqref{eff-} and \eqref{eff0} \cite{rashba2001},  and their onset is  clearly enforced 
by  the  combined effect of the two terms. However,
in the absence of the $V$ term  in Eq. \eqref{hamilpair}
(therefore with only an Hubbard attraction), mean-field decoupling does not yield triplet components 
{\color{black} (indeed, recover these terms requires resorting to alternative approaches, such }as Monte-Carlo simulations, see e.g. \cite{rosenberg2017}).  
Therefore, we expect that  the mean-field  approach  overestimates the singlet components, 
while the {\color{black} emergence} of triplet components appears even more substantiated.

In momentum space,  we describe the set of pairing configurations in Eq. \eqref{hamilbogo} 
within the general matrix parametrization \cite{annett}
\beq
\Delta^{\tau} ({\bf k}) =   \left(\begin{array}{c} c_{\uparrow}  ({\bf k}) \\ c_{\downarrow}  ({\bf k}) \end{array} \right)^{\dagger}  \, \tilde{\Delta}^{\tau} ({\bf k}) \, 
\left(\begin{array}{c} c_{\uparrow}  (-{\bf k}) \\ c _{\downarrow}  (-{\bf k}) \end{array} \right)^* \,  ,
\label{pairing0}
\eeq
with
\beq
\tilde{\Delta}^{\tau} ({\bf k}) =  i \,  \Big(\Delta_{s}^{\tau} ({\bf k}) \, {\bf I}_{2 \mathrm{x} 2} + {\bf d}^{\tau} ({\bf k}) \cdot {\bf \sigma} \Big) \, \sigma_y \, .
\label{pairingmat}
\eeq
In particular, Eq. \eqref{pairingmat} can be rewritten in components as (see {\color{black} Appendix 1 for more details}): 
\begin{align}
\label{pairing}
\color{black}
&\Big(\Delta_{s}^{\tau} \, , \, \Delta_{t , \uparrow \downarrow}^{\tau}  \Big[ \alpha \, (s_x + i \, s_y) + \beta\,  (s_x - i \, s_y) \Big] \,  ,  \nonumber \\
{}  \\
&  \Delta_{t , \uparrow}^{\tau} (s_x + i \, s_y)  \, , \, \Delta_{t , \downarrow}^{\tau}\, (s_x - i \, s_y) \Big) \, , \nonumber
\end{align}
with $\Delta_{t , \sigma}^{\tau} = -  i V F_{\tau \sigma}$ {\color{black} and $s_{\{x,y\}} \equiv \sin k_{\{x,y\}}$. Since no interaction term $\propto V$ between {\color{black} electrons with } opposite spins
is present in Eq. \eqref{hamilpair}, then $\Delta_{t , \uparrow \downarrow}^{\tau} = 0$ in Eq. \eqref{pairing}.}

{\color{black} In principle, all  the parameters in Eq. \eqref{pairing} may have nonzero} phases. However, 
time-reversal invariance, with the same $U_T$ for the positive-energy sector as in Eq. \eqref{defUT}, 
{\color{black} sets the phases to 0} (see \cite{brydon2015} and Appendix 1).
The ansatz in Eq. \eqref{pairing} perfectly reproduces the superconducting solutions known in the literature as $p_x+ip_y$ and $p_x-ip_y$, which are known to support topological 
 helical superconductivity at zero magnetic field \cite{qi2009,sato2017}. Moreover, from Eq. (\ref{pairingmat}), the triplet pairing vector ${\bf d}(\bf{k})$ (giving the 
 superconducting excitation gap, in the absence of the singlet pairing) is given by following components: 
\begin{equation}
d_x({\bf k})=  2 V F_{r\sigma} \sin(k_y a), \ \ d_y({\bf k})= - 2 V F_{r\sigma} \sin(k_x a),    \label{dvector} 
\end{equation}
with $d_z({\bf k})=0$. Indeed, in \cite{frigeri2004} it has been shown that 
the superconducting transition
temperature is maximized when the spin-triplet pairing vector $d({\bf k})$ is aligned with the polarization vector ${\bf g}({\bf k})$ (essentially two-dimensional) parametrizing  the spin-orbit coupling. {\color{black} Additional details about the properties of the pairing vector  $d({\bf k})$ will be provided  in Appendix 4.} 

{\color{black} We comment finally that the possibility of helical phases \cite{samokhin}, with Cooper pairs with nonzero momentum, has not been considered in the present work.
Similar phases are expected to be suppressed in the regime of small filling for the $\eta_0$ band, where, 
around ${\bf k} = 0$, effective Rashba terms are not linear but at least cubic in the momentum. We point out that the most interesting results discussed in the next sections focus just on this parameter regime.}

\section{Superconducting solutions at zero magnetic field}
\label{MFsol}

 In  this  Section,  we  analyze the interplay between the  triplet and the singlet 
order parameters, in the absence  of an external magnetic field. 

\subsection{Emergence and stability of the superconducting solution}

To assess the presence and {\color{black} the} stability of the superconducting states, we solve the self-consistent equations by using a variational method. 
Specifically,  we study variationally the zero-temperature mean-field grand-canonical free-energy $\Omega$ determined by the pairing in Eq. \eqref{pairing}  plus 
 the single-particle Hamiltonian in Eq. \eqref{e.1}. In particular, we
 numerically  minimize $\Omega$, 
varying $D_{i \tau}$, $F_{i \tau \sigma}$.  In the minimization procedure, we vary the superconductive amplitudes, the 
interaction strength $U$ in Eq. \eqref{hamilpair} in the interval $\Delta U = [0, 500]$ meV, and the 
interaction strength $V$  in the interval $\Delta V = [0, 1000]$ meV. We consider  a square lattice with side-length $L$, 
with $L = [40, 240]$ {\color{black} sites}, finding a satisfying convergence starting from $L = 120$. Furthermore, 
we set $\mu = E_F$, $E_F$ being the Fermi energy at a fixed $\nu$ and, typically, between the minima of the $\eta_0$ and $\eta_+$ bands. 
%%%%%%%%%%%%%%%%%%%%%%%%%%%%%%%%%%%%%%%%%%%%%%%%%%%%%%%%%%%
\begin{figure} [t]
\includegraphics[scale=0.295]{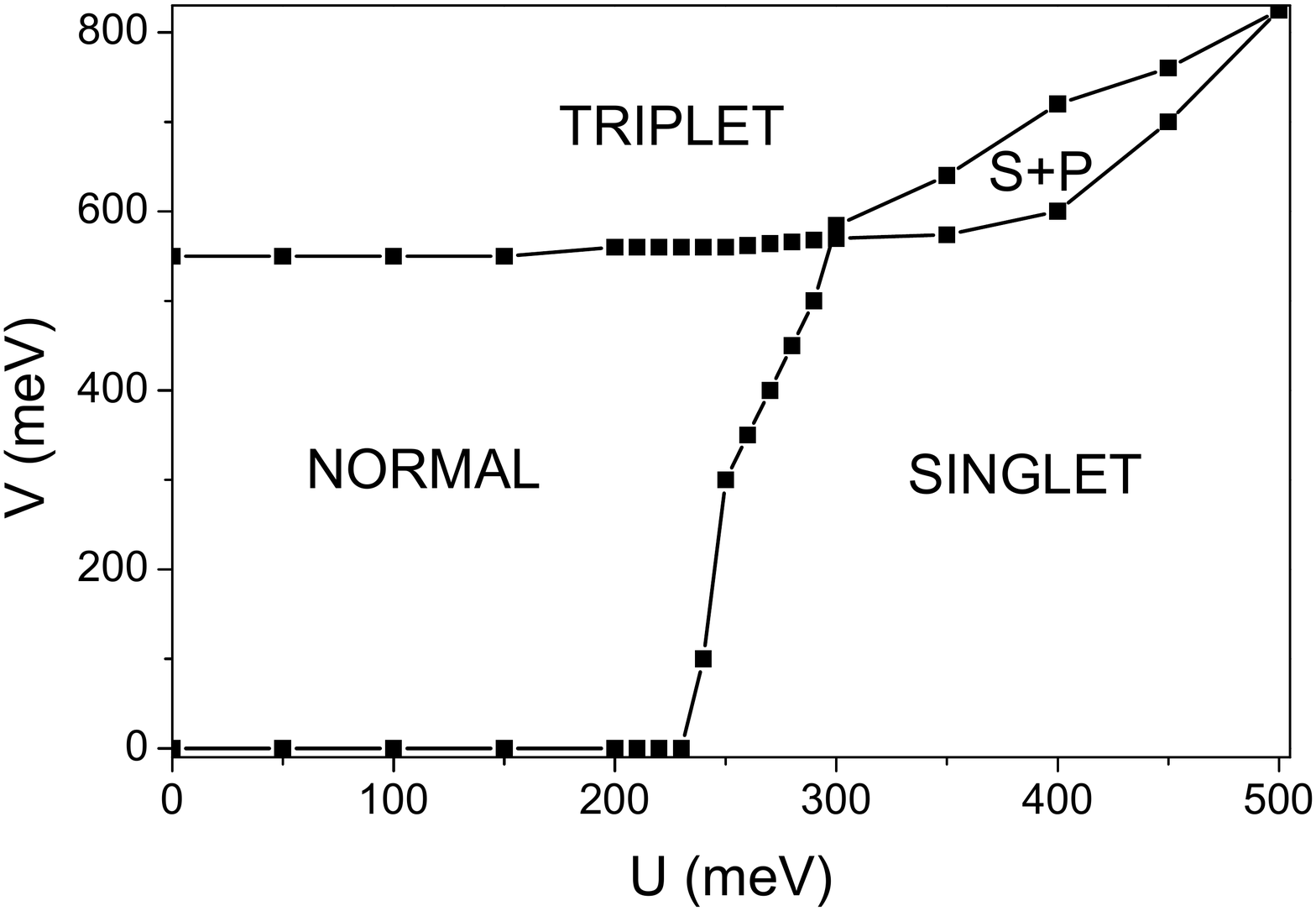}
\includegraphics[scale=0.29]{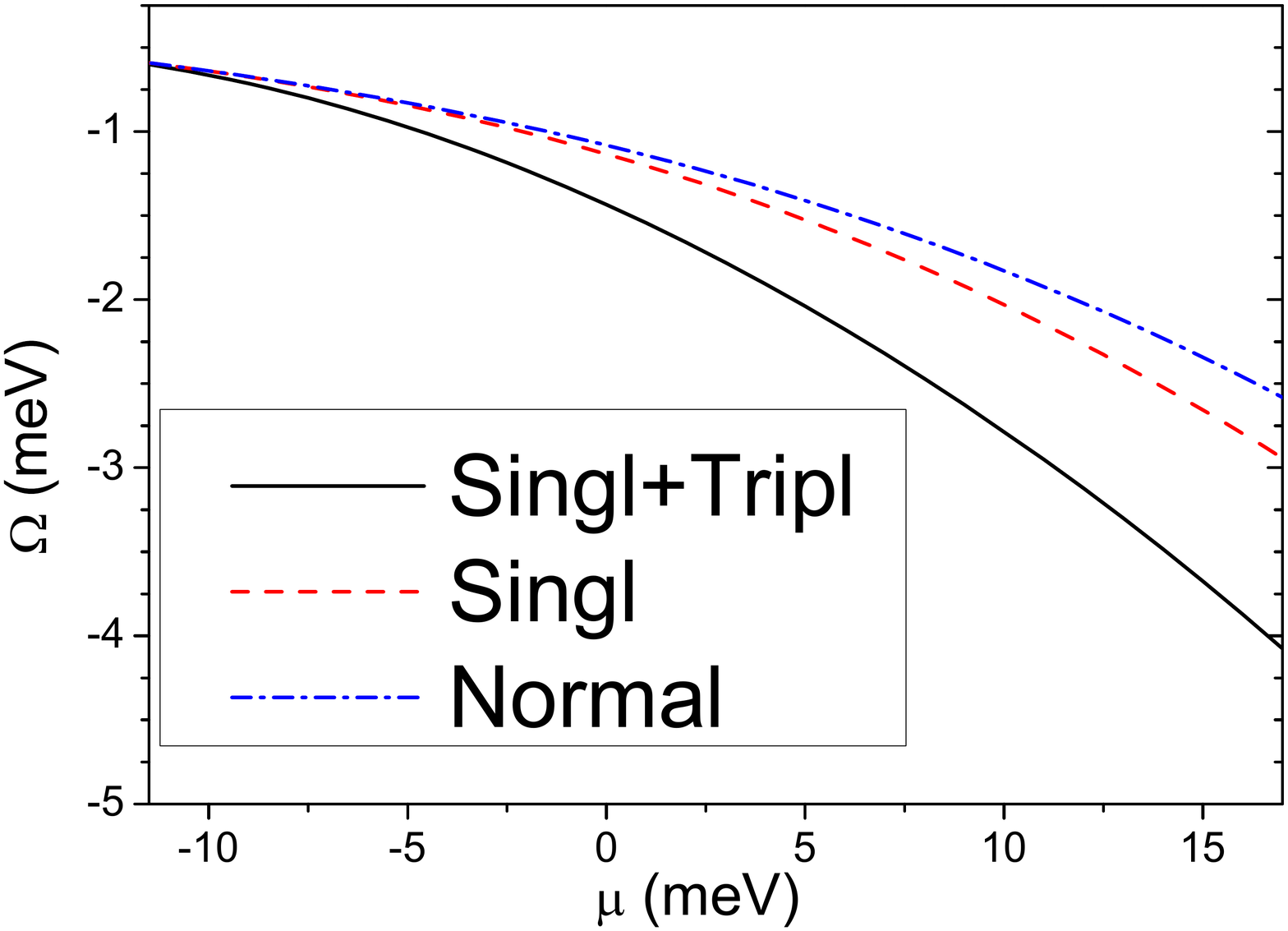}
\caption{Upper panel: Phase diagram in the plane $U-V$ to distinguish the normal state (absence of superconductivity), 
singlet, triplet and singlet+triplet (S+P) superconducting phases at the chemical potential $\mu=-9$ meV ( close to the
onset of intermediate electronic bands).  Lower panel: Comparison between the 
grand-canonical free energies of the normal state, of the singlet-pairing and the triplet-singlet superconductive ground-states, 
at  $U = 350$ meV and $V = 600$ meV.}
\label{phasefig}
\end{figure}
%%%%%%%%%%%%%%%%%%%%%%%%%%%%%%%%%%%%%%%%%%%%%%%%%%%%%%%%%%%%
%%%%%%%%%%%%%%%%%%%%%%%%%%%%%%%%%%%%%%%%%%%%%%%%%%%%%%%%%%%
\begin{figure*} [t!]
\includegraphics[scale=0.28]{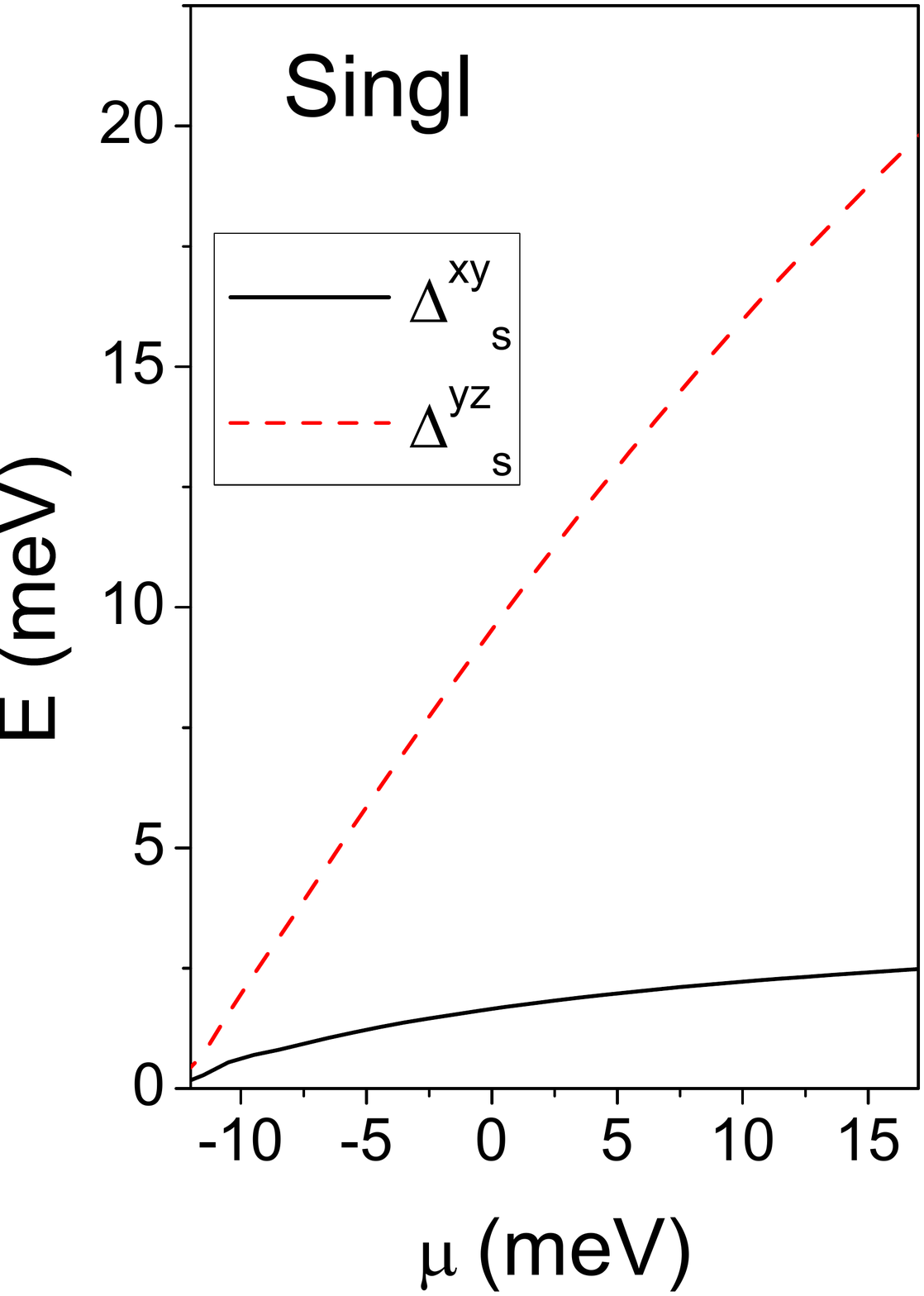}
\includegraphics[scale=0.28]{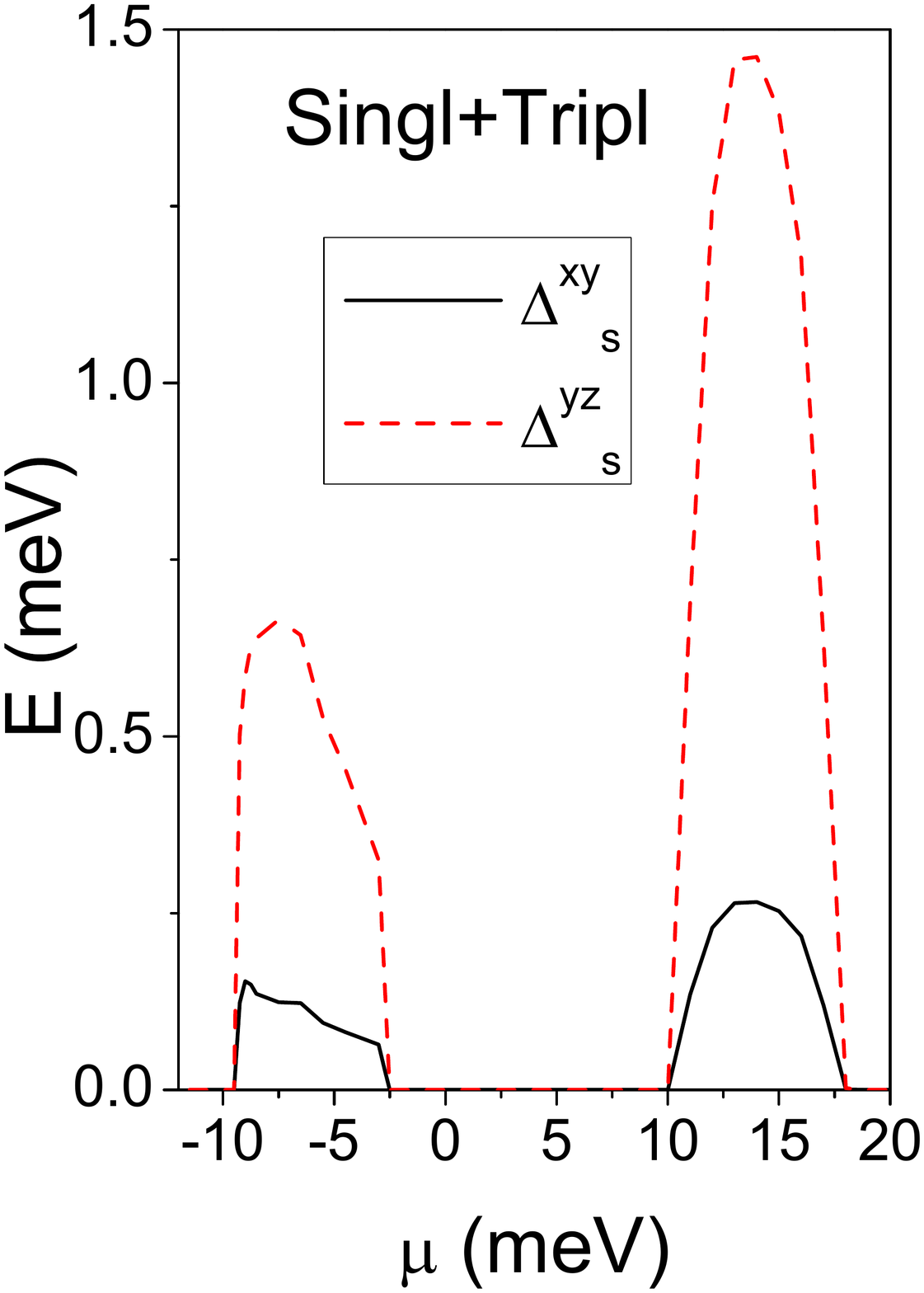}
\includegraphics[scale=0.28]{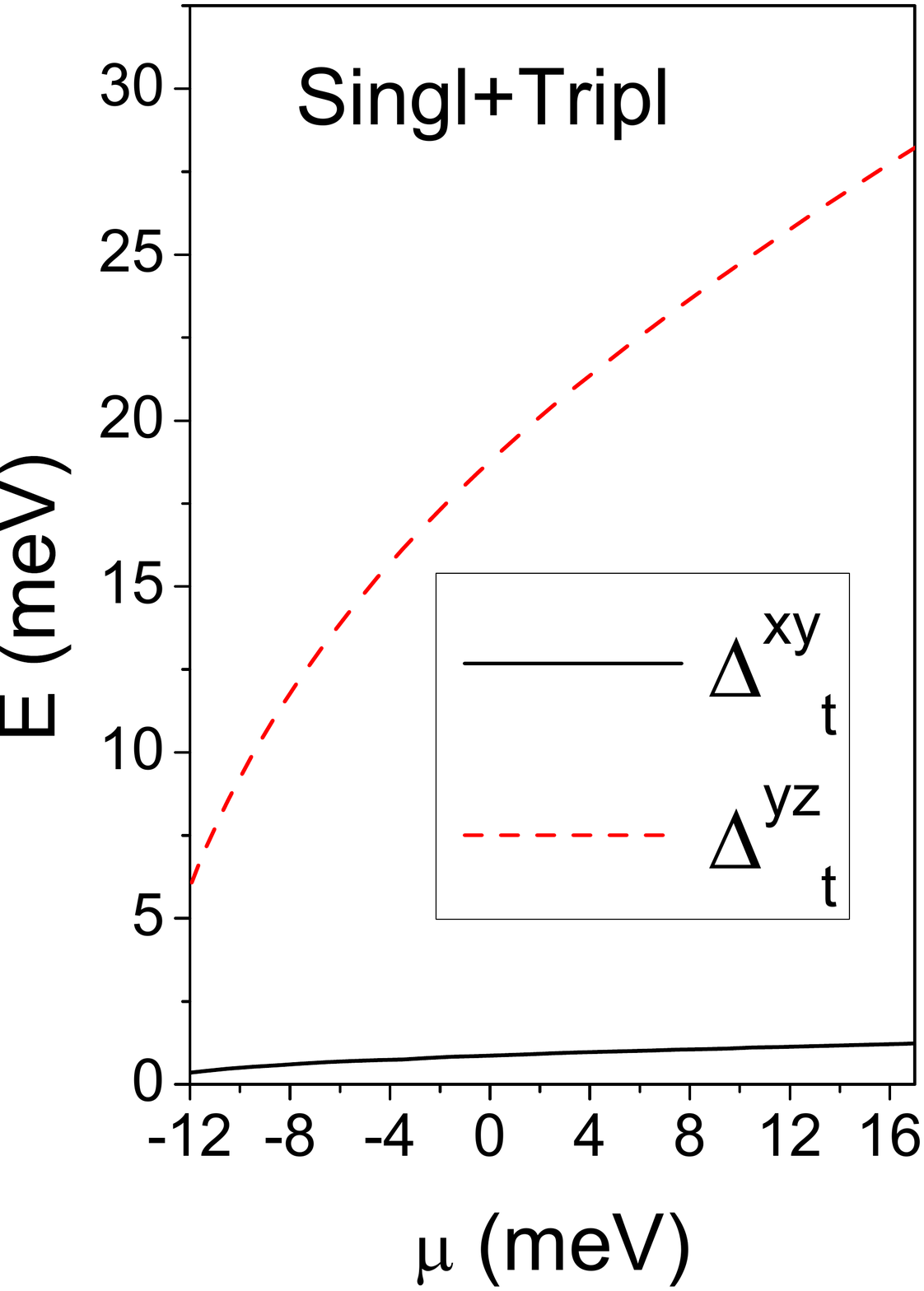}
\caption{{\color{black} Left panel: Zero temperature singlet pairing amplitudes, for $U =350$ meV and $V = 0$ meV.  
Middle panel: The same as in the left panel, but for $U = 350$ meV and   $\frac{V}{2} = 300$ meV. Right panel: Zero temperature triplet pairing amplitudes, 
again for $U = 350$ meV, $\frac{V}{2} = 300$ meV. Notice that, in the absence of magnetic field, $\Delta^{\tau}_{t}=\Delta^{\tau}_{t,\uparrow}=\Delta^{\tau}_{t,\downarrow}$.}}
\label{pairingpl}
\end{figure*}
%%%%%%%%%%%%%%%%%%%%%%%%%%%%%%%%%%%%%%%%%%%%%%%%%%%%%%%%%%%%

{\color{black} By means of our variational procedure}, 
beyond the normal state and a singlet-pairing regime, we find a range of values for $U$ 
and $V$ where the singlet and triplet pairings coexist. {\color{black} Instead, as pointed out in the previous subsection, no triplet 
pairing between different spins is found {\color{black} (we will comment later on on this result).}  When the  coexistence} 
takes place, the ratio between $U$ and $V$ can be tuned such that the triplet component of the order parameter is not negligible 
{\color{black} or even dominant} (further details and insight on the solutions described below will be given in Section \ref{topnoB}, 
via the analysis of the effective Hamiltonian in Eq. \eqref{eff0}).

First, we focus on the regime of values for $U$ and $V$ where only the singlet pairing is stable. {\color{black} As shown 
in Fig. \ref{phasefig}, this occurs for {\color{black} $U = [300, 400] \:{\rm meV}$  and for $V < 600$ meV.} Superconductivity becomes stable 
once the chemical potential $\mu$ is set above the minimum of the rotated band $\eta_0$, around $-54$ meV. Conversely, no pairing 
is observed for lower values of $\mu$. As reported in Fig. \ref{phasefig}, for $U<300$ meV, superconductivity is not found  
(that is true at any $\mu$).}

In the left panel {\color{black}  of Fig. \ref{pairingpl}}, corresponding to {\color{black} $U = 350$ meV and $V=0$,} we plot the singlet order parameters,
showing that they are different from zero, starting from the minimum of intermediate pair of bands ($\mu$ goes from about $-10$ meV to about $20$ meV).
{\color{black} The obtained starting point for superconductivity is in agreement with the appearance of a finite density of states, shown in the previous Section, and with the experimentally measured behavior of the superconducting pairing \cite{gariglio2016,singh}.} 
We remark that we always find ${\color{black} \Delta_{s}^{yz}}={\color{black} \Delta_{s}^{zx}}$ in the self-consistent superconducting solutions. 
Moreover, as shown in the left panel of Fig. \ref{pairingpl}, not only ${\color{black} \Delta_{s}^{yz}}$, but also ${\color{black} \Delta_{s}^{xy}}$ is different from zero, 
starting from the the minimum of the $\eta_0$  band.
Due to the difference of mass and density of states in the normal state, ${\color{black} \Delta_{s}^{xy}}$ is smaller than ${\color{black} \Delta_{s}^{yz}}$. 
{\color{black} We also point out that, increasing the carrier density, the order parameters 
 get enhanced continuously: apparently, there is no dome, as a function of the density. {\color{black} This finding matches the general 
 result in \cite{paramekanti2020}, relating the presence of superconducting  domes with finite-range potentials.}}    

Then, we consider the effect of a {\color{black} nonzero attractive term $ \propto V$.}  
As reported in Fig. \ref{phasefig},
for $V \gg U$, the triplet is the dominant pairing, while the singlet pairing tends to vanish. However,  Fig. \ref{phasefig} 
shows  an intermediate regime, from values of $V$ slightly smaller than $2U$, where  the order parameters coexist. 
In the middle panel of Fig. \ref{pairingpl}, we plot the singlet order parameters for the different orbitals in the same regime, 
as a function of the chemical potential.  At finite $V$, the appearance of a triplet pairing  parallels a reduction of the s-wave order parameters.
In particular, the smallest one, $ \Delta_{s}^{xy}$ assumes values in agreement with experimental estimates, around $0.1$ meV. Therefore, there 
is a destructive interference between singlet and triplet amplitude pairings. Moreover, this interplay is also able to induce a dome, as a function of 
the density, {\color{black} beginning from the place where the density of states of the normal phase shows} a step. 
We notice that a similar behavior takes place at higher densities, corresponding to the minimum  of the band $\eta_+$, where 
the density of states has another step. Therefore, our theoretical calculation predicts that, beyond the over-doped regime of the first dome,  
corresponding to the band $\eta_0$, there is a possibility of a second dome with similar extent, related to the higher energy band $\eta_+$. 
{\color{black}
We observe finally the presence of a regime where $U \neq 0$ and $V \neq 0$ but the normal state is favoured. 
This effect is due to a low  density of states in the analyzed range of chemical potential, much lower (at least two orders of magnitude) than in a standard metal.
{\color{black} To rule out the possibility that this would be due to finite-size effects, in our calculation, we analyzed $L \times L$ square lattices up to $L = 160$, checking numerical convergence of the superconductive gap parameters.}
Also a MF analysis on the $\eta_-$ and $\eta_0$ bands, starting from the effective theory in Eq. \eqref{eff0}, led to the same conclusion.  
}
%%%%%%%%%%%%%%%%%%%%%%%%%%%%%%%%%%%%%%%%%%%%%%%%%%%%
\begin{figure} [h!]
\includegraphics[scale=0.3]{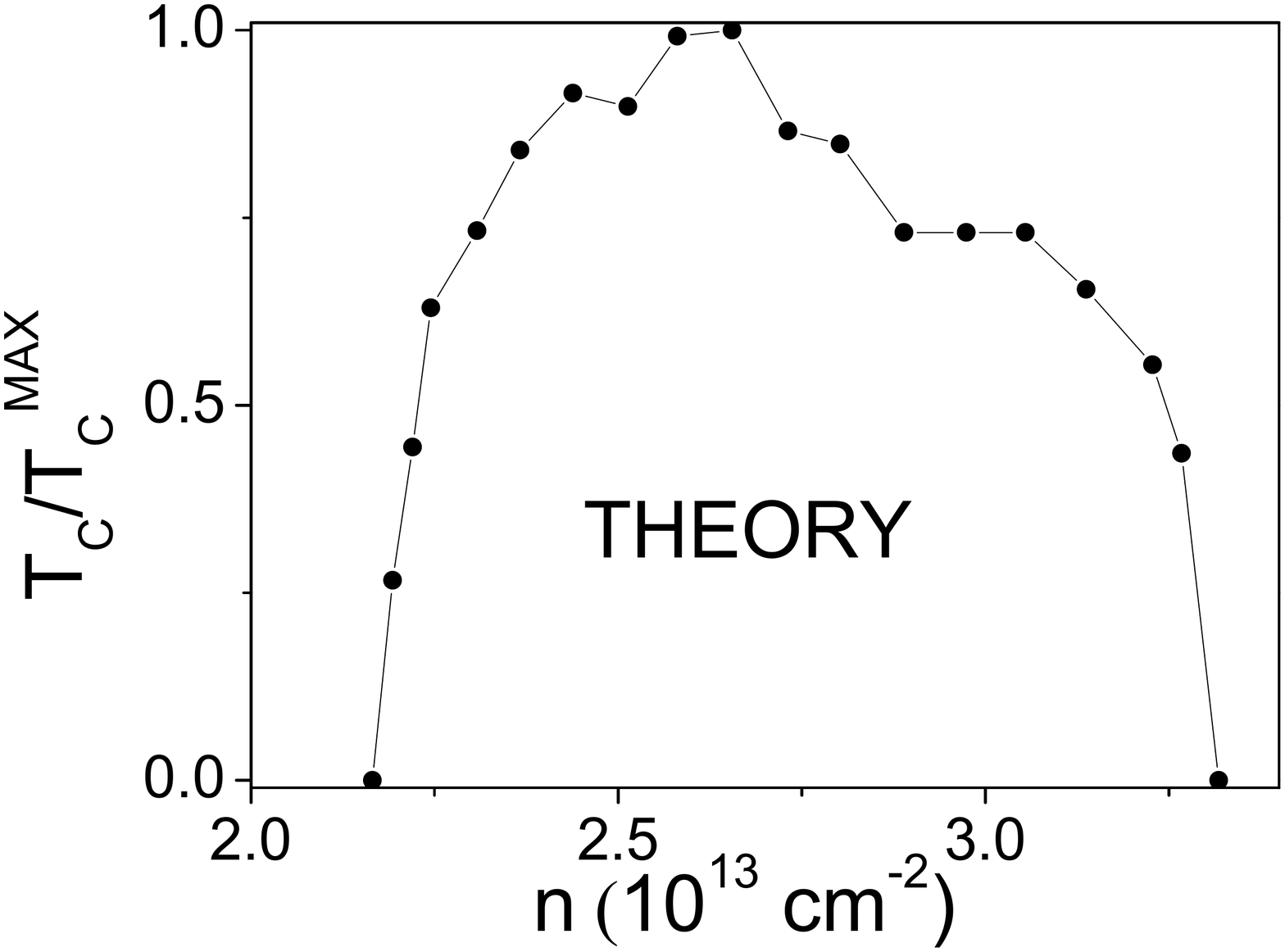}
\includegraphics[scale=0.3]{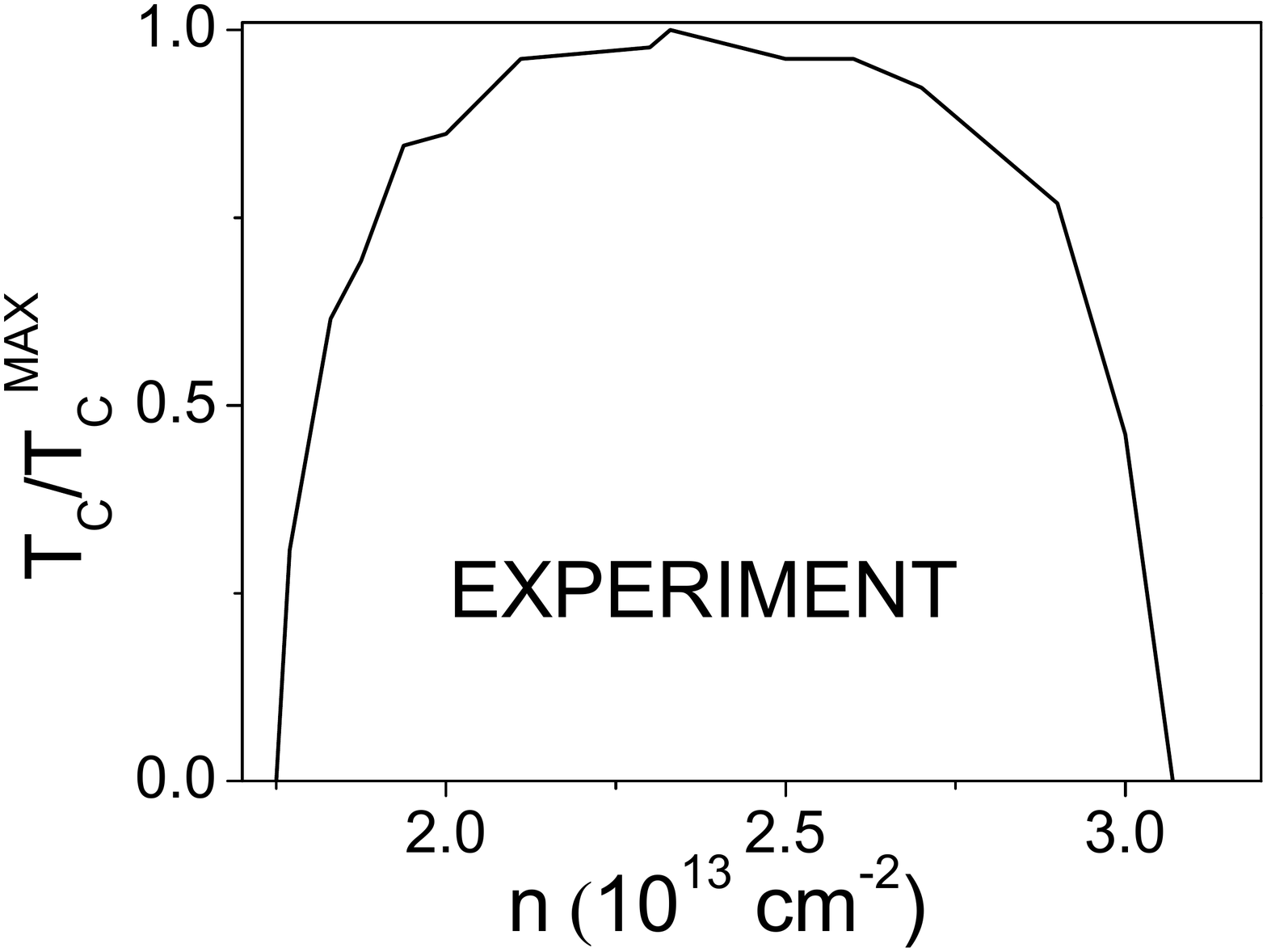}
\caption{Upper Panel: Relative critical temperature for singlet order parameters, at $U = 350$ meV, $\frac{V}{2} = 600$ meV, and $\mu$ varying, starting from the bottom of $\eta_0$ as calculate in this paper.
Lower Panel: Relative critical temperature taken from Ref. \cite{Stornaiuolo}}
\label{temp}
\end{figure}
%%%%%%%%%%%%%%%%%%%%%%%%%%%%%%%%%%%%%%%%%%%%%%%%%%%%%%%%%%%%%%%

Finally, in the right panel of Fig. \ref{pairingpl}, we plot 
the {\color{black} $\uparrow$ triplet amplitudes for the orbital $\tau$, again at $U = 350$ meV and $\frac{V}{2} = 300$ meV. 
In analogy with the singlet pairing, we always find $\Delta_{t , \sigma}^{yz}=\Delta_{t , \sigma}^{zx}$}, therefore the orbital $yz$ and $zx$ are  strongly
 coupled also in the triplet channel. {\color{black}  Moreover, since  there is no applied magnetic field,  
 $\Delta_{t, \uparrow}^{\tau}=\Delta_{t, \downarrow}^{\tau}$}. We notice that the triplet pairings always increase as a function 
 of the density,  then the reduction of the singlet pairing is compensated by an enhancement of the triplet channel, for all the values of the chemical potentials. 
%%%%%%%%%%%%%%%%%%%%%%%%%%%%%%%%%%%%%%%%%%%%%%%%%%%%%%%%%%%%%%%%
\begin{figure} [t]
\includegraphics[scale=0.27]{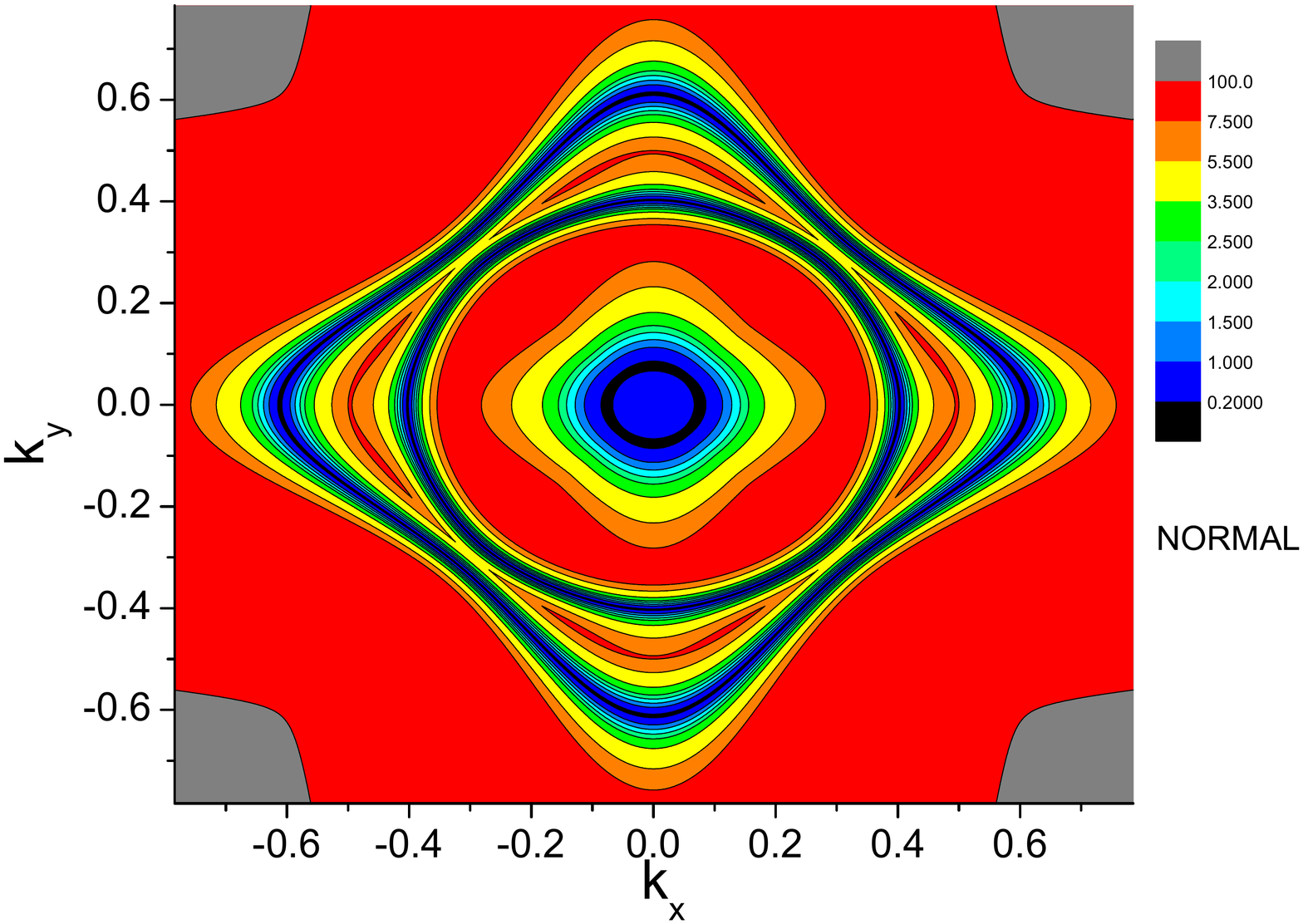}
\includegraphics[scale=0.27]{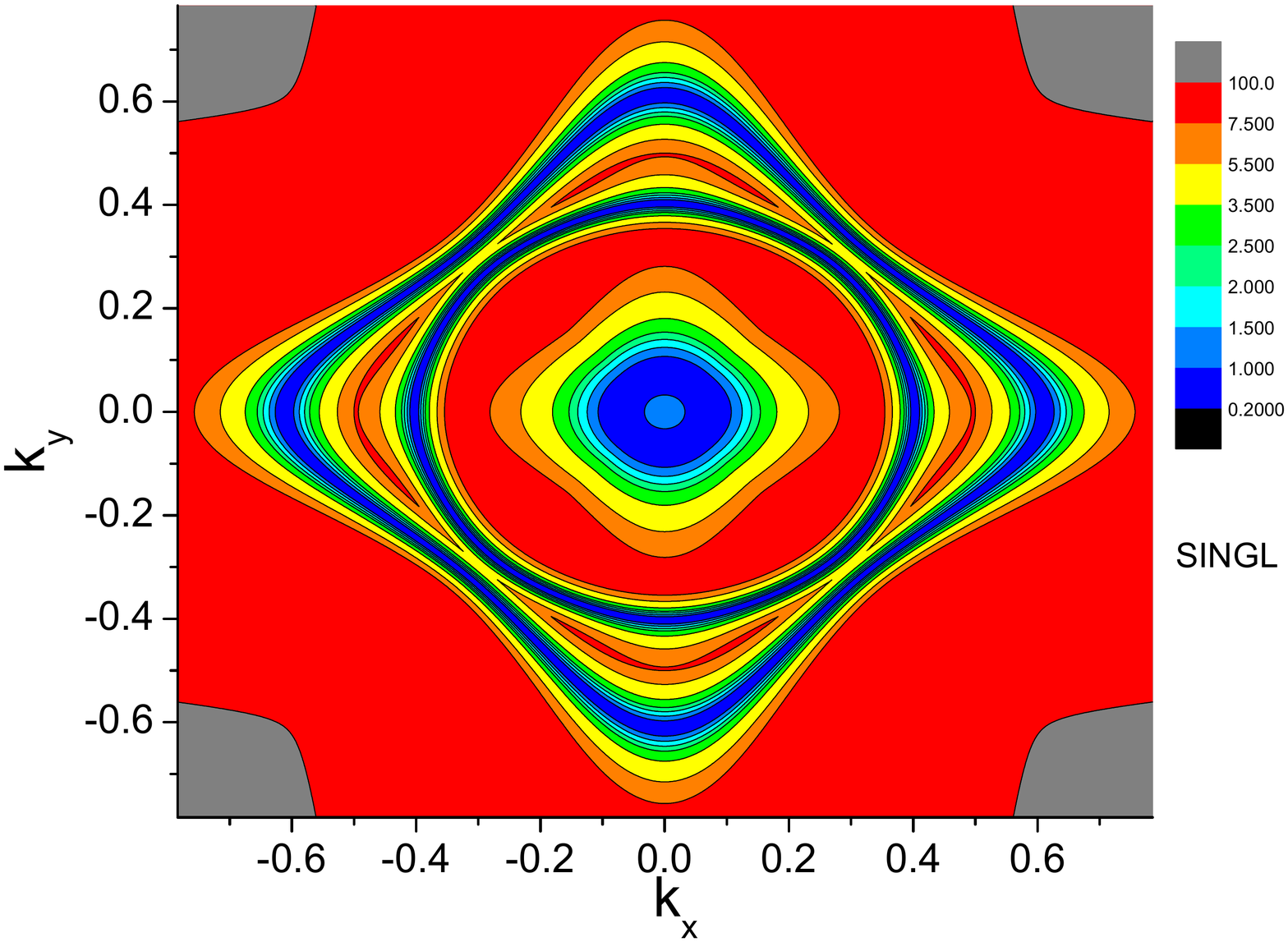}
\includegraphics[scale=0.27]{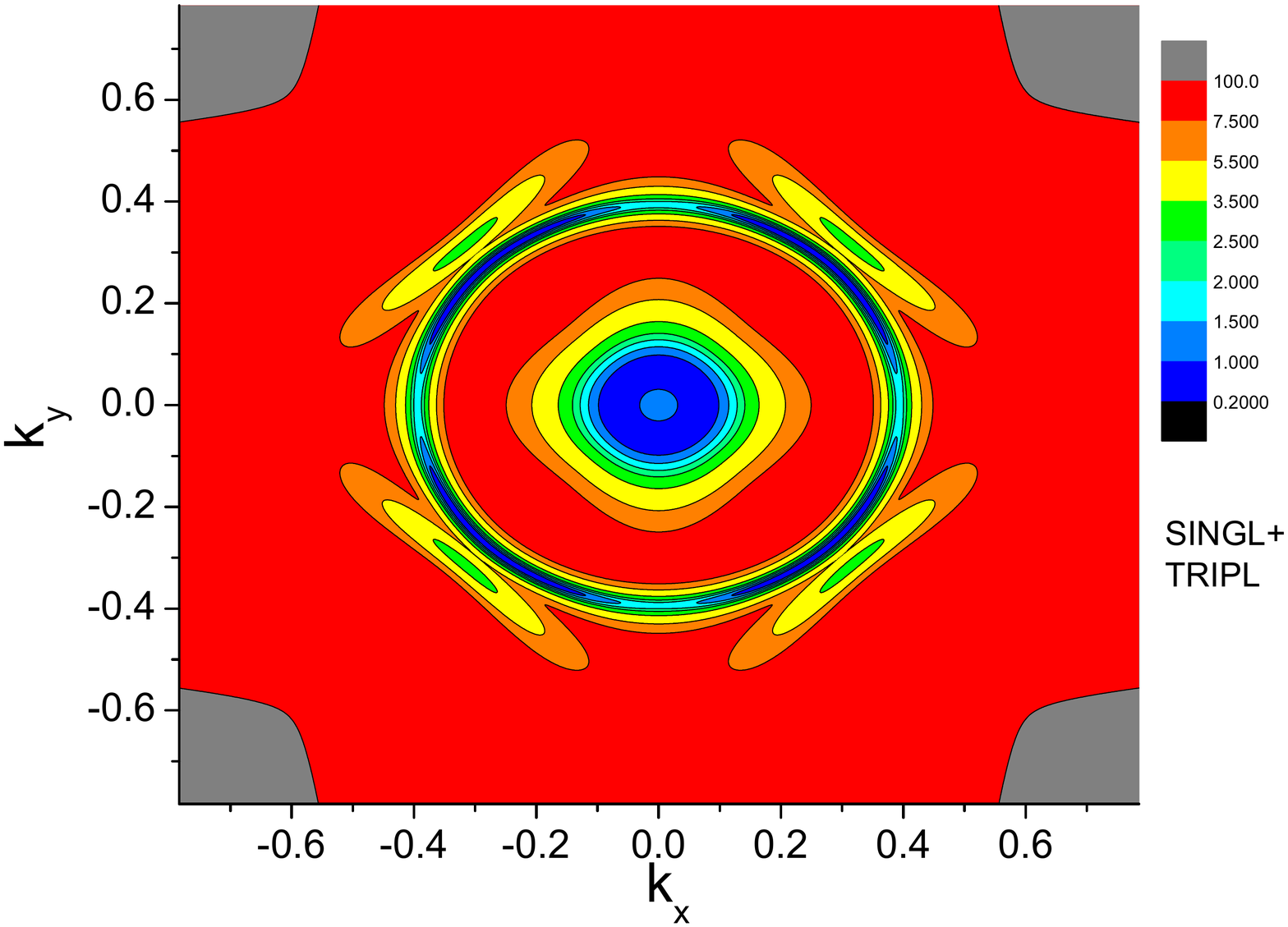}
\caption{{\color{black} Upper panel: Bogoliubov-de Gennes spectra at $\mu=-9$ meV, for $U = V = 0$ meV. Central panel: The same as 
in the upper panel, but for $U = 350$ meV, $V = 0$ meV. Lower panel: The same as in th upper panel, but for $U = 350$ meV 
and  $\frac{V}{2} = 300$ meV.}}
\label{contour}
\end{figure}
%%%%%%%%%%%%%%%%%%%%%%%%%%%%%%%%%%%%%%%%%%%%%%%%%%%%%%%%%%%%%%%%

{\color{black} In the lower panel of Fig. (\ref{phasefig}), we perform a comparison between the grand-canonical 
free energies of the normal state, of the singlet-pairing and the triplet-singlet superconductive ground-states, 
at  $U = 350$ meV and $V = 600$ meV. We point out that  the grand-canonical potential can be lowered in a significant
way if the triplet component is included in the energy balance.}

{\color{black}
In Fig. \ref{temp} (upper panel), we plot the relative critical temperature for the singlet order parameters, obtained by employing the mean-field approach, and by varying the particle density and by setting  $U = 350$ meV and $\frac{V}{2} = 300$ meV. Moreover, the range for the single-particle energy to 
starts from the bottom of the $\eta_0$ band.  We recover a dome-shaped plot, with the maximum critical temperature  
$T_c^{\mathrm{max}} \simeq 1$ K.  At this critical temperature, the triplet order parameters are reduced, in comparison to their 
value at zero temperature, however they are not necessarily zero. }

{\color{black}
More in detail, we locate the beginning the first dome at energy $E = \mu  = -10$ meV and electronic density $n = 2.14 \cdot 10^{13}$ $\mathrm{cm}^-2$, 
the maximum of $T_c$ at E = - 7 meV and $n = 2.6 \cdot 10^{13}$ $\mathrm{cm}^{-2}$, and finally
the end of the first dome at E = -2.5 meV and  $n = 3.3 \cdot10^{13}$ $\mathrm{cm}^{-2}$. These values 
can be compared e.g. with experimental data in \cite{Stornaiuolo}, summed up in the lower panel of Fig. \ref{temp}. There the same dome is measured
between $n = 1.8 \cdot 10^{13}$ $\mathrm{cm}^{-2}$ and  $n = 3.1 \cdot 10^{13}$ $\mathrm{cm}^{-2}$,
the top of the dome being around $n = 2.4 \cdot 10^{13}$ $\mathrm{cm}^-2$.  
Therefore,   we find a  reasonable agreement between theory and experiment. 
We remark that experimental data \cite{joshua2012universal,Stornaiuolo} have identified  a critical value $n_c$ of the density of the order of $1.8-1.9$ $10^{13}$ $\mathrm{cm}^{-2}$, which characterizes  the underdoped superconducting regime.  
Actually, in comparison with experimental results, as reported in Fig. \ref{temp}, the theoretical data show only a slight shift of the entire phase diagram. In our approach, this effect depends only on the energy $E_t$ which measures the energy distance between the band $xy$ and $yz$ in Hamiltonian \ref{e.5}. If a  slightly smaller value than 50 meV had been chosen,  a perfect agreement with the experimental critical temperature would be obtained}.

Since the triplet order parameter depends on the momentum ${\bf k}$, it is interesting to plot the Bogoliubov-de Gennes spectrum  obtained
by using the superconducting order parameters obtained 
from the minimization of the free energy $\Omega$. In Fig. \ref{contour}, we focus onto 
$\mu=-9$ meV, that is, a chemical potential slightly higher than the minimum of the band $\eta_0$. First, in the upper panel of 
Fig.  \ref{contour}, we plot the spectrum corresponding to the normal state.  The zeros correspond to the the center zone $\eta_0$ 
and the finite momentum zone $\eta_{-}$ whose origin is $xy$-like.  At finite momentum,
there is a small splitting of the zeroes, due to the fact that this value of the chemical potential is  close to the avoided crossing of the electronic bands.   
In the middle panel of  Fig. \ref{contour}, we plot the spectrum corresponding to the singlet pairing. The gaps correspond to 
the center zone $\eta_0$ and the finite momentum zone $\eta_{-}$ whose origin is $xy$-like. Therefore, the gaps at finite momentum are
smaller than those at center zone. However, when only the singlet is present,  along the $\Gamma X$ ($k_y=0$ or $k_x=0$) and $\Gamma M$ 
($k_x=k_y$) directions, the gaps are perfectly symmetric. Finally,  {\color{black} in the left panel of Fig. \ref{contour},} we consider the
combined effect of singlet and triplet pairings. We notice that the spectrum  drastically changes with respect to the case where only 
the singlet is present. In particular, the gaps get reduced along $\Gamma M$ direction. This behavior can be ascribed to the
{\color{black} $s_x$ and $s_y$ dependence of the order parameter given in Eq. (\ref{pairing}): along the $\Gamma M$ direction,} at finite wave-vector, 
the gap is mainly given by singlet channel.

To conclude, we mention the interesting possibility to consider possible superconductive solutions from the continuum effective Hamiltonians in 
Eqs. \eqref{eff-} and \eqref{eff0}. We analyzed their presence, by adding an attractive potential as in Eq. \eqref{hamilpair}. Working with lattice sizes up
to $L = 256$, we found only normal solutions in large ranges for $U$ and $V$ (as large as in Section \ref{super}). We ascribe this outcome to 
relevant finite-size effects at the considered low fillings $\nu < 0.1$, required for the validities of the effective theories
in Eqs. \eqref{eff-} and \eqref{eff0}.

%%%%%%%%%%%%%%%%%%%%%%%%%%%%%%%%%%%%%%%%%%%%%%%%%%%%%%%%%%%%

\subsection{Clues of emerging topology at zero magnetic field}
\label{topnoB}

The pairing in Eq. \eqref{pairing},  preserving time-reversal invariance,
leads to a Bogoliubov Hamiltonian in the class DIII of the classification for topological insulators and superconductors 
\cite{ludwig2009, ryu2016, Bernevigbook}. Indeed $U_T = {\bf I}_{3 \mathrm{x} 3} \otimes  {\bf I}_{2 \mathrm{x} 2} \otimes \sigma_y$, 
$U_T \,  U_T^* = {\bf I}_{12 \mathrm{x} 12}$ and $U_C = {\bf I}_{3 \mathrm{x} 3} \otimes \sigma_y \otimes \sigma_y$, 
$U_C \,  U_C^* =-{\bf I}_{12 \mathrm{x} 12}$ (the second matrices in the Kronecker products acting on the Nambu-Gorkov indices). The chiral symmetry $U_S = U_T \, U_C = U_C \, U_T$,  $U_S^2 = {\bf I}_{12 \mathrm{x} 12}$, 
is also realized, however not affecting the topology class (see Appendix 1).
In this class,  nontrivial topological configurations are possible. {\color{black} Therefore, in the following we are going to {\color{black} explore}
this possibility.}
%%%%%%%%%%%%%%%%%%%%%%%%%%%%%%%%%%%%%%%%%%%%%%%%%%%%%%%%%%%
\begin{figure*} [t!]
\includegraphics[scale=0.28]{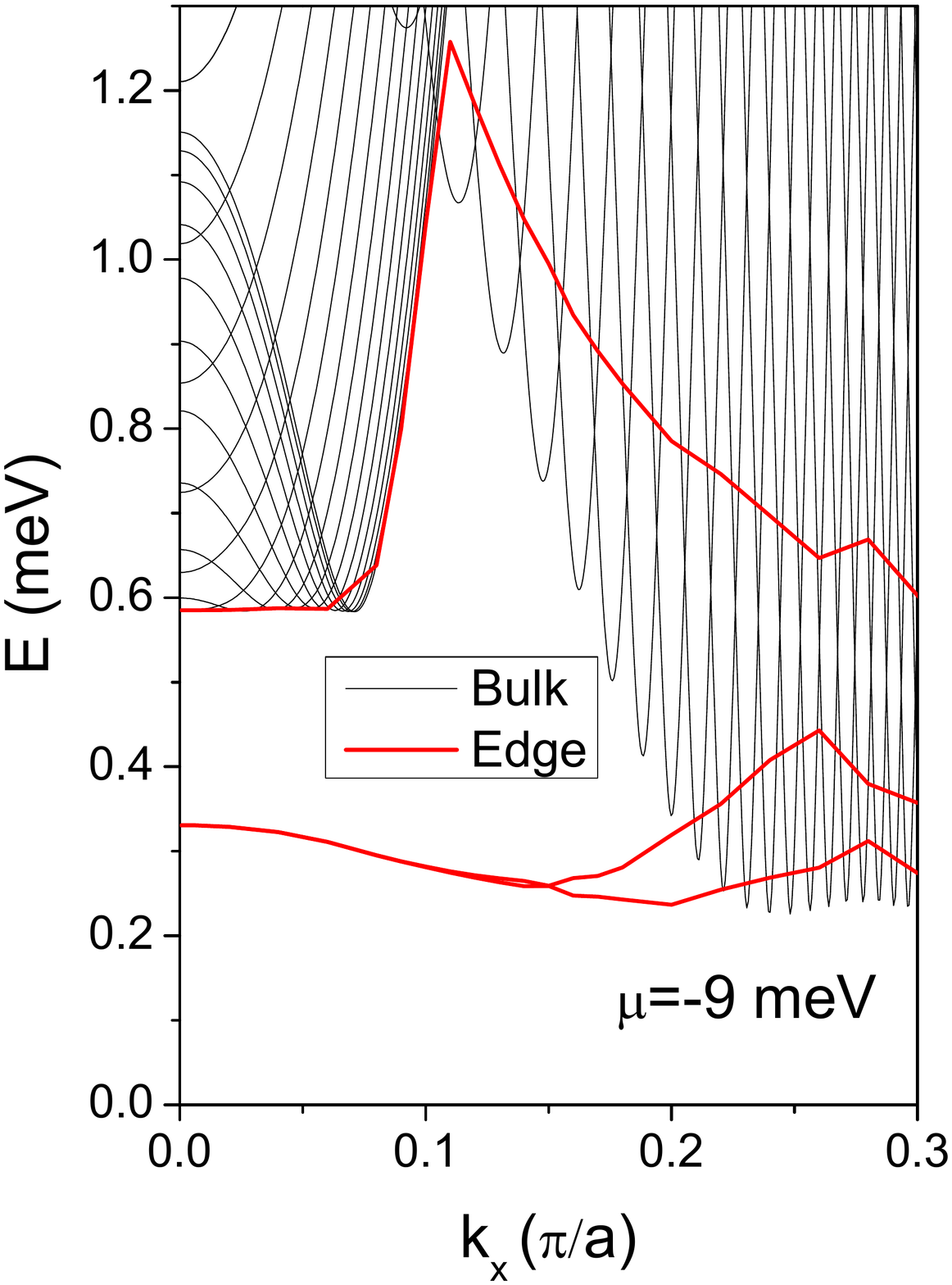}
\includegraphics[scale=0.28]{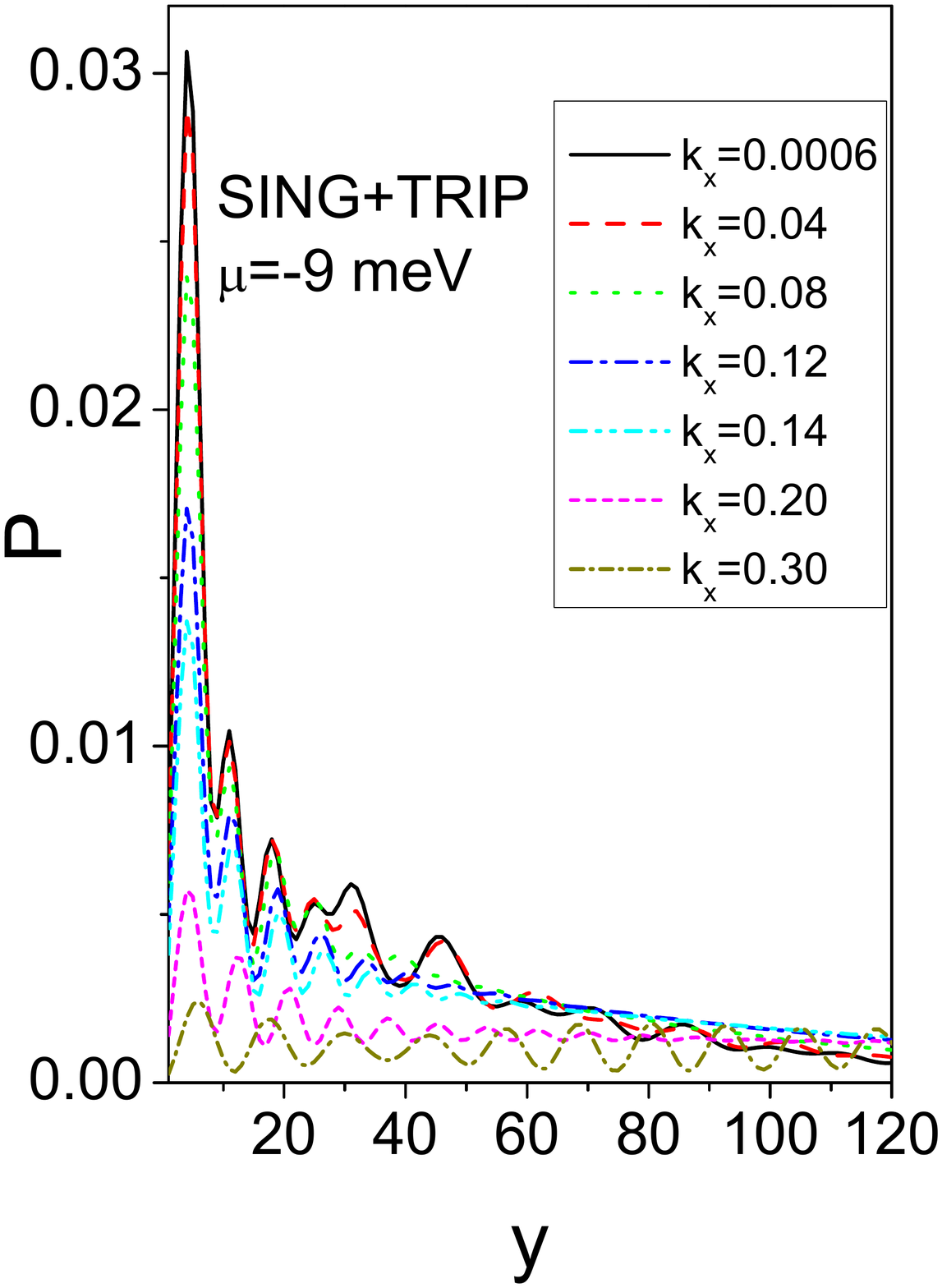}
\includegraphics[scale=0.28]{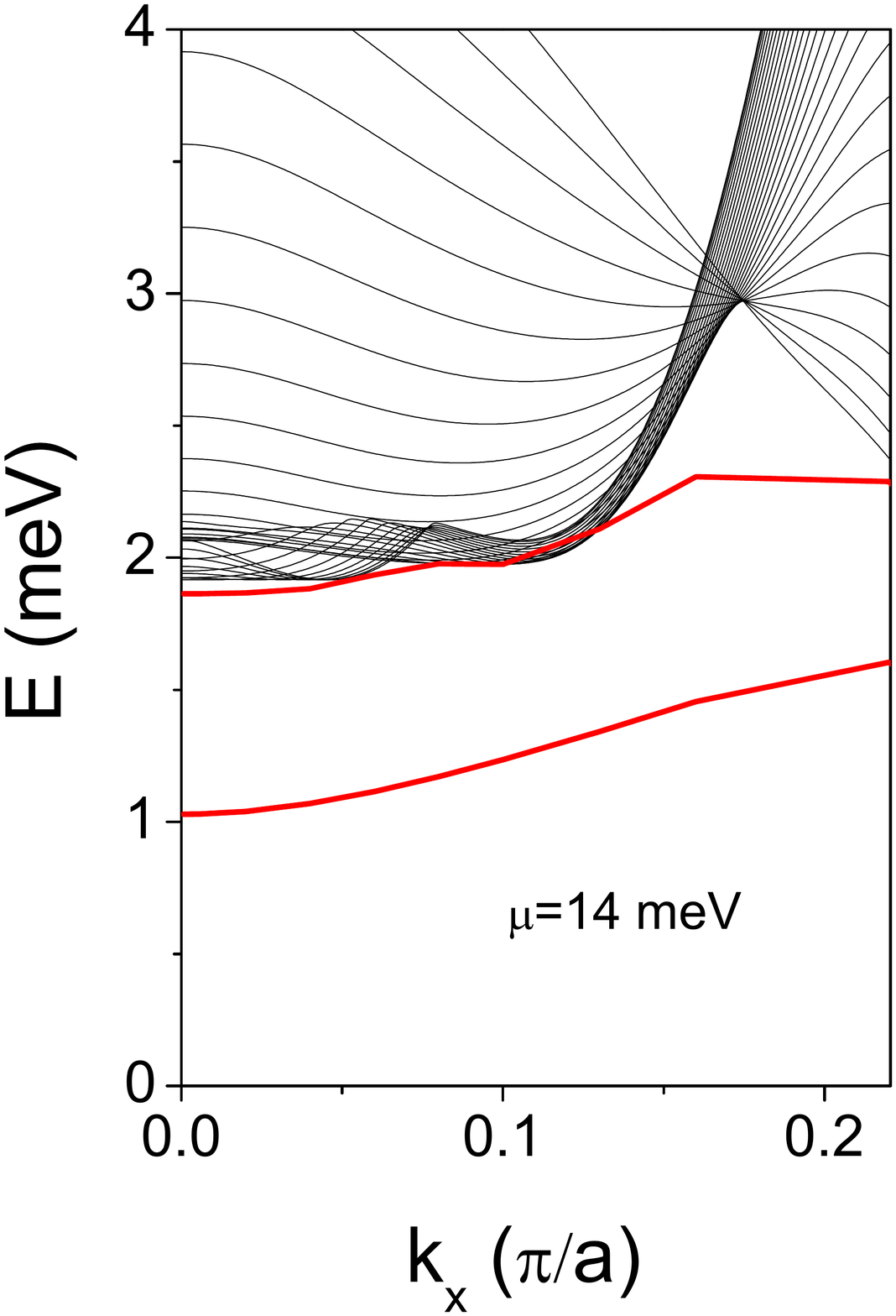}
\caption{{\color{black} Bogoliubov spectrum for varying $k_x$, $U = 350$ meV, $V = 600$ meV, at $\mu = -9$ meV (first dome) in the left panel, 
at $U = 310$ meV, $V = 520$ meV $\mu = 14$ meV (second dome) in the right panel. The probability P of the first edge state corresponding to the 
spectrum of the left panel as a function of the position y (in units of the lattice parameter $a$) for different values of $k_x$. For all the plots, 
a finite size $L_y = 950$ along the $y$-axis is adopted.}}
\label{topofig1}
\end{figure*}
%%%%%%%%%%%%%%%%%%%%%%%%%%%%%%%%%%%%%%%%%%%%%%%%%%%%%%%%%%%%

{\color{black} We have analyzed the topological features of the (minimal energy) mean-field superconducting solutions,
within the tight-binding scheme exposed in the above}. 
In particular, we have examined different values of chemical potential. {\color{black} Correspondingly, we found signatures of helical superconductivity, 
both in the first and second dome. As an example, in the left and middle panels of Fig. \ref{topofig1},  we show the Bogoliubov spectrum at $\mu=-9$ meV 
(that is,  in the correspondence of the second dome, 
that means for the $\eta_0$ bands),  for a lattice that is infinite along the  $x$-axis, but is finite along $y$-axis, 
thus breaking  translational invariance in that direction. 
In particular, along the $y$-axis, we have considered a finite size of $L_y = 950$ sites, with hard-wall boundary conditions. Then,} we 
have calculated the excitation spectrum at fixed $k_x$, finding that two {\color{black} finite-energy edge states are present   within  the bulk gap,
which become degenerate  at $k_x=0$}.  This can be ascribed to the fact that, at
the onset of the first  dome ($\mu \simeq -10$ meV), as shown in Fig. \ref{bande}, the energy spectrum in the normal 
state does not present a linear Rashba coupling. 
Therefore, in the limit of small wave-vectors, the dispersion of the edge states is not linear, as a function of $k_x$. 
Rather, it is nearly flat. As shown in the middle panel of Fig.  \ref{topofig1}, for small values of $k_x$, the wavefunctions 
corresponding to the in-gap edge states are well localized close to an edge of the system.  With increasing $k_x$, their wave-function tends to spread over all the bulk.

We point out that the observed  behavior is mostly related to the triplet pairing. Indeed, at zero magnetic field, the singlet pairing term  
alone is not able  to provide nontrivial topological phases. However, at  $\mu=-9$ meV we have checked a topological phase transition, induced by 
the attractive term $V$ in Eq. \eqref{hamilpair} when the triplet order parameters are stable (see Fig. \ref{phasefig}).  In particular, the results 
shown in Fig. \ref{topofig1} are obtained when singlet and triplet pairings coexist above a critical value of the attractive potential $V$.   
{\color{black} In Section \ref{emtop}, we discuss in detail the nature of those phase  transitions.}

We have also analyzed some features, possibly topological, at higher values of the carrier density. In the left panel of 
Fig.  \ref{topofig1}, we report the excitation spectrum, with  finite size along $y$-axis and at $\mu=14$ meV. Even if
the excitation spectrum is more complex in comparison with that shown at lower densities,  at zero magnetic field, we still
find the presence of in-gap edge states, which gradually merge in the bulk continuum, with increasing $k_x$.\\

\subsection{Further insights from the  effective theory}
\label{emtop}

Starting from the full multiband Hamiltonian in Eq. \eqref{e.1}, it is rather cumbersome to characterize entirely 
the full topology content of our system, since {\color{black} it is tough to recover the required analytical form
of the band wave-functions in momentum space \cite{ludwig2009, ryu2016, Bernevigbook}.} However, this 
task can be achieved, at least partially, starting from the effective Hamiltonians in Eqs. \eqref{eff-} and especially \eqref{eff0}.
This approaches demonstrates to be valuable also for a better characterization {\color{black} of the entire phase diagram in the considered range of fillings,} 
as well as of the nature of the first superconducting dome.

Since in most of the available literature \cite{gariglio2016}, the superconducting dome is located approximately in correspondence  of 
the $\eta_0$ band, we focus mainly on $H_0^{(\mathrm{eff})} ({\bf k})$ in Eqs. \eqref{eff0}, {\color{black} with superconducting
pairings added. We recall that $H_0^{(\mathrm{eff})} ({\bf k})$ is expected to described properly at least the lower-density half of 
the dome. Moreover, the resulting Bogoliubov Hamiltonian shares the same (time-reversal  and charge-conjugation) symmetry content as the 
multiband one in Eq. \eqref{hamilbogo}, therefore allowing (but not guaranteeing) the same topological phase structure. 
Indeed, a similar discussion  can be performed directly for the $\tau$ bands, obtaining the same qualitative results: for a given pair content,
the effective spin-orbit ${\bf g} ({\bf k})$ (absent in the  $\tau$ bands basis) has only quantitative effects on the phase diagram.

We assume again  a pairing ${\bf \Delta } ( \bf{k} )$, in the form of
 Eq. \eqref{pairingmat} ({\color{black} at this stage, we do not need to specify  the precise form of the corresponding 
 attractive potential, which instead we generically denote as $G$)}. 
The resulting Bogoliubov Hamiltonian} is of the general form
$H_0^{\mathrm{BG}} = \frac{1}{2} \sum_{\bf{k}}
\psi^{\dag}_{\bf{k}} \mathcal{H} (\bf{k}) \psi^{\ }_{\bf{k}}$, 
with
\beq 
\label{bog0}
\mathcal{H} ( \bf{k} )
=
\left(\begin{array}{cc}
h_0 ( \bf{k} ) & {\bf \Delta } ( \bf{k} )
 \\
{\bf \Delta}^{\dag} ( \bf{k} ) & - h_0^{T} ( - \bf{k} ) \, 
\end{array}\right)
\eeq
$\psi_{\bf{k} } = ( c^{\ }_{{\bf k} \uparrow}, c^{\ }_{\bf{k} \downarrow}
c^{\dag}_{- \bf{k} \uparrow}, c^{\dag}_{- \bf{k} \downarrow} )^{\mathrm{T}}$
and 
\beq
h_0 ( {{\bf k}} )
=
\xi_0 ({\bf k} )  \, {\bf I}_{2 \mathrm{x} 2} + {\bf g} ({\bf k}) \cdot \bm{\sigma } \, .
\label{eq:HnormalNCS}
\eeq
with $\xi_0 ({\bf k} ) = \epsilon_0 ({\bf k} ) -  \mu$ ($\mu$ measured from $\epsilon_0 ({\bf k} = 0)$).
By time-reversal-symmetry, it turns out  that $\xi_0 ({\bf k})$
and ${\bf g}({\bf k})$ are symmetric and antisymmetric in ${\bf k}$,
respectively \cite{sato2006}. 
The Hamiltonian in Eq. \eqref{bog0} has the spectrum \cite{sato2006}
\beq
E_{\pm}({\bf k}) = \sqrt{\Big(\xi_0 ({\bf k} ) \pm |{\bf g} ({\bf k})|\Big)^2 + \Big( \Delta_{s} \pm  |{\bf d} ({\bf k})|\Big)^2}  
\label{spettrototale}
\eeq 
if  ${\bf d}({\bf k})$ (see Eq. \eqref{pairingmat}) is aligned with ${\bf g}({\bf k})$, which is 
the most likely  possibility, as we noted above.
In three dimensional systems, this spectrum can give rise to topologically-protected nodal lines, where $E_-({\bf k}) = 0$,
leading to nodal (non-centrosymmetric) superconductors. These systems require a partially different topological classification from the ten-fold way for fully gapped ones {\color{black}\cite{brydon2015,samokhin2015}, as well as for their edge modes \cite{samokhin2016}.}
{\color{black} However, in two dimensions, only isolated zeros for  $E_{\pm}({\bf k}) = 0$ can occur, not topologically protected, and identified as transition points.}

We analyze separately {\color{black} three configurations for ${\bf d}({\bf k})$:} {\it i)} pure s-wave (spin singlet) pairing, {\it ii)} pure p-wave (triplet) state, {\it iii)} 
mixed s-p waves pairing. In the first configuration, no topology is realized: the phase is continuously connected with a purely BCS one, where ${\bf g} ({\bf k}) = 0$. Indeed,
in spite of 
${\bf g} ({\bf k}) \neq 0$, since $\Delta_{s} \neq 0$, no zeros of $E_{\pm}({\bf k})$ can occur in between {\color{black} and for a certain ${\bf k}$. This regime clearly 
corresponds to that in Section \ref{MFsol} for the multiband model in Eq. \eqref{hamilbogo}, at $\mu = -9$ meV, $U<300$ meV and $V = 0$.

In the normal state, due to the terms in Eqs. \eqref{e.7} and \eqref{e.8}, mixing the spins, only the sum of the spin currents
is conserved, related to the global symmetry group
$U(1)_V$, in turn defined by the transformations:  
$c_{\uparrow} ({\bf k}) \to e^{i \theta} \, c_{\uparrow} ({\bf k})$, $\theta = [0, 2 \pi)$, and the same for $c_{\downarrow} ({\bf k})$. 
Therefore, due to the bilinear pairing condensate $\langle c_{\uparrow} ({\bf k}) c_{\downarrow} (-{\bf k}) \rangle = \Delta_s$, 
the spontaneous symmetry breaking
\beq
U(1)_{V}  \to Z_{2 \, V}
\label{pat1}
\eeq
occurs, where for $Z_{2 \, V}  \in U(1)_{V}$, $\theta = \{ 0, \pi \}$.}

Similarly, in the configuration {\it ii)}, when $\Delta_{s} = 0$, two phases can be realized a  priori,  {\color{black} in this scheme discerned by 
the sign of $\mu$ \cite{readgreen,Bernevigbook}.  These two phases, one with non trivial and one with trivial topologies, are continuously connected with those 
at ${\bf g} ({\bf k}) = 0$.}  This because the unique zeros of $E_{\pm}({\bf k})$ can occur where ${\bf d} ({\bf k}) = 0$, that means at ${\bf k} = 0$. In this
point, also ${\bf g} ({\bf k}) = 0$ and $\epsilon_0 ({\bf k}) = 0$, therefore $E_{\pm}({\bf k}) = \mu$. Now, as {\color{black} $G \to 0$,} then $\mu \approx 
k_F >0$ (since the filling of the $\eta_0$ band is positive), {\color{black} while the condition $\mu = 0$, separating the two phases, occurs for  $G$ large 
enough (similarly as in the previous Section).  Again, the breaking pattern in Eq. \eqref{pat1} is realized.}

The data from the mean-field analysis on the lattice suggests that the topological phase sets in for
$U \geq 250$ and $V \geq 550$ meV,  where we find the triplet pairing (see Fig. \ref{phasefig}).   
The same phase, a superconducting counterpart of a quantum spin Hall phase, is the unique possible {\color{black} topological} one in the DIII class 
{\color{black} and in two dimensions} \cite{ludwig2009, ryu2016, Bernevigbook}, and it is also known as helical superconducting phase. Moreover,
this is labelled by a topological index $n = 1$, witnessing the number of {\it pairs} of edge states, related by the time-reversal conjugation 
\cite{qi2009,Bernevigbook}. This index (written for the general case e.g. in \cite{sato2017}) becomes, in the limit of decoupled spin sectors,
${\bf g} ({\bf k}) \to 0 , \, \, \forall \, {\bf k}$, \cite{Bernevigbook}:
\beq
n  = \frac{1}{2} \, (n_{\uparrow} - n_{\downarrow}) \quad \mathrm{mod} \, 2 \, ,
\label{phase}
\eeq
where $n_{\uparrow} = 1$ and $ n_{\downarrow} = -1$ label the topological phases (with broken time-reversal invariance, then in the D class of the ten-fold way 
classification \cite{ludwig2009, ryu2016, Bernevigbook}) described in \cite{readgreen}.  The numbers $n_{\uparrow / \downarrow} = \pm 1$  correspond to nontrivial 
element of the first homotopy class \cite{nakahara,coleman1,coleman2}, $\pi_1(\tilde{O)}$, on a circle $\tilde{O}$ around the  high symmetry node at ${\bf k} = 0$.
In particular, the nontrivial homotopies are related to the phase factors $e^{\pm i \, \phi_{\bf k}}$ parametrizing as follows the nonvanishing p-wave gaps
(connected by time reversal conjugation): 
\beq
\Delta_{t , \uparrow  /  \downarrow} \,  (k_x \pm  i \, k_y) =   \Delta_{t , \uparrow / \downarrow} \, |{\bf k}| \, e^{\pm i \, \phi_{\bf k}} \, .
\label{phase2}
\eeq  
Therefore, for instance, $n = 0$ for the extended s-wave case, considered in \cite{zeg2020}.
In the more general case where the spin sectors are coupled together (as when ${\bf g} ({\bf k}) \neq 0$, or in the presence of a superconducting pairing
between the  two spin-species, as in the case {\it iii)}), the topological index {\color{black} $n = 1$} can be expressed in terms of Chern numbers of positive- 
and negative-energy eigenstates \cite{sato2017}. Alternatively, the same index can be defined as a winding number of a phase \cite{yokoyama2014}, exploiting 
directly the full Bogoliubov Hamiltonian $H_{\mathrm{BG}}({\bf k})$, instead of its bands ({\color{black} see \cite{notatop} for more details)}.

Finally, in the configuration {\it iii)}, where singlet and triplet pairings coexist, a trivial and a topological phase can be again realized a priori. 
Referring to the Hamiltonian in Eq. \eqref{eq:HnormalNCS}, they are separated by the condition (on $G$, 
$\mu$, $\Delta_{s}$, and  ${\bf d} ({\bf k})$) that $E_-({\bf k}) = 0$, for a certain $|{\bf k}|$. This is expected for $|\Delta_{s}| \sim |\Delta_{t , \uparrow}| = |\Delta_{t , \downarrow}|$.
 Again, the breaking pattern in Eq. \eqref{pat1} is realized.
Being again in the DIII class of the ten-fold way classification, the two phases belong respectively to the same topological classes of the phases in the case {\it ii)},
then they are discerned by the same topological index \cite{yokoyama2014,sato2017}. 

This situation corresponds to the singlet-triplet coexisting regime, found in Section \ref{MFsol}. 
Correspondingly, the presence of edge states  indicates
 that between the two possible phases described above,  
the topological one is realized. In the following Section, we will analyze its response to an applied magnetic field.
%%%%%%%%%%%%%%%%%%%%%%%%%%%%%%%%%%%%%%%%%%%%%%%%%%%%%%%%%%%%%%%%
\begin{figure} [t!]
\includegraphics[scale=0.32]{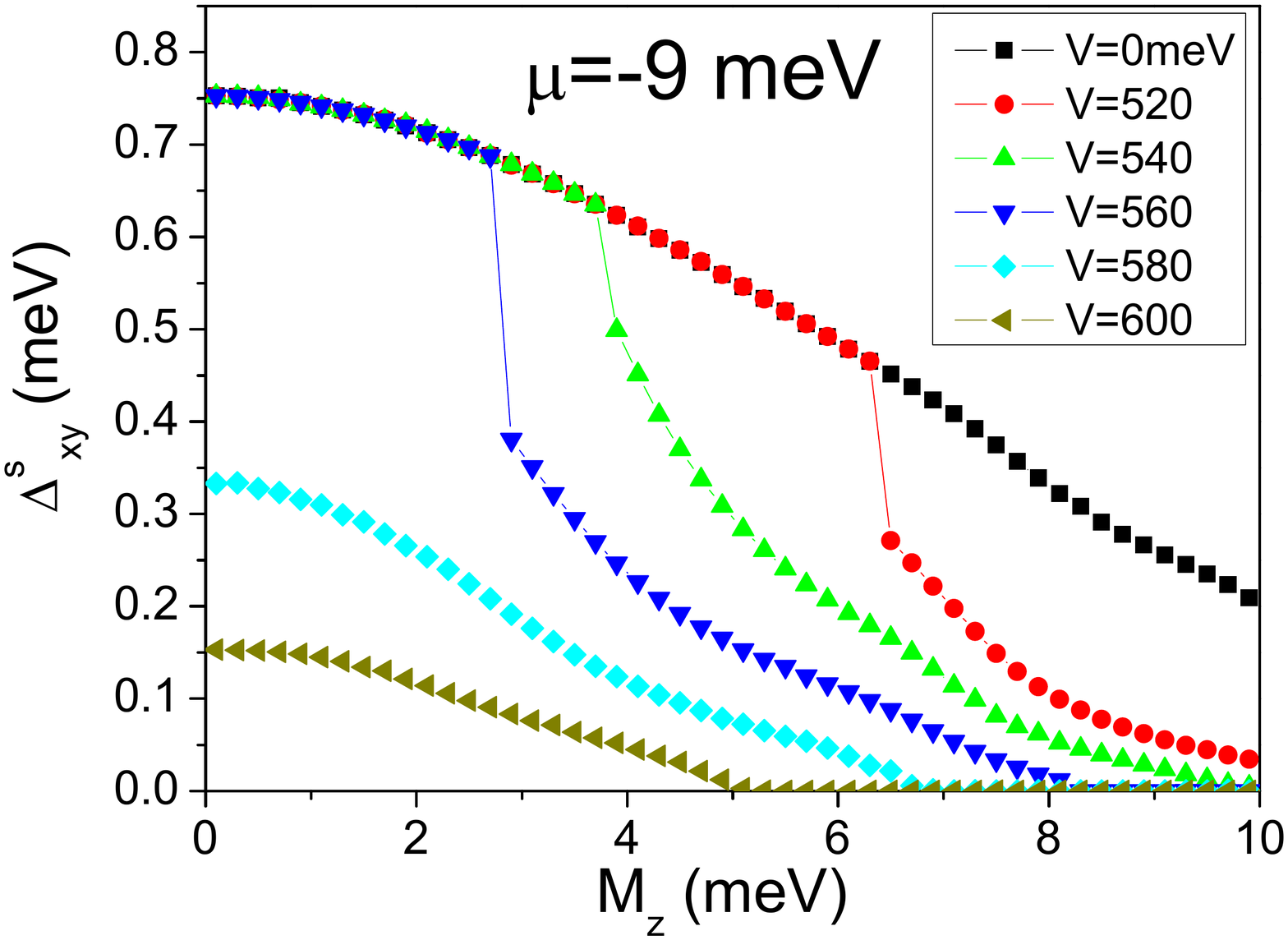}
\includegraphics[scale=0.32]{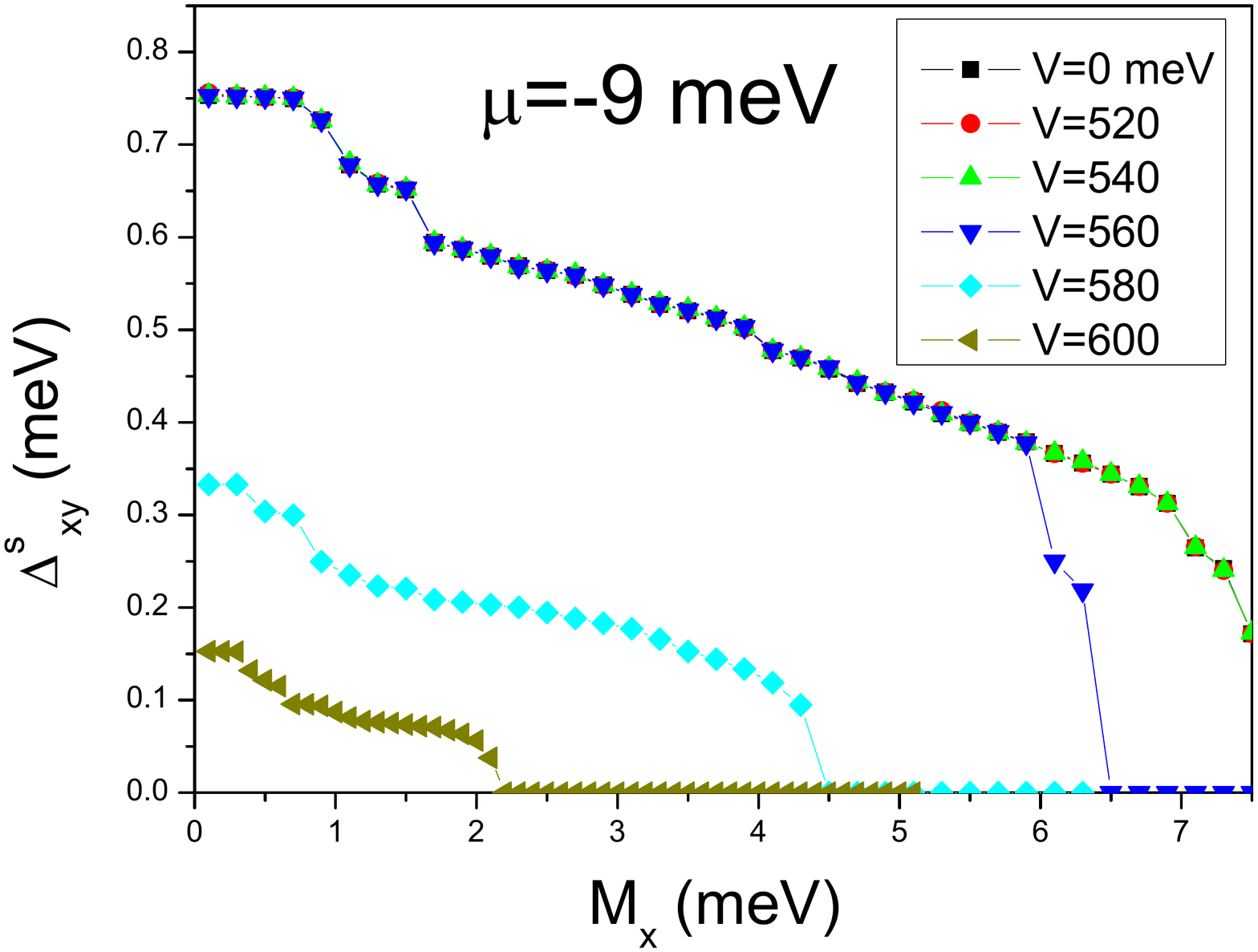}
\caption{{\color{black} Upper panel: Singlet order parameter $\Delta^{xy}_s$, as a function of the magnetic field along the $z$-axis 
for $U = 350$ meV and for  different values of the attractive term $V$ and at $\mu=-9$ meV.
Lower panel: Same as in the upper panel, but with the magnetic field along the $x$-axis (lower panel).
We recall that in Eq. \eqref{hamilpair} the nearest neighbour pairing has coupling $\frac{V}{2}$.}}
\label{magnfig}
\end{figure}
%%%%%%%%%%%%%%%%%%%%%%%%%%%%%%%%%%%%%%%%%%%%%%%%%%%%%%%%%%%%%%%%

\section{Superconducting solutions in the presence of a magnetic field}
\label{supermagnetic}

{\color{black} It is important, on the experimental point of view as well, to consider how  the  above scenario is
modified under the application of a uniform  magnetic field, inducing  the Zeeman term in Eq. \eqref{e.9}. For this purpose, we start with 
discussing how an applied magnetic field modifies the superconducting order parameter.} 

\subsection{Effects of the magnetic field on the superconducting order parameter}

The first macroscopic effect 
is the breakdown of the time-reversal symmetry. {\color{black} Along this direction,} 
 we analyze the effects of an external magnetic field on the superconducting solutions, considering field orientations 
both along the $z$-axis (out-plane) and the $x$-axis (in-plane). 

First, we focus on  the singlet order parameter related to the orbital $xy$, $\Delta_{xy}^s$. We recall that this order parameter 
provides an estimate of the minimal superconducting gap in the excitation spectrum. In the upper panel of Fig. \ref{magnfig},  
we plot {\color{black} this quantity}, as a function of $M_x$, 
and for different values of $V$.  
In particular, we focus on the under-doped regime of the first dome, at $\mu=-9$ meV.
For $V=0$, there is a continuous curve, as a function of the magnetic field.  However, there is a range of values for $V$
where the order parameter shows a discontinuity. Actually, this discontinuity marks the {\color{black} onset} of the triplet pairing, 
which, on the other hand, start weakening the singlet order parameter (see Fig. \ref{phasefig}). As shown below,  {\color{black} the
triplet pairing can lower the ground-state energy, at increasing magnetic field and fixed $V$.}  Therefore, there is a transition between 
a phase with only the singlet pairing to a phase with combined singlet and triplet {\color{black} pairings}.  With increasing $V$, the triplet pairing 
becomes stable at lower strengths of the magnetic field (Indeed,  
{\color{black}  we have shown above that, for  $V > 560$ meV, there is the formation of the triplet pairing even 
at $M_z=0$).  For these values of $V$, the reduction of the singlet order parameter is quite rapid as a function of the strength of $M_z$.} 

In the lower panel of Fig. \ref{magnfig},   we plot $\Delta_{xy}^s$ as a function of  
$M_x$, for different values of $V$. There is a different coupling between the in-plane magnetic field and the triplet order parameter. 
{\color{black}  Therefore, the role of the triplet pairing is affected  not only by the density, but also by the orientation of the magnetic 
field. Indeed, it can be responsible of the experimentally measured magnetic field anisotropy, in some 
superconducting properties, between the in-plane and out-of-plane field configurations \cite{caviglia2009}. In particular, in the under-doped regime of the
first dome, at $\mu=-9$ meV, the anisotropy emerging from the comparison between the upper and lower panels of Fig. \ref{magnfig} is present
but not marked. Indeed, the behavior of the order parameter, as a function of $M_x$, follows that as a function of $M_z$, the first behavior being characterized
only by slightly smaller values of critical fields (at different $V$).}
%%%%%%%%%%%%%%%%%%%%%%%%%%%%%%%%%%%%%%%%%%%%%%%%%%%%%%%%%%% 
\begin{figure*} [t!]
\includegraphics[scale=0.28]{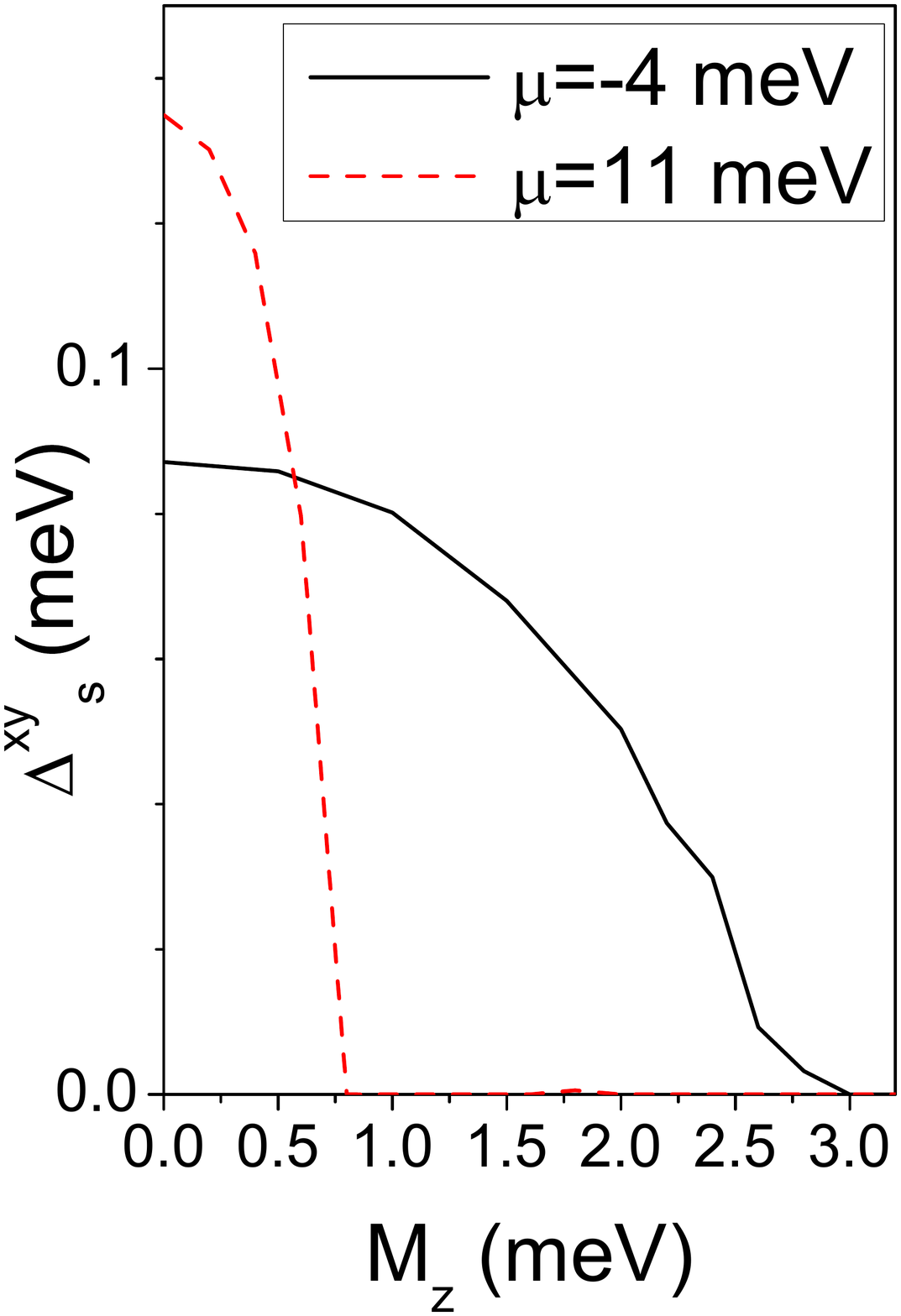}
\includegraphics[scale=0.28]{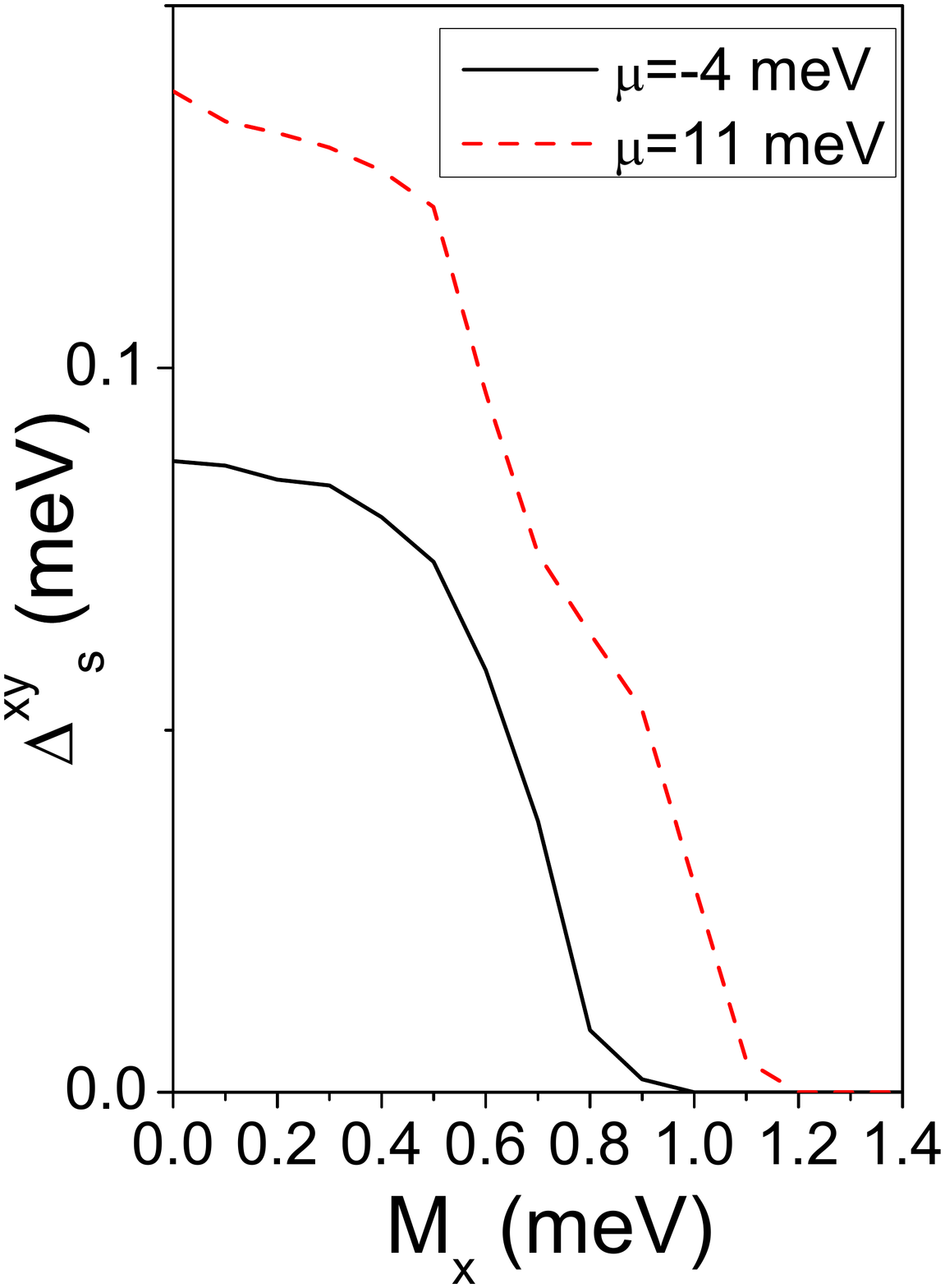}
\includegraphics[scale=0.28]{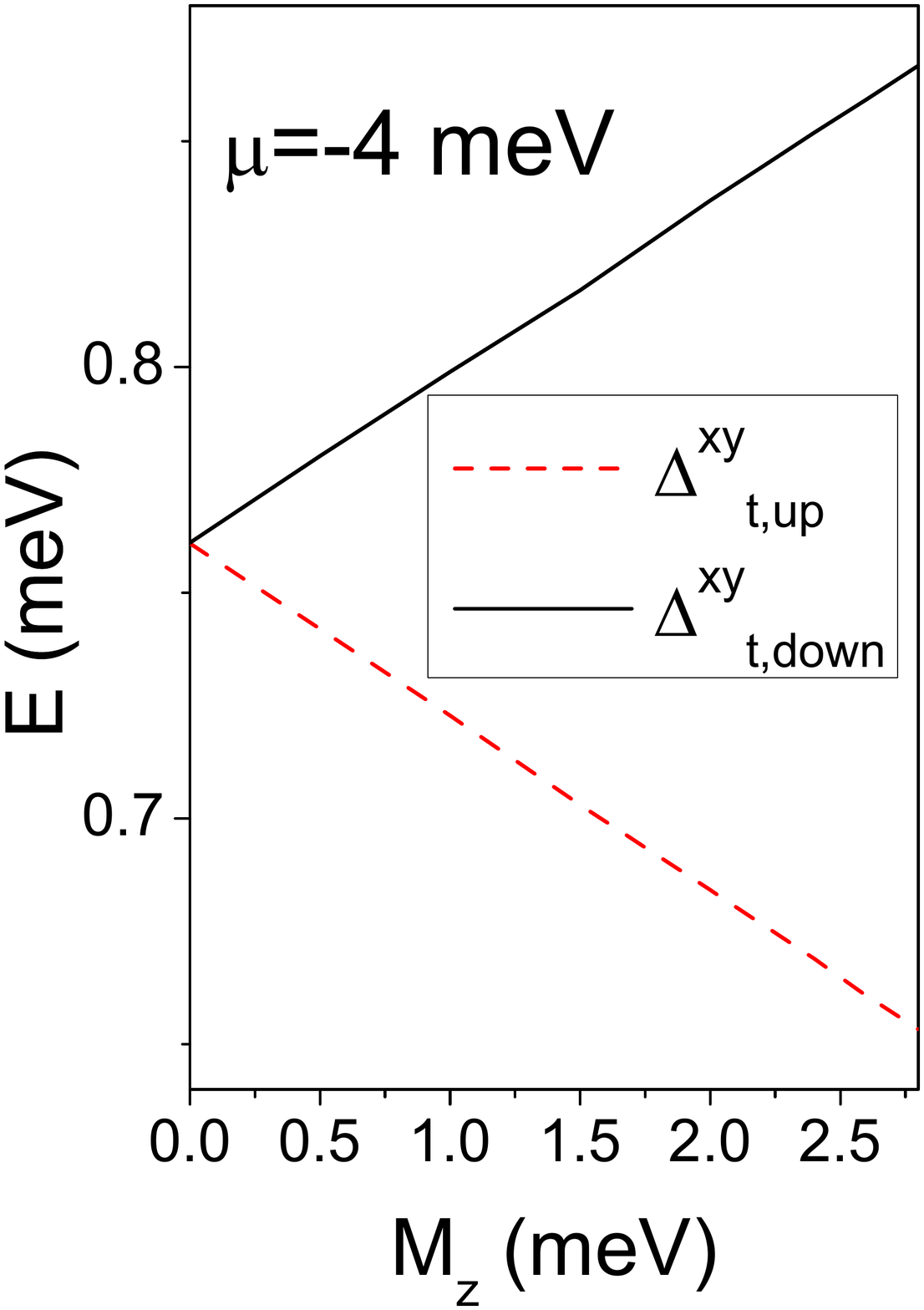}
\caption{Singlet (left panel) and triplet (right panel) pairing amplitudes, as a function of a constant magnetic field ${\bf M}$ along the $z$-axis (out of plane) with 
 $U =350$ meV and $\frac{V}{2} = 300$ meV.
Middle panel: same as in the left panel, but with the magnetic field along $x$-axis (in-plane).}
\label{magnfig2}
\end{figure*}
%%%%%%%%%%%%%%%%%%%%%%%%%%%%%%%%%%%%%%%%%%%%%%%%%%%%%%%%%%%%

In order to deeply analyze the role of the anisotropy of the superconducting solutions, we consider different fillings, comparing the behaviors of the order
parameters between the first dome, beginning at the minimum of the $\eta_0$ band  ($\mu \approx -10$ meV), and the second one, related to the $\eta_+$ band 
($\mu \approx 10$ meV). Therefore, in the left and the middle panel of Fig. \ref{magnfig2}, we consider two values of $\mu$:  $\mu=-4$ meV, 
corresponding to the over-doped regime of the first dome, and $\mu=-11$ meV, corresponding to the under-doped regime of the second dome. In the 
left panel we analyze the behavior of $\Delta_{xy}^s$, as a function of 
$M_z$,  and in the middle panel  we draw a similar plot, 
evidencing the  dependence on  $M_x$. Actually, as shown in Fig. \ref{magnfig2}, at $\mu=-4$ meV, there is an anisotropy favoring the stability along the $z$-axis.  O
n the other hand, as discussed above, the triplet pairing  becomes stronger with increasing density, and it systematically weakens 
the singlet amplitudes. Moreover, the triplet order parameters are more sensitive to the magnetic field  along the $z$-axis (as shown in the right panel of 
Fig. \ref{magnfig2})  than to the one  along the $x$-axis. Therefore, with increasing density, the anisotropy changes drastically. In particular, 
at $\mu=11$ meV, the critical field along $z$  is less than 1 meV, while that along $x$  becomes larger than unity. 
Therefore, the results are in qualitative agreement with experimental  data, \cite{caviglia2009} where the anisotropy is more marked.
Indeed, even if the role of the triplet pairing can be relevant to inducing anisotropy in the superconducting order parameters,  
{\color{black} it is not enough} to explain all the relevant features found in {\color{black} LAO/STO-001} as a function of in-plane and out-of-plane magnetic fields.   

Finally, in the right panel of Fig. \ref{magnfig2}, we have carefully analyzed also the behavior of the triplet pairing,  as a function of $M_z$,
{\color{black} finding a remarkable linear dependence of the triplet} order parameters, opposite for the two spin components: one is enforced, the other one is weakened.

\subsection{Clues of emergent topology in the presence of a magnetic field}

%%%%%%%%%%%%%%%%%%%%%%%%%%%%%%%%%%%%%%%%%%%%%%%%%%%%%%%%%%%%
\begin{figure} [t]
\includegraphics[scale=0.32]{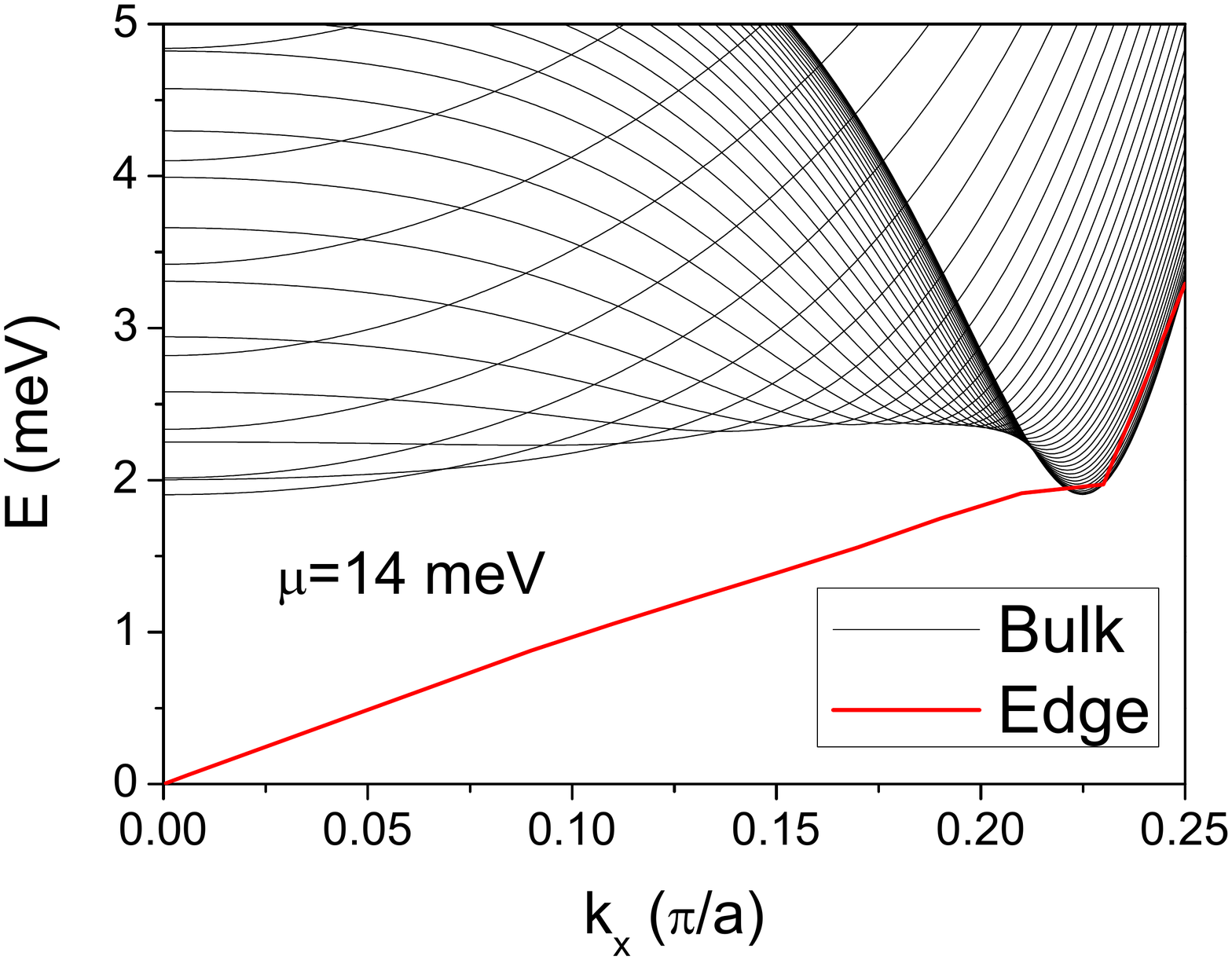}
\includegraphics[scale=0.32]{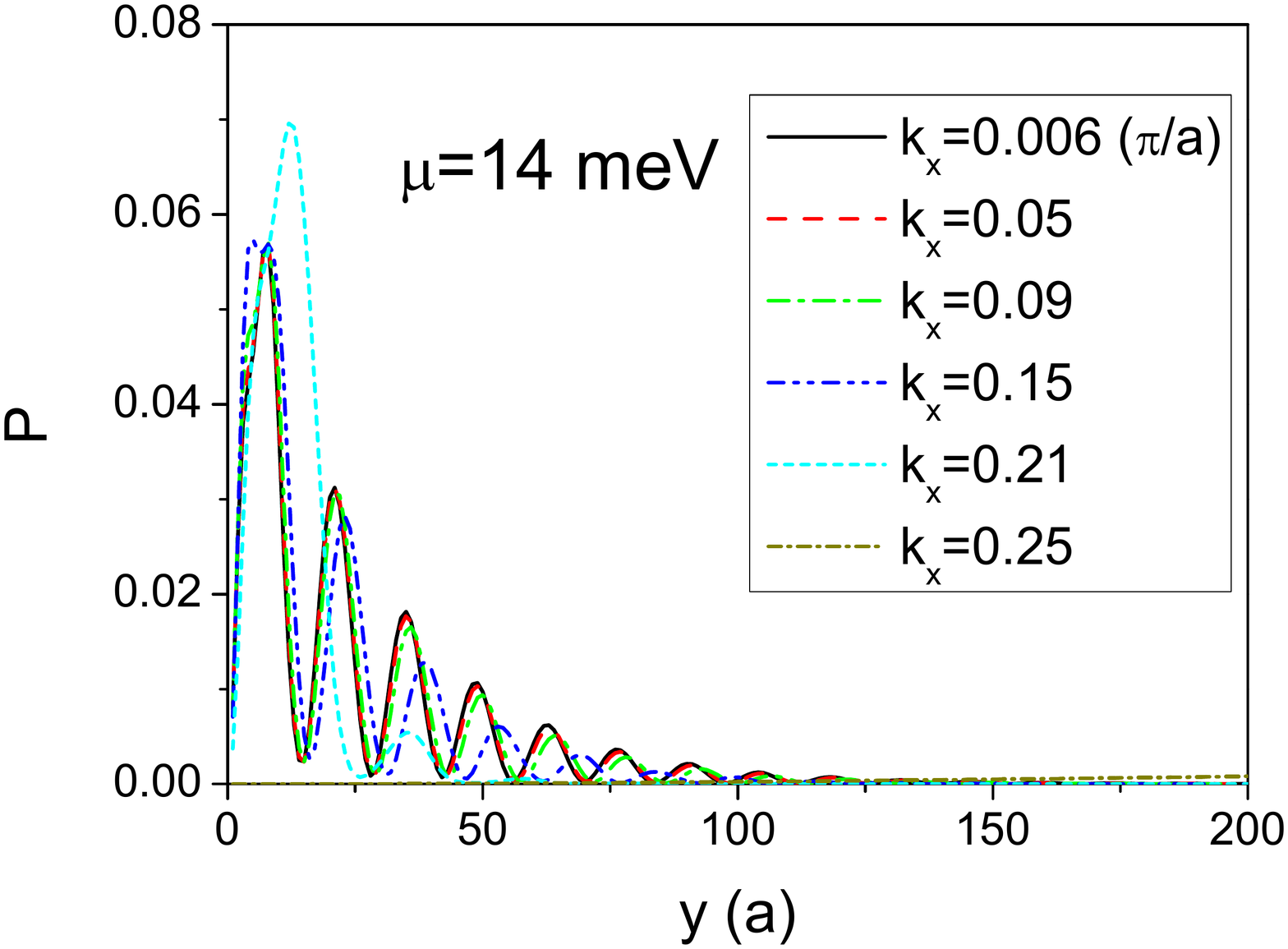}
\caption{Upper panel: Bogoliubov spectrum for varying $k_x$, $V = 0$ meV,  $\mu = 14$ meV and finite size $L_y = 950$
sites along the $y$-axis. Lower panel: probability of the edge state as a function of the position (in units of the lattice parameter $a$)
for different values of $k_x$ (in units of $\pi/a$).  
The value of $M_z$, inducing a topological phase transition (see the main text), is smaller than the critical value suppressing the self-consistent superconducting solutions.}
\label{topofig2}
\end{figure}
%%%%%%%%%%%%%%%%%%%%%%%%%%%%%%%%%%%%%%%%%%%%%%%%%%%%%%%%%%%%%

We now focus on the possibility  of topology in the presence of a magnetic field. In these conditions, the breakdown 
of the time-reversal symmetry makes  the topology class of the system to change from DIII to D. In two dimensions, 
the latter class still allows nontrivial topology, still labelled by a $Z$ index ($Z_2$ for 1D systems; in this case 
the precise calculation of the related invariant has been performed in \cite{perroni2019}). In this way, a magnetic field
can preserve the edge states {\color{black} that we discuss} {\color{black} in Section \ref{topnoB}.

 Still analyzing the possible presence of edge states, we} find {\color{black} that topological properties are favored 
  by an out-of-plane magnetic field, but disfavored by an
 in-plane one}. Indeed, they are weakened with changing the orientation of the magnetic field from the $z$-axis to $x$-axis,
 and there here are characteristic angles (around $\frac{\pi}{4}$) where a topological phase transition takes place. Eventually,
 {\color{black} for a sufficiently intense} in-plane magnetic field, one only gets trivial  topological phases, in agreement with \cite{Loder}.
  
In order to detect the chiral topological superconducting phase when $M_x$ is sub-relevant, we {\color{black} consider again} a lattice with 
a finite size along the  $y$-axis. In this case, the increase of $M_z$ is able to induce a phase transition, {\color{black} as 
soon as it takes a value of the order of the chemical potential, as measured from the bottom of the sub-band: $M_z > \sqrt{\tilde{\mu}^2 + \Delta^2}$, where $\tilde{\mu}$ is the chemical potential measure from the bottom of the $\eta_+$ sub-bands, and $\Delta$ is the effective superconductive gap. 
Thus, the critical values to enter   the topological phase depend in general on $U$, $V$, and $\mu$}. In the upper and lower panels of 
Fig. \ref{topofig2}, we show the results obtained at $V=0$, {\color{black} with only the singlet pairing present}. We remark that, in contrast with
the results of the previous Section at zero magnetic field, now the topological phases can be driven by the singlet order parameters only. 
We find that the chiral Majorana edge states have a linear dispersion as a function of $k_x$, and are quite localized at the edge for small values of $k_x$,
similarly as in  \cite{Loder}.  As usual, with increasing $k_x$, these states merge into the bulk continuum.

Again the described picture can be understood in better qualitative details by analyzing the effective Hamiltonians in Eqs. \eqref{eff-} and \eqref{eff0}, 
{\color{black} in analogy with the discussion at   ${\bf M} = 0$ in Section \ref{supmulti}.}
Since the Zeeman term in Eq. \eqref{e.9} has the form  $H_{M} = {\bf I}_{3 \mathrm{x} 3} \otimes  {\bf M} \cdot {\bf \sigma} $,  it is sufficient to add this 
term to $H_0^{(\mathrm{eff})}$, as described in the Appendix 3 (before Eq. \eqref{e.4}), in spite of the fact  that 
 the $zx$ and $yz$ bands are mixed to yield the $\eta_0$ band: this mixing is diagonal on the spin $\sigma$ index. In Eq. \eqref{eq:HnormalNCS}, 
 the same added term results in the shift ${\bf g} ({\bf k}) \to {\bf g} ({\bf k}) + {\bf M}$, losing explicitly the antisymmetry required by time-reversal invariance \cite{sato2006}.

A Zeeman term, $H^{(z)}_{M} = M_z \sigma_z$, along the $z$-axis, orthogonal to the plane of our system,  
creates an effective unbalance in the chemical potentials {\color{black} with respect to $\sigma$}. This 
imbalance is known to spoil the s-wave pairing {\color{black} $\Delta^{\tau}_{s}$ (as well as $\Delta^{\tau}_{p , \uparrow \downarrow}$},
if present), leading to the normal phase, or at most {\color{black} to a normal-superconductive mixed one} (a secondary possibility that we neglect, 
in the light of the stability of the p-wave pairing, see below) \cite{pethick,marchetti2007,rad2007}. This transition occurs, {\color{black} in the low-pairing limit,}
around the value {\color{black} $M_z = \frac{\mu_{\uparrow} - \mu_{\downarrow}}{2} = \frac{ \Delta^{\tau}_{s}}{\sqrt{2}}$,} the so called
Chandrasekhar-Clogston limit  (strictly valid for perturbative Hubbard attractions) \cite{pethick}.
Concerning the triplet pairings  {\color{black} $\Delta_{t, \uparrow}$ and $\Delta_{t, \downarrow}$,}
they are spoiled asymmetrically by the  $M_z$ term, mainly via the shift of the chemical potentials $\mu_{\uparrow , \downarrow}$.
In particular, if $M_z > 0$, $\mu_{\downarrow}$ is decreased (at fixed {\color{black} $\Delta_{t, \downarrow}$}). At a critical value
of $M_z$, the condition $\mu_{\downarrow} = 0$ is realized. Then, the topological phase with $n_{\downarrow} = -1$ 
is driven to a topologically trivial phase, still with p-wave pairing (the strongly-coupled phase in \cite{readgreen}). At the point  $\mu_{\downarrow} = 0$,
a zero in the Bogoliubov gap is reached. The residual topology from the $\uparrow$ contribution is encoded in the $n_{\uparrow} = 1 \in Z$
phase, belonging to the D class of the ten-fold way classification.

A qualitatively similar phase evolution, determined by the effective unbalances induced by $M_z$, is valid if the phases do not host 
a nontrivial topology, a possibility {\color{black} allowed} by symmetry considerations only, and possibly occurring  
in certain coupling regions, as mentioned in Section \ref{topnoB}.

\section{Conclusions and outlook}
\label{conclusion}

In this paper we have discussed  the onset and the physical consequences of a singlet-triplet  pairing in  the two dimensional 
electron gas at LAO/STO-001 interfaces. This configuration looks rather natural a priori, due to the  inversion symmetry breaking term 
in the tight-binding Hamiltonian of the system. We have made an extended superconducting mean-field analysis of this multi-band 
tight-binding Hamiltonian, as well as of effective electronic bands in the limit of low values of the momentum. We have included 
static on-site (favoring spin-singlet pairings) and inter-site (promoting spin-triplet order parameters) intra-band attractive 
potentials under applied magnetic fields. It is interesting to notice that a singlet-triplet mixed pairing here results 
robust for {\color{black} extended} regions of the analyzed space of parameters.

We have found various interesting features, as a reduction of the singlet order parameter, as a function of
the charge density, an asymmetric response to in-plane and out-of-plane magnetic fields (a fact already observed
experimentally \cite{caviglia2009}), and the possibility on nontrivial topology and protected edge states.
In particular, not-linear spin-orbit couplings and inter-site attractive interactions make stable a time-reversal
invariant topological helical superconducting phase in the absence of a magnetic field.

In this paper, we have discussed the interplay between singlet and triplet order pairings in the the clean limit. Effects of dilute nonmagnetic impurities 
can affect the properties of non-centrosymmetric superconductors \cite{samokhin}. For example, the impurity effects on the critical temperature are similar to those in
multi-band centrosymmetric superconductors. Moreover, Anderson's theorem holds for singlet pairing \cite{samokhin,anderson1959}. Indeed, 
we expect that disorder effects can induce a reduction of the triplet order parameters changing only quantitatively the interplay between singlet and triplet pairings.
This is the reason why in the paper the focus has been on the properties related to the singlet order parameters.

A natural development of the present work is the study of the effective shape of the dome, that means
the finite-temperature dependence. We mention here that we also performed mean-field calculations at finite temperature,
finding a first dome for the singlet order parameter, corresponding with the $\eta_0$ band, and with a  shape qualitatively very 
similar to that reported in \cite{gariglio2016}. However, to achieve reliable quantitative details,  the inclusion of
fluctuations beyond mean field is required. On the same regard, we also mention a very recent paper \cite{jouan2021},
where two-different purely $s$-wave phases have been predicted around the optimal doping, where the critical temperature is maximized.
The relation with the present work surely deserves future attention.
Other future generalizations of the present work are aimed to include the effect of the second term on the multiband potential
in Eq. \eqref{potgen}, that is inter-orbital attractive potentials. Closely related, a relevant issue consists of the possibility of
nonzero-momentum superconductive couplings between different $\eta$ bands. This feature extremely 
unlikely in translational invariant systems \cite{pethick,marchetti2007,rad2007}, is known {\color{black} to be} 
more stable in lattice systems, see e. g. \cite{moore2012,shi2013,mannarelli2018}.

 {\color{black}
Finally, we point out that effects due to unconventional superconductivity can be confirmed not only by thermodynamic properties but also by response properties such as the I-V characteristics of Josephson junctions \cite{tafuri2017,alidoust2021}. In quasi one-dimensional systems,     
 notable properties can be deduced from anomalous Josephson effects \cite{tagliacozzo2015,tagliacozzo2018} and local spectroscopic measurements \cite{perroni_1d}.  These probes could be also used to disentangle superconducting topological features from trivial ones in the actual two dimensional case. }

\section*{Acknowledgements}
The authors thank Michele Burrello, Simone Paganelli, Marco Salluzzo, and Andrea Trombettoni for useful discussions.
L. L., D. G. and C. A. P. acknowledge  financial support  from Italy's MIUR  PRIN project  TOP-SPIN (Grant No. PRIN 20177SL7HC).
C.A.P. acknowledges funding from the project QUAN-TOX (QUANtum Technologies with 2D-OXides) of Quan-tERA ERA-NET Cofund in 
Quantum Technologies (GrantAgreement No. 731473) implemented within the European Union's  Horizon 2020 Programme. 
A. N. was financially supported  by POR Calabria FESR-FSE 2014/2020 - Linea B) Azione 10.5.12,  grant no.~A.5.1.

\newpage

\onecolumngrid

\section*{APPENDIX 1: Hamiltonian symmetries}
In this Appendix, we discuss the symmetries of the normal Hamiltonians in Eqs. \eqref{e.3} and \eqref{eff0}, as well as the corresponding 
Bogoliubov Hamiltonians with the pairing in Eq. \eqref{pairing} added.

The normal Hamiltonians in Eqs. \eqref{e.3} (here considered at vanishing Zeeman coupling $H_M$) can be expressed in terms of 
{\color{black} Gell-Mann} matrices $\{ \lambda_i \}$,
{\color{black}  acting on the on the $\tau$ index, times the Pauli matrices acting on the spin index $\sigma$:}
\beq
\Big(a ({\bf k}) \, {\bf I}_{3 \mathrm{x} 3} + b ({\bf k})  \, \lambda_3 + c ({\bf k}) \,  \lambda_8 \Big) \otimes 
{\bf I}_{2 \mathrm{x} 2} + w_{\mathrm{SO}} \, \Big( - \lambda_2 \otimes \sigma_z +  \lambda_5 \otimes \sigma_y -
\lambda_7 \otimes \sigma_x  \Big) + \gamma \, \Big( \sin k_x \, \lambda_5 \otimes  {\bf I}_{2 \mathrm{x} 2} + 
\sin k_y \, \lambda_5 \otimes  {\bf I}_{2 \mathrm{x} 2} \Big) \, ,
\eeq
the precise expressions for $a ({\bf k})$, $b ({\bf k})$, and $c ({\bf k})$ being unimportant here.
Exploiting now the property for the Pauli matrices, $\{\sigma_i , \sigma_j\} = \delta_{ij}$, it is know immediate to prove (Eq. \eqref{defUT}):
\beq
 H ( {\bf k} ) = U_T^{-1}  H^* (- {\bf k} ) U_T \, , \quad  \quad \quad U_T = {\bf I}_{3 \mathrm{x} 3} \otimes \sigma_y \, .
 \label{defUT2}
\eeq 
In the Nambu-Gorkov basis $\Psi({\bf k}) = \left(\begin{array}{c} {\bf c}({\bf k}) \\ i \sigma_y \, {\bf c}^{\dagger}(-{\bf k})
\end{array} \right)$,  $c({\bf k}) = \left(\begin{array}{c}  c_{\uparrow}({\bf k}) \\ c_{\downarrow} ({\bf k}) \end{array} \right)$,  
$H ( {\bf k} )$ is recast as follows:
\beq
 H ( {\bf k} ) \to  H_{\mathrm{BG}} ( {\bf k} ) = \left(\begin{array}{cc} H({\bf k}) & {\bf 0}_{2 \, \mathrm{x} \, 2} 
 \\ {\bf 0}_{2 \, \mathrm{x} \, 2} & -H (-{\bf k}) \end{array} \right)
 \label{HBog}
\eeq
so that $U_T$ becomes $U_T = {\bf I}_{3 \mathrm{x} 3} \otimes {\bf I}_{2 \mathrm{x} 2} \otimes \sigma_y $,
the second identity in the Kronecker product acting on the Nambu-Gorkov indices.
The normal effective Hamiltonian in Eq. \eqref{eff0} for the $\eta_0$ band, neglecting the out-of-diagonal 
contribution with higher momentum power than one, reads (see the derivation in the Appendix 3):
\beq
H_0^{(\mathrm{eff})} ({\bf k}) =  
\Big( -10.8 +157.2 \, (k_x^2 + k_y^2) \Big) \, {\bf I}_{2 \mathrm{x} 2} - a_3 \, k_x \, \sigma_y + a_3 \, k_y \,  \sigma_x \, .
\eeq
with $\epsilon_0 ({\bf k}) = \big(-10.8 +157.2 \, (k_x^2 + k_y^2) \big)$ meV (so that $t_0^{(\mathrm{eff})} = 157.2$ meV) and $a_3= 0.8$ meV.
It holds:
\beq
H_0^{(\mathrm{eff})} ({\bf k}) = \tilde{U}_T^{-1}  H_0^{(\mathrm{eff}) *} (- {\bf k} ) \tilde{U}_T \, , \quad  \quad \quad \tilde{U}_T = \sigma_y \, ,
 \label{defUTeff}
\eeq
$\tilde{U}_T = \sigma_y$ acting on the spin index, as $\sigma_y$ in Eq. \eqref{defUT2}. In the Bogoliubov form, obtained as 
above for $H ( {\bf k} )$, $\tilde{U}_T =  {\bf I}_{2 \mathrm{x} 2} \otimes \sigma_y$.

The next step is to include the pairing in Eqs. \eqref{pairingmat} and \eqref{pairing}. We have to consider the three spin channels:
\beq
\langle c^{\tau}_{\downarrow} ({\bf k})  c^{\tau}_{\downarrow} (- {\bf k}) \rangle =  e^{i\, \phi^{\tau}_{\downarrow \downarrow}}
\,  {\color{black} \Delta_{t , \downarrow}^{\tau} }(k_x - i \, k_y) \, ,
\label{vev1}
\eeq
\beq
\langle c^{\tau}_{\uparrow} ({\bf k})  c^{\tau}_{\uparrow} (- {\bf k}) \rangle =  e^{i\, \phi^{\tau}_{\uparrow \uparrow} } 
\,  {\color{black} \Delta_{t , \uparrow}^{\tau}} (k_x + i \, k_y) \, ,
\label{vev2}
\eeq
and
\beq
\langle c^{\tau}_{\uparrow} ({\bf k})  c^{\tau}_{\downarrow} (- {\bf k}) \rangle =  e^{i\, \phi^{\tau}_{\uparrow \downarrow}} 
\, \Big({\color{black} \Delta_{s , \uparrow}^{\tau} + \Delta_{t , \uparrow}^{\tau}}  \Big[ \alpha \, (k_x + i \, k_y) + \beta\,  (k_x - i \, k_y) \Big] \Big) \, .
\label{vev3}
\eeq
Two global phase factors between $ e^{i\, \phi^{\tau}_{\uparrow \downarrow}}$, $e^{i\, \phi^{\tau}_{\uparrow \uparrow}}$,
and $e^{i\, \phi^{\tau}_{\downarrow \downarrow}} $ can be reabsorbed overall via a phase redefinition of the  fermionic annihilation 
operators. However, even in this way the third phase cannot be  reabsorbed. This phenomenon a condensed matter counterpart to 
(minimal model for) the CP$\sim$T violation in particle physics, via chiral fermion condensation \cite{weinberg2}.  
We choose to keep the phase factor $e^{i\, \phi^{\tau}_{\uparrow \downarrow}}$. 

{\color{black} We notice that} time-reversal invariance, interchanging the spins, imposes {\color{black} $\Delta_{t , \uparrow}^{\tau} = 
\Delta_{t , \downarrow}^{\tau}$}, a result also found by the minimization procedure of the mean-field free energy in the main text.
{\color{black} Dividing all the parameters in Eq. \eqref{vev3} in real an imaginary parts, it is useful to parametrize them as follows:
\beq
\langle c^{\tau}_{\uparrow} ({\bf k})  c^{\tau}_{\downarrow} (- {\bf k}) \rangle =  \Big(\Delta_{1}^{\tau} + i  \, \Delta_{2}^{\tau}
\Big)+ \Big(\Delta_{3}^{\tau} + i \,  \Delta_{4}^{\tau} \Big)  k_x + \Big(\Delta_{5}^{\tau} + i \,  \Delta_{6}^{\tau} \Big)  k_y     \, .
\label{veve4}
\eeq
Therefore, for each $\tau$ band, we obtain for the $4 \, \mathrm{x} \, 4$ pairing block Hamiltonian $H^{\tau}_{{\bf \Delta}}({\bf k})$ 
of the total one  $H_{{\bf \Delta}}({\bf k})$:
\beq
H^{\tau}_{{\bf \Delta}}({\bf k}) = \Delta_{1}^{\tau} \, \big(\sigma_x \otimes {\bf I}_{2 \mathrm{x} 2} \big) + \Delta_{2}^{\tau} \, 
\big(\sigma_y \otimes {\bf I}_{2 \mathrm{x} 2} \big) + \Big(\Delta_{1}^{\tau} \, k_x +\Delta_{1}^{\tau} \, k_y\Big) \, \big(\sigma_x \otimes \sigma_z \big)  
- \Big(\Delta_{2}^{\tau} \, k_x +\Delta_{4}^{\tau} \, k_y\Big)  \, \big(\sigma_y \otimes \sigma_x \big) \, .
\eeq
Imposing now time reversal invariance of the total multiband Hamiltonian $H_{\mathrm{BG}} ( {\bf k} ) + H_{{\bf \Delta}}({\bf k})$ (see Eq. \eqref{HBog}),  
\beq
H_{\mathrm{BG}} ( {\bf k} ) + H_{{\bf \Delta}}({\bf k}) = V_T^{-1} \big(H^{*}_{\mathrm{BG}} (- {\bf k} ) + H^{*}_{{\bf \Delta}}(- {\bf k}) \big) V_T \, ,
\eeq
we immediately obtain that it must hold
\beq
 V_T =  U_T = {\bf I}_{3 \mathrm{x} 3} \otimes {\bf I}_{2 \mathrm{x} 2} \otimes \sigma_y \, ,
\eeq
and importantly $\Delta_2 = \Delta_4 = \Delta_6 = 0$. That means that all the pairing couplings in Eqs. \eqref{vev1}-\eqref{vev3} must be 
real, as claimed in the main text. At this point is straightforward to obtain that $H_{\mathrm{BG}} ( {\bf k} ) + H^{\tau}_{{\bf \Delta}}({\bf k})$
is also invariant under charge conjugation:
 \beq
H_{\mathrm{BG}} ( {\bf k} ) + H_{{\bf \Delta}}({\bf k}) = - U_C^{-1} \big(H^{*}_{\mathrm{BG}} (- {\bf k} ) + H^{*}_{ {\bf \Delta} }
(- {\bf k}) \big) U_C \, ,  \quad  \quad {\mathrm{with}} \quad \quad  U_C = {\bf I}_{3 \mathrm{x} 3} \otimes \sigma_y \otimes \sigma_y \, .
\eeq
A further symmetry of the Bogoliubov Hamiltonian $\big(H_{\mathrm{BG}} ({\bf k} ) + H_{{\bf \Delta}}({\bf k}) \big)$ is the chiral symmetry
\beq
H_{\mathrm{BG}} ( {\bf k} ) + H_{{\bf \Delta}}({\bf k}) = U_S^{-1} \big(H^*_{\mathrm{BG}} (- {\bf k} ) + H^{*}_{{\bf \Delta}}(- {\bf k}) \big) U_S \,
, \quad \quad U_S = U_T \, U_C = U_C \, U_T \, , \quad \quad  U_S^2 = {\bf I}_{12 \mathrm{x} 12} \, ,  
\eeq 
composed by the product of time- and charge-conjugation symmetries. Chiral symmetry does not change the topology class of the  Bogoliubov Hamiltonian, remaining DIII.
}

\section*{APPENDIX 2: setting the pairing ansatz}

{\color{black}
On a general lattice, the mean field approach to superconductivity proceeds identifying all the possible pairings corresponding to the irreducible representations of the point-group 
symmetry of the lattice: the ansatzs for the pairings are expressed in terms of them \cite{annett}. This procedure is analogous to the expansion in terms of
spherical harmonics in the isotropic, $SO(3)$  invariant,  free space.

The spectra in Eq. \eqref{e.5}, in the absence of the spin-orbit and inversion terms in Eqs. \eqref{e.7} and \eqref{e.8}, display a $D_2$ point group symmetry \cite{dress_book}.
However, due to the presence of the latter two terms, in order to implement the mean field approach, we have still to assume all the pairings allowed 
by the symmetry of the isotropic square lattice. This lattice has (finite) point group symmetry $D_4$, containing $D_2$ and composed by the group $C_4$ of the rotation 
of angles $\theta_n = n \, \frac{\pi}{2}, \, \, n = 0, \dots,3$, and by the reflections around one (vertical or horizontal) axes (see e.g. \cite{dress_book}).
}

 In this case, the basis functions labelling the irreducible representations which the eight-fold regular representation decomposes in, are $k_x^2 + k_y^2$, $k_x^2 -
 k_y^2$, $k_x k_y$, $\Big(\begin{array}{c} k_x \\ k_y \end{array}\Big)$. The first three (parity even)  functions, together with the constant $c$,  can be assumed as
 a basis for the spin-singlet pairings, while the (parity odd) doublet can be assumed as a basis for the spin-triplet pairing. Summing up, the superfluid pairing can be generally assumed as:
\beq
\Delta = \Delta_{s,0} + \Delta_{s,1} \, m(|{\bf k}|) + \Delta_{s,2} \, f(|{\bf k}|) \, \frac{k_x^2 - k_y^2}{|{\bf k}|^2} + \Delta_{s,3} \,  g(|{\bf k}|) \, 
\hat{k}_x \hat{k}_y + {\color{black} \Delta_t} \,  h(|{\bf k}|) \,  (\hat{k}_x + i \hat{k}_y) \, ,
\label{totpar}
\eeq
plus possible powers of these terms (required e. g. in the presence of a strong cubic Rashba-coupling, {\color{black} as in \cite{nakamura2012}}). 
Notice however that pairings involving higher powers than 1 should be less relevant around $k_x = k_y = 0$.
Both in the effective theories for $\epsilon_-$ and $\epsilon_0$ bands, {\color{black} it is expected that the spin-orbit interaction (including 
the relevant not-linear corrections) plays a critical role in determining the  pairings that set in}. The effect of a linear spin-orbit coupling has been studied in \cite{rashba2001}, where a
mixing (singlet-triplet and s-p wave) has been identified. The same situation could be expected here. However, the not-linear spin-orbit couplings could change the picture.

\section*{APPENDIX 3: derivation of the effective theories in Eqs. \eqref{eff-} and  \eqref{eff0}}

In order to probe the presence of non-BCS pairings, suggested by recent works, it is useful to restart considering the structure of the 
$\epsilon_{yz}$, $\epsilon_{zx}$, and $\epsilon_{xy}$  bands (in the absence of Zeeman couplings), as well as the real bands  $\epsilon_{+,i}$,
$\epsilon_{0,i}$, and $\epsilon_{-,i}$ (in the following we will neglect the degeneracy index $i$, for sake of brevity) resulting from the mixing 
of them, according to the Hamiltonian in Eq. \eqref{e.3}. The plots of them are given in Fig. \ref{bande}.

It is clear that the mixing of the  $\epsilon_{yz}$, $\epsilon_{zx}$, and $\epsilon_{xy}$ is relevant only around their band-touching points, 
around the momenta $k_x = k_y = 0$ and 
$k_x = k_0 \approx \pm 0.35 , k_y = 0$, as expected from the low ratios $\frac{\gamma}{t_1}$ and $\frac{\Delta_{so}}{t_1}$ in Eq. \eqref{e.3}. 
Clearly, the importance of the same mixing relatively to the three bands depends on which ones are touching each others: around $k_x = 0$, 
the bands $\epsilon_{yz}$, $\epsilon_{zx}$ undergo an important mixing, while around the point  $k_x = k_0 , k_y = 0$ ($k_x = 0 , k_y \approx \pm 0.35$),
the bands $\epsilon_{zx}$, and $\epsilon_{xy}$ ($\epsilon_{yz}$, and $\epsilon_{xy}$) do.

We focus first around the point $k_x = k_y = 0$. In this region, due to the relatively small ratios $\lambda_1 = 
\frac{\gamma}{{\color{black} E_t}} = 0.4$ and $\lambda_2 =\frac{\Delta_{so}}{{\color{black} E_t}} = 0.2$ effective expressions 
can be obtained for $\epsilon_+$, $\epsilon_0$, and $\epsilon_-$, exploiting second-order perturbation theory (in standard notation):
\beq
E_n (\lambda) = E_n^{(0)}  +
 \lambda^2 \sum_{k \neq n} 
 \frac{| \langle k^{(0)}  | V | n^{(0)} \rangle |^2}
 {E_n^{(0)} - E_k^{(0)}} 
+ O(\lambda^3) \, ,
\label{pert}
\eeq
and, in the presence of two perturbations $\lambda_1 \, V_1$ and  $\lambda_2 \, V_2$:
\beq
E_n (\lambda) = E_n^{(0)}  +
{\color{black} E_t}^2 \, \sum_i \lambda_i^2 \sum_{k \neq n} 
 \frac{| \langle k^{(0)}  | V_i | n^{(0)} \rangle |^2}
 {E_n^{(0)} - E_k^{(0)}} + {\color{black} E_t}^2 \, \lambda_1 \lambda_2 \, \sum_{k \neq n} 
\Bigg( \frac{\langle k^{(0)}  | V_1 | n^{(0)} \rangle \langle n^{(0)}  | V_2| k_0^{(0)} \rangle}
 {E_n^{(0)} - E_k^{(0)}} + \mathrm{H. c.} \Bigg)
+ O(\lambda^3) \, .
\label{pert2}
\eeq

\subsection*{Lower band}
We start with the derivation of the effective theory for the  $\eta_-$ band. Exploiting Eq. \eqref{pert2}, the couplings of  $\epsilon_{xy}$ 
with $\epsilon_{yz}$ and  $\epsilon_{zx}$, included perturbatively, yields to
\beq
H_-^{(\mathrm{eff})} = \Big(\epsilon_{xy} + {\color{black} E_t}^2 \, \frac{\lambda_1^2 \sin^2  k_x+   \lambda_2^2}{\epsilon_{xy} - \epsilon_{yz}} 
+{\color{black} E_t}^2 \,  \frac{\lambda_1^2 \sin^2  k_y+  \lambda_2^2}{\epsilon_{xy} - \epsilon_{zx}}\Big) \, {\bf I} + 2 \, {\color{black} E_t}^2 \, 
\lambda_1 \, \lambda_2 \, \Big( \frac{\sigma_y \sin k_x}{\epsilon_{xy} - \epsilon_{yz}} - \frac{\sigma_x \sin k_y}{\epsilon_{xy} - \epsilon_{zx}}  \Big)+  O(\lambda_i^3) \, ,
\eeq
and, expanding in powers of $k_x$ and $k_y$:
\beq
H_-^{(\mathrm{eff})}  ({\bf k}) = \epsilon_- ({\bf k}) \, {\bf I}  -  (a_1 \, k_x + a_2 \,  k_x^3)  \, \sigma_y +  (a_1 \, k_y + a_2 \, k_y^3)  \, \sigma_x \, ,
\label{eff-bis} 
\eeq 
with $a_1 = 8$ meV (in units of the lattice step $a \equiv 1$), $a_2 = 43.46$ meV, $ \epsilon_- ({\bf k}) = \big(- 54. + 280.8 \, (k_x^2 + k_y^2) \big)$
meV (so that  $t_-^{(\mathrm{eff})} = 280.8$ meV). 
The spectrum of the effective Hamiltonian in Eq. \eqref{eff-bis}, compared with the exact one, is shown in Fig. \ref{bande2}. The agreement is excellent around $k_x = k_y = 0$.

\subsection*{Central band}

The derivation of the effective theory for the  $\eta_0$ band is more subtle. Indeed, the  $\epsilon_{yz}$, $\epsilon_{zx}$ are degenerate at  $k_x = k_y = 0$, 
then before applying second-order perturbation theory, some elaboration of the Hamiltonian \eqref{e.3} is required. In particular,  we notice that $H_Z ( {\bf k} )$ in  Eq. \eqref{e.8} do not mix 
$\epsilon_{zx}$  with $\epsilon_{yz}$. Therefore we start diagonalizing {\it exactly}  the partial Hamiltonian  $H_0 ( {\bf k} ) + H_{SO}$ 
in the subspace $\epsilon_{zx}$  with $\epsilon_{yz}$, by a unitary transformation $O = \mathrm{diag} (O_{2 \mathrm{x} 2} , 1)$). This transformation 
does not mix the spins, {\color{black} therefore $O$ can ne also written as $O = \tilde{O}_{2 \mathrm{x} 2} \otimes {\bf I}_{2 \mathrm{x} 2}$.} In this way, we obtain
\beq
H_0^{\prime} + H_{SO}^{\prime \, (\mathrm{rid})}  = \left[ \begin{array}{ccc}
\epsilon_{a} \, {\bf I} & {\bf 0} & {\bf 0 } \\
{\bf 0} & \epsilon_{b} \, {\bf I} & {\bf 0} \\
{\bf 0} & {\bf 0} & \epsilon_{xy} \, {\bf I} 
                         \end{array} \right] 
\;\;\;\; ,
\label{e.4t}
\eneq
\noindent\, 
with $\epsilon_a < \epsilon_b$ around $k_x = k_y = 0$. {\color{black} At this level, the spins get mixed each others. The bands $\epsilon_{a}$,
$\epsilon_{b}$, and $\epsilon_{xy}$ are coupled further by the terms from $H_{SO}$ and $H_Z({\bf k})$, rotated by $O$:
\beq
H_0^{\prime} + H_{SO}^{\prime} + H^{\prime}_Z  = \left[ \begin{array}{ccc}
\epsilon_{a} \, {\bf I} & {\bf 0} &  V  \\
{\bf 0} & \epsilon_{b} \, {\bf I} & W \\
 V^{\dagger}  &  W^{\dagger} & \epsilon_{xy} \, {\bf I} 
                         \end{array} \right] \, ,
\label{e.4bis}
\eneq
where
\beq
 V ( {\bf k})  \equiv V = \left[ \begin{array}{cc}
 a( {\bf k})  &  b( {\bf k})   \\
 {} \\
 -  b^*( {\bf k})   &  a( {\bf k}) \\
 \end{array} \right] 
\eeq
is the matrix mixing the $\epsilon_a$ and  $\epsilon_{xy}$ bands, while $W$ is the matrix mixing $\epsilon_{b}$ and $\epsilon_{xy}$.
The precise expressions for its elements are rather involved and not immediately relevant here. No further coupling
between $\epsilon_{b}$ and $\epsilon_{a}$ occurs from $H_{SO}$ and $H_Z({\bf k})$.

The effect of $V$ and $W$ is introduced perturbatively, as above, obtaining for $\epsilon_{a}$:
\beq
H_0^{(\mathrm{eff})}  = \epsilon_a \, {\bf I} + \frac{1}{\epsilon_a - \epsilon_{xy}} \, 
\left[ \begin{array}{cc}
 |a( {\bf k})|^2 + | b( {\bf k})|^2  & - a( {\bf k})  b( {\bf k}) + a^*( {\bf k})  b( {\bf k})   \\
 {} \\
 - a^* ( {\bf k})  b^*( {\bf k}) + a( {\bf k})  b^*( {\bf k})   &  |a( {\bf k})|^2 + | b( {\bf k})|^2 \\
 \end{array} \right] \,  \, 
\eeq}

\begin{figure}[h!]
\includegraphics[scale=0.4]{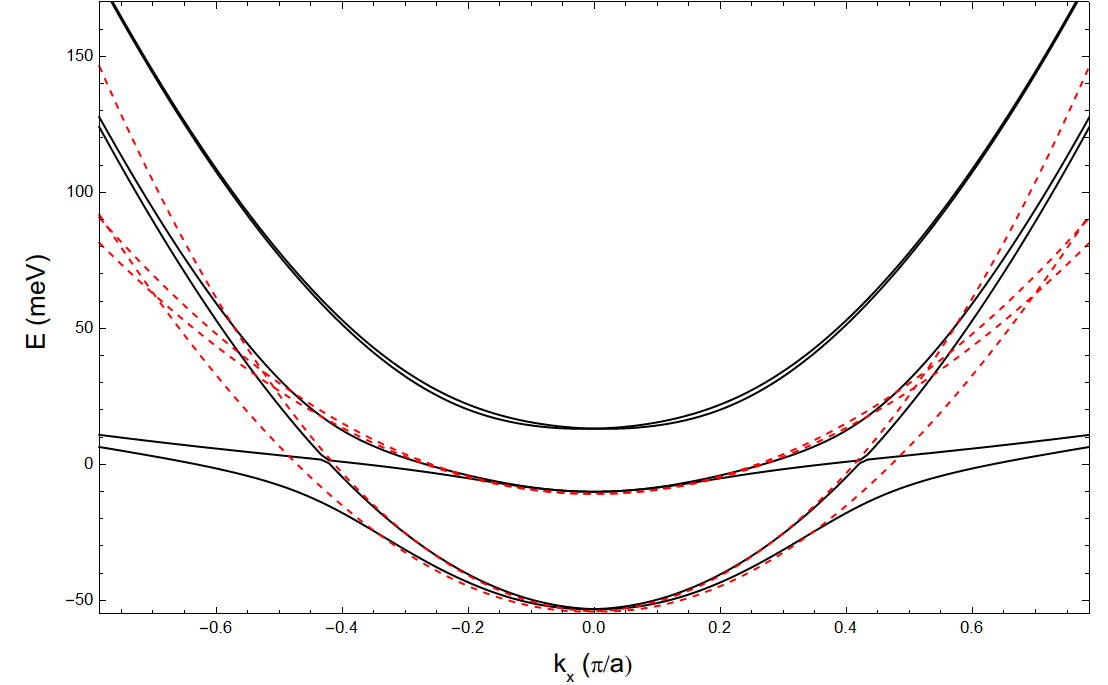}
\caption{Comparison of the exact bands from Eq. (\ref{e.1}) (black lines) with the bands of the effective Hamiltonians in Eqs. \eqref{eff-} and \eqref{eff0} (dashed lines). 
}
\label{bande2}
\end{figure}

Expanding all these expressions around $k_x = k_y = 0$, we obtain approximately (and up to momentum powers bigger than 3):
\beq
H_0^{(\mathrm{eff})} ({\bf k}) =  
\left[ 
\begin{array}{c}
\epsilon_0 ({\bf k})  \quad  \quad a_3 \, (i k_x+ k_y)  - a_4 \, (i k_x -k_y)^3 - a_5 \, k_x k_y (k_x + i k_y)  \\
 {} \\
a_3\, (- i k_x+ k_y)  + a_4 \, (i k_x + k_y)^3 - a_5 \, k_x k_y (k_x -i k_y)   \quad \quad \epsilon_0({\bf k}) \\
 \end{array} 
 \right] \, ,
 \label{eff0bis}
\eeq
with $a_3= 0.8$ meV, $a_4 = 8.627$ meV, $a_5 = 22.8$ meV and  $\epsilon_0 ({\bf k}) = \big(-10.8 +157.2 \, (k_x^2 + k_y^2) \big)$ meV (so that $t_0^{(\mathrm{eff})} = 157.2$ meV).
We see that the dispersion is very close to that before the perturbative mixing. The spectrum of the effective Hamiltonian in Eq. \eqref{eff0}, 
compared with the exact one, is shown in Fig. \ref{bande2}. The agreement is excellent around $k_x = k_y = 0$.\\

Adding a Zeeman term as in Eq. \eqref{e.9} is rather straightforward. Indeed, since the Zeeman term in the same equation has the form 
{\color{black}$H_{M} = {\bf I}_{3 \mathrm{x} 3} \otimes  {\bf M} \cdot {\bf \sigma} $,  it is sufficient to add the term  $\bf{M} \cdot \bf{\sigma}$ } to $H_-^{(\mathrm{eff})}$.

The same situation is realized for  $H_-^{(\mathrm{eff})}$, since, as we described before Eq. \eqref{e.4t}, the matrix $O$, mixing the $zx$ 
and $yz$ bands, is diagonal in the spin index $\sigma$, $O = \tilde{O}_{2 \mathrm{x} 2} \otimes {\bf I}_{2 \mathrm{x} 2}$.
In Fig. \ref{bande2}, we compare the exact bands from Eq. (\ref{e.1}) with the bands of the effective Hamiltonians in Eqs. (\ref{eff-},\ref{eff0}).

{\color{black}
\section*{APPENDIX 4: Additional pairings}
 In this Appendix , we discuss additional pairings which could be considered in the system.

In Eq. \eqref{hamilpair},  an  inter-site attraction between particles with opposite spins can be included, as well. Correspondingly,
we considered also a triplet term related to a pairing between opposite spins.
However, the self-consistent solution of the superconducting phase showed that this term is always zero in the range of $U$ and $V$ that we have considered in this paper. 
Actually, in the  triplet pairing  representation, this term would correspond to a vector $d_z ({\bf k})$, which  comes out to  vanish. This is 
quite typical of two-dimensional superconductors  with in-plane spin-orbit coupling \cite{frigeri2004,sato2010}, such as the one we consider here, while an  out-of-plane spin-orbit coupling tends to favour a nonzero $d_z$.
That  $d_z({\bf k})=0$ can be  already inferred from Eq. \eqref{eff0}. Indeed, in \cite{frigeri2004} it has been shown that 
the superconducting transition
temperature is maximized when the spin-triplet pairing vector $d({\bf k})$ (see Eq. \eqref{pairing}) is aligned with the polarization vector ${\bf g}({\bf k})$ parametrizing 
the spin-orbit coupling \big($H_{\mathrm{SO}} ({\bf k}) = {\bf g}({\bf k}) \cdot {\bf \sigma}$\big). For $H_0^{(\mathrm{eff})}  ({\bf k})$ in  Eq. \eqref{eff0} 
(neglecting the out-of-diagonal terms with power in the momenta higher than one, subleading around ${\bf k} = 0$), we have 
\beq
H_0^{(\mathrm{eff})} ({\bf k}) \approx \epsilon_0 ({\bf k})  \, {\bf I}_{2 \mathrm{x} 2}  + g_1 ({\bf k}) \, \sigma_x + g_2 ({\bf k}) \, \sigma_y \, ,
\eeq
with $g_1({\bf k}) = 0.8 \, k_x$ and $g_2({\bf k}) = - 0.8 \, k_y$, and $g_3({\bf k}) = 0$. Therefore it is expected that  $d_z({\bf k}) = 0$.

Finally, it is worth commenting on the fact that, strictly speaking, we make the ansatz in Eq. \eqref{hamilbogo}  
for the unrotated  $\tau$-bands.  However, the same ansatz can be adopted  at least for the low-density regime 
of the $\eta_-$ and especially $\eta_0$ bands, the latter doublet being the regime where superconductivity is postulated \cite{gariglio2016}. 
 Indeed, around ${\bf k} = 0$, one gets (see Appendix 3 for details),
\beq
\eta_{0 , \sigma} ({\bf k}) = \alpha_{(yz) , \sigma} ({\bf k}) \, c_{(yz) , \sigma} ({\bf k}) + \alpha_{(zx) , \sigma} ({\bf k}) \, c_{(zx) , \sigma} \, ({\bf k}) \, ,
\eeq 
 with $\big( \tau = (yz , zx)$\big):
\beq
|\alpha_{\tau , \sigma} ({\bf k} \to 0) | = \frac{1}{\sqrt{2}} + O(k_x^2) + O(k_y^2) \, .
\eeq 
 Therefore, since the same mapping is also diagonal in $\sigma$, it preserves, up to phases, the structure of the s-p pairing 
(at most linear in the momenta), at least around ${\bf k} = 0$.  This behaviour is even strengthened for the $\eta_-$ band, that results
from the mixing around ${\bf k} = 0$ of the $xy$ bands with the others:  in fact,
this mixing is suppressed by the energy gap $E_t$ in Eq. \eqref{e.5}.

It is also worth stressing that we do not add a  spin-singlet intersite contribution. Indeed, our analysis focuses on the regime of low filling, where the occupied electronic states are close to ${\bf k} = 0$.
In the small momentum limit, the spin-singlet intersite term would give rise to  extended s-wave order parameter (ruled by a sum of cosines of the momentum) which would result in just a subleading additive contribution to the standard s-wave parameter induced by on-site attractive terms. Therefore, we neglected it, adding instead the leading nonzero contribution with equal spins (that is also more relevant once one turns on a magnetic field).

\section*{APPENDIX 5: Effects of inversion-symmetry breaking term}
\label{APP_SPIN}
}

In Section \ref{MFsol} of the main text, we have found a large regime for $U$ and $V$, where triplet and singlet pairings coexist in the first dome.
Then we inferred that this mixing, allowed by $V$, stems from the parallel
contributions of the spin-orbit term in Eq. \eqref{e.7}  and of the inversion breaking one in Eq. \eqref{e.8}. Indeed, at the effective level, 
they result together in the Rashba-like coupling of Eq. \eqref{eff0}, known to induce a mixing of pairings with different parity \cite{rashba2001}.
More in detail, in the same regime of energies, the contribution of Eq. \eqref{e.7} is dominant on that of Eq. \eqref{e.8} (vanishing at ${\bf k} = 0$), a fact can be inferred also in 
the direct construction of Eq. \eqref{eff0}.

\begin{figure}[h!]
\includegraphics[scale=0.30]{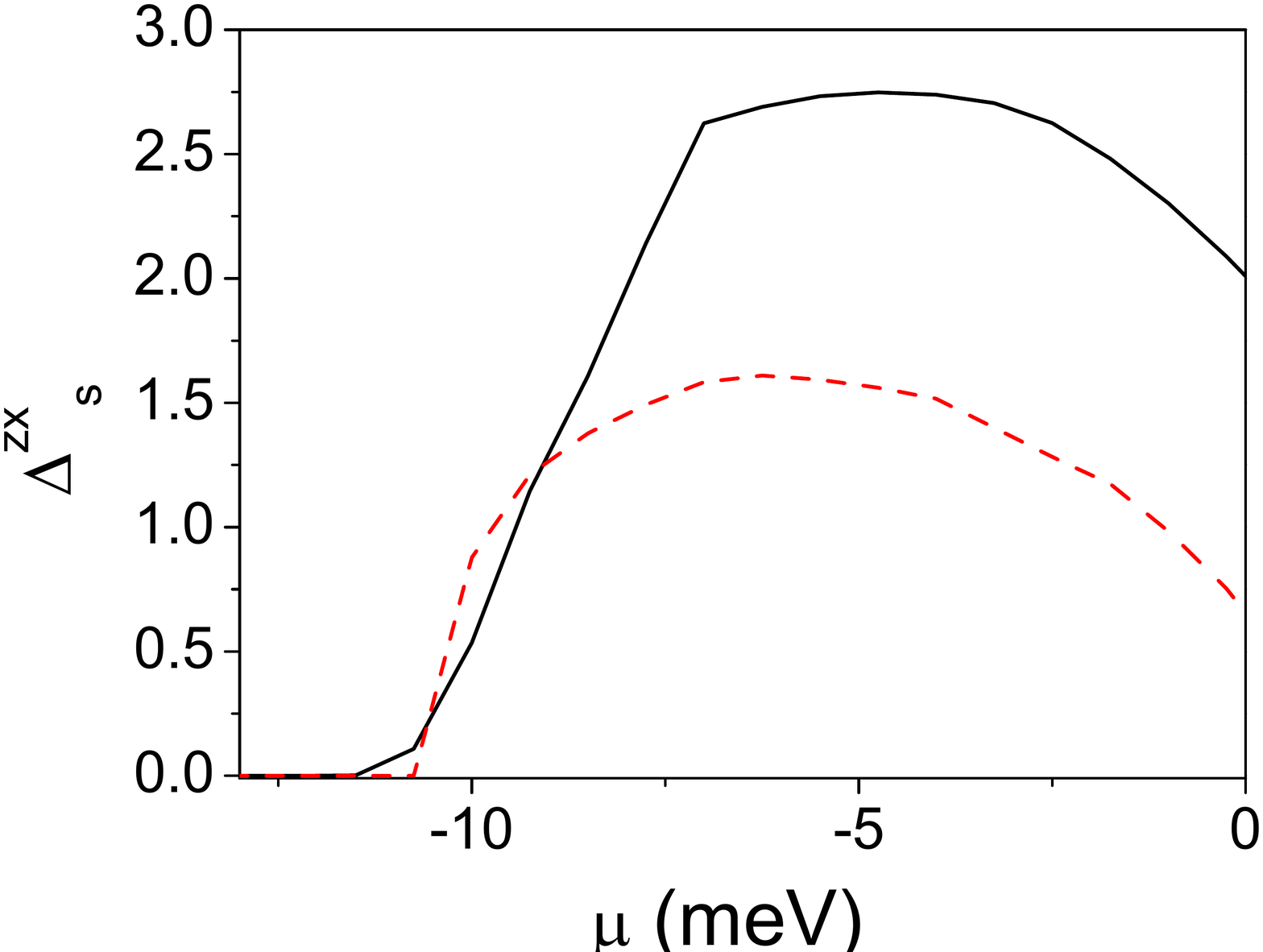}
\includegraphics[scale=0.30]{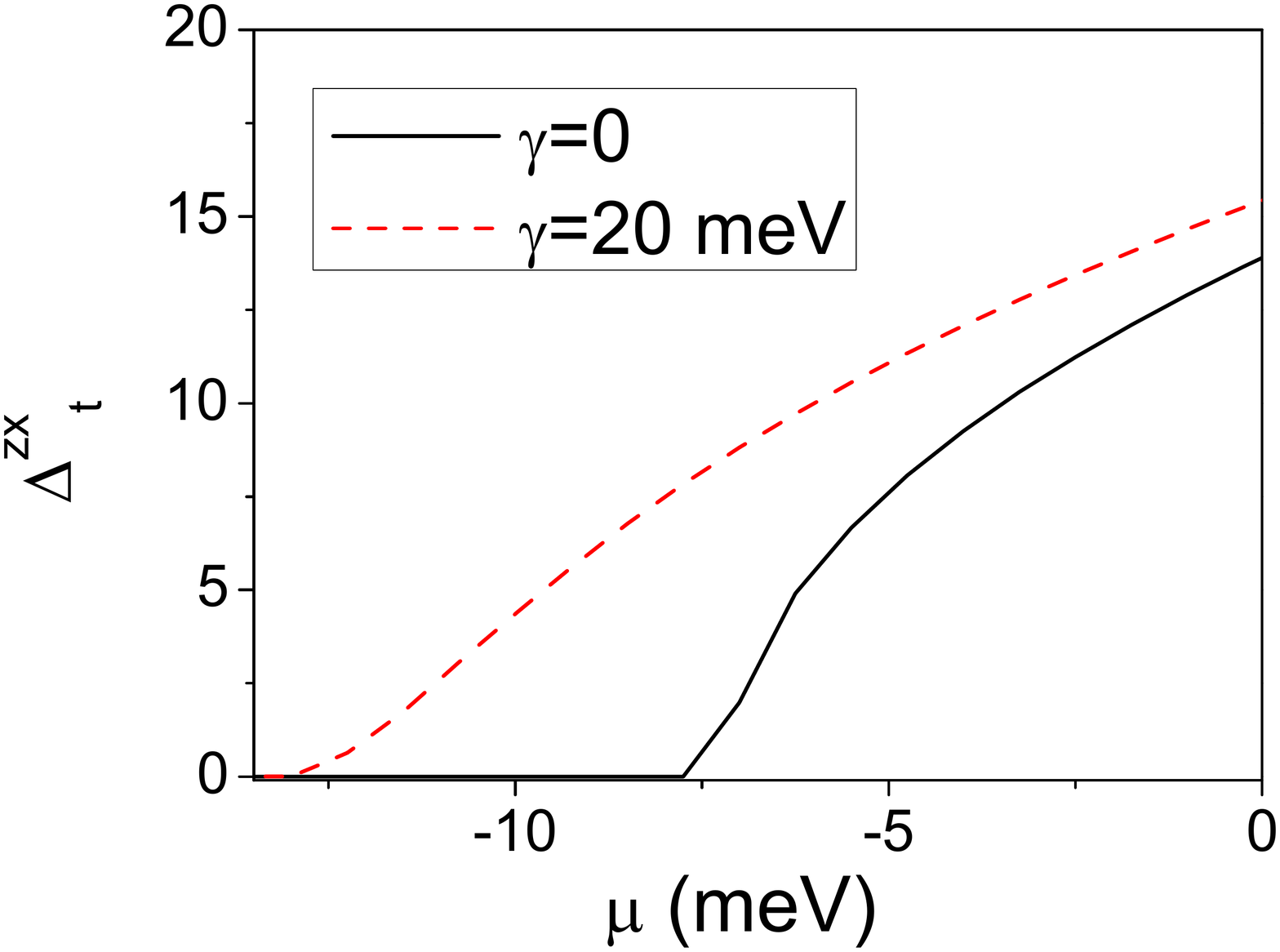}
\caption{{\color{black} Singlet (Left panel) and triplet (right panel) pairing amplitudes (in meV) for the $\tau=zx$ bands, as a function of the chemical potential (in meV). 
The attractive couplings in  Eq. \eqref{hamilpair} are set as $U =350$ meV and $\frac{V}{2} = 290$ meV.}}
\label{SOC}
\end{figure}

It is important to investigate directly this effect of the  spin-orbit term in Eq. \eqref{e.7} on the mixing of the pairings. For this purpose, we
repeat the mean field procedure performed above, switching off the same term.
In the resulting Fig. \ref{SOC}, again at fixed $U = 350$ meV, $\frac{V}{2} = 290$ meV, and $\mu = -9$ meV, it appears clear that Eq. \eqref{e.7}
collaborates to enforce the triplet pairing, correspondingly lowering the singlet component. However, this effect does not look critical, changing the 
previously found behaviours only quantitatively.
Therefore, the major role to the singlet-triplet mixing seems, in the analyzed regimes, to result from the $V$ term of the potential in Eq. \eqref{hamilpair},
even for chemical potentials close to the Lifshitz point where the first dome starts.  In turn, the $V$ term can be ascribed to the relatively 
low charge densities at the location of the dome. For these reasons, similar results are obtained at higher chemical potentials going from the first to the second dome. 
We ascribe the present result to the ability, described in Section \ref{super}, of the mean-field approach to grasp the interplay between singlet and triplet components.
%%%%%%%%%%%%%%%%%%%%%%%%%%%%%%%%%%%%%%%%%%%%%%%%%%%%%%%%%%% 
%%%%%%%%%%%%%%%%%%%%%%%%%%%%%%%%%%%%%%%%%%%%%%%%%%%%%%%%%%%%

\newpage

\twocolumngrid

\end{document}